\RequirePackage[british]{babel}
\pdfoutput=1
\documentclass[reqno,a4paper,12pt]{amsart}
\usepackage[a4paper,hmargin=2cm,tmargin=3cm,bmargin=3cm]{geometry}
\usepackage[utf8x]{inputenc}
\usepackage{amsmath,amssymb,amstext,amsthm,amscd,mathrsfs,eucal,bm}
\usepackage{graphicx,color}
\usepackage{array}
\usepackage{cite}
\usepackage{hyperref}
\hypersetup{%
  pdftitle   = {On the structure of quadrilateral brane tilings},
  pdfkeywords = {Calabi-Yau, toric variety, moduli space, supersymmetry, gauge theory, brane tilings, quiver, eulerian, graph theory},
  pdfauthor  = {Paul de Medeiros},
  pdfcreator = {\LaTeX\ with package \flqq hyperref\frqq}
}
\PrerenderUnicode{éÉ}
\newcommand{\cG}{\mathcal{G}}
\newcommand{\cM}{\mathcal{M}}

\newcommand{\fd}{\mathfrak{d}}

\newcommand{\fF}{\mathfrak{F}}
\newcommand{\fG}{\mathfrak{G}}

\newcommand{\fsu}{\mathfrak{su}}
\newcommand{\fu}{\mathfrak{u}}

\newcommand{\RR}{\mathbb{R}}
\newcommand{\CC}{\mathbb{C}}

\newcommand{\ZZ}{\mathbb{Z}}

\newcommand{\eQ}{\mathscr{Q}}

\theoremstyle{plain}

\theoremstyle{definition}

\newcommand{\MUNCH}[1]{\relax}

\allowdisplaybreaks
\begin{document}
\title[On the structure of quadrilateral brane tilings]{On the structure of quadrilateral brane tilings}
\author[de Medeiros]{Paul de Medeiros}
\address{School of Mathematics and Maxwell Institute for Mathematical Sciences, University of Edinburgh, Scotland, UK}
\email{p.demedeiros@ed.ac.uk}
\date{\today}
\begin{abstract}
Brane tilings provide the most general framework in string and M-theory for matching toric Calabi-Yau singularities probed by branes with superconformal fixed points of quiver gauge theories. The brane tiling data consists of a bipartite tiling of the torus which encodes both the classical superpotential and gauge-matter couplings for the quiver gauge theory. We consider the class of tilings which contain only tiles bounded by exactly four edges and present a method for generating any tiling within this class by iterating combinations of certain graph-theoretic moves. In the context of D3-branes in IIB string theory, we consider the effect of these generating moves within the corresponding class of supersymmetric quiver gauge theories in four dimensions. Of particular interest are their effect on the superpotential, the vacuum moduli space and the conditions necessary for the theory to reach a superconformal fixed point in the infrared. We discuss the general structure of physically admissible quadrilateral brane tilings and Seiberg duality in terms of certain composite moves within this class.
\end{abstract}
\maketitle

\clearpage

\vspace*{.2cm}

{\small{\tableofcontents}}

\clearpage

%%%%%%%%%%%
\section{Introduction}
\label{sec:introduction}

Many interesting supersymmetric quantum field theories in four dimensions undergo non-trivial renormalisation group flow to interacting superconformal fixed points in the infrared. A large class of such theories describe the low-energy behaviour of D3-branes in type IIB string theory. For D3-branes probing a conical Calabi-Yau singularity, the bijection between strongly coupled superconformal fixed points and weakly curved background geometries in IIB supergravity represents by far the most exhaustively studied and best understood class of AdS/CFT dualities \cite{Malda,Gubser:1998bc,Witten:1998qj,KlebanovWitten,AFHS,Morrison:1998cs,AdSCFTReview}. Near the singularity, the transverse space to the D3-branes is an affine Calabi-Yau cone over a Sasaki-Einstein 5-manifold $X$ in the dual $AdS_5 \times X$ supergravity background which emerges in the near-horizon limit. For a single D3-brane, the gauge group in the field theory is abelian and holography identifies the Calabi-Yau cone with a particular branch in the space of gauge-inequivalent superconformal vacua. In general, there can exist a number of different supersymmetric field theories in the ultraviolet which flow in the infrared to the same superconformal fixed point dual to a given Calabi-Yau singularity. These different phases of the dual field theory are related by Seiberg duality \cite{Seiberg:1994pq}. 

The most extensive checks of various aspects of the correspondence in this context have been achieved in cases where $X$ is toric \cite{Malda,KlebanovWitten,Uranga:1998vf,Martelli:2004wu,Benvenuti:2004dy,Herzog:2004tr,Hanany:2005hq}. In such cases, the techniques developed in \cite{Douglas:1996sw,Douglas:1997de,Beasley:1999uz,Feng:2000mi,Feng:2001xr} can be utilised to determine from the data for the toric Calabi-Yau singularity both the superpotential and the gauge-matter couplings in the dual field theory in terms of a quiver representation of the gauge group. The quiver representation is specified by a directed graph with each vertex $i$ assigned a positive integer $N_i$. If the directed graph has $n$ vertices then the gauge group is $\prod_{i=1}^n U( N_i )$ and an arrow in the directed graph pointing from vertex $i$ to $j$ corresponds to a chiral matter superfield in the bifundamental representation of $U( N_i ) \times U( N_j )$. Seiberg duality for supersymmetric quiver gauge theories has been investigated in \cite{Feng:2002zw,Berenstein:2002fi,Franco:2003ja,Franco:2005rj} and identified with a certain geometric duality for toric Calabi-Yau manifolds in \cite{Feng:2000mi,Beasley:2001zp,Feng:2001bn,Cachazo:2001sg}. As will be recalled in section~\ref{sec:branetilings}, it is defined with respect to a given vertex and specifies an involution between two quivers such that the underlying directed graphs are related by a mutation. If $A_{ij}$ denotes the number of arrows pointing from vertex $i$ to $j$ in the directed graph underlying the quiver representation of $\prod_{i=1}^n U( N_i )$, Seiberg duality with respect to $i$ also maps $N_i \mapsto -N_i + \sum_{j\neq i} A_{ij} N_j$ with all other $N_j$ the same in the new quiver.

A convenient framework for matching toric Calabi-Yau singularities with superconformal fixed points for supersymmetric quiver gauge theories is provided by brane tilings \cite{Hanany:2005ve,Franco:2005rj,Feng:2005gw,Franco:2006gc,Franco:2005sm,Hanany:2005ss,Kennaway:2007tq,Yamazaki:2008bt,Jejjala:2010vb,Hanany:2011ra}. As will be reviewed in section~\ref{sec:branetilings}, a brane tiling can be thought of as a particular supersymmetric configuration of NS5- and D5-branes in IIB string theory whose intersection describes a bipartite tiling of the torus. As will be discussed in section~\ref{sec:bipartitetilingsT2}, this tiling can be thought of as a particular kind of graph drawn on the torus with no edges crossing. Each edge in the tiling encodes an arrow for the quiver. Each tile forms a face bounded by edges and encodes a vertex for the quiver (such that edges bounding a given face encode arrows connected to the corresponding vertex). Bipartite means that each vertex in the tiling is coloured either white or black such that no edge connects two vertices with the same colour. Being bipartite defines an orientation for the quiver and implies that the underlying directed graph must be connected with precisely the same number of incoming and outgoing arrows at each vertex. Any such directed graph is necessarily eulerian and this useful characterisation will be recalled in section~\ref{sec:graphtheory}. The circulation of edges around a vertex in the tiling encodes a circuit in the directed graph and thus a gauge-invariant operator in the field theory. The superpotential is formed as a sum of such operators associated with each vertex in the tiling, such that vertices with opposite colours contribute terms with opposite signs to the superpotential. When all $N_i$ equal the same number $N$, the space of vacua associated with critical points of this superpotential assumes a toric structure. 

A brane tiling therefore contains all the data required to specify the lagrangian of a supersymmetric quiver gauge theory in four dimensions. However, this lagrangian is generally not invariant under the classical superconformal algebra $\fsu (2,2|1)$ and the field theory must undergo a non-trivial renormalisation group flow to reach a superconformal fixed point in the infrared. The existence of such a fixed point puts further constraints on the class of admissible brane tilings. When all $N_i =N$, bipartition of the tiling ensures that there are no gauge anomalies at one-loop in the field theory. At a superconformal fixed point, satisfying this requirement is equivalent to the first Chern class of the toric Calabi-Yau cone vanishing. Furthermore, the exact NSVZ $\beta$-functions for the gauge and superpotential couplings must vanish at the superconformal fixed point \cite{Novikov:1983uc,Leigh:1995ep}.  As will be reviewed in more detail in section~\ref{sec:branetilings}, these conditions give non-trivial relations amongst the positive charges for the matter fields under the exact $\fu (1) < \fsu (2,2|1)$ R-symmetry at a unitary superconformal fixed point. If a superconformal fixed point does exist then the exact R-symmetry can be determined via a-maximisation \cite{Intriligator:2003jj} (or via Z-minimisation in terms of the holographically dual toric Calabi-Yau cone \cite{Martelli:2005tp,Martelli:2006yb}). The necessary conditions for the existence of a superconformal fixed point have been determined for a given brane tiling in \cite{Franco:2005rj,Hanany:2005ss} and the admissible ones are characterised by the existence of an isoradial embedding for the tiling in a torus with flat metric. The effect of Seiberg duality on a brane tiling is described in \cite{Franco:2005rj}. Since the rank of each factor in the gauge group is $N$ initially, an important observation is that Seiberg duality with respect to face $i$ in a brane tiling maps the rank of the $i$th $U(N)$ factor $N \mapsto ( -1 + \sum_{j\neq i} A_{ij} ) N$. It therefore defines an operation within this class of brane tilings only when applied to a quadrilateral face which encodes a vertex with precisely two incoming and two outgoing arrows in the directed graph.   

The aim of this paper is to explore the structure of quadrilateral brane tilings wherein each face is bounded by exactly four edges. The directed graph encoded by any quadrilateral brane tiling is connected, contains no loops and has exactly two incoming and two outgoing arrows at each vertex. The field content has exactly $2N$ flavours and $N$ colours associated with each vertex in the corresponding quiver representation. Being within the range $\tfrac{3}{2} N < 2N < 3N$ \cite{Seiberg:1994pq}, the expectation is that the supersymmetric quiver gauge theory based on an admissible quadrilateral brane tiling should always flow to an interacting superconformal fixed point in the infrared. The renormalisation group flow and consistency conditions for more general supersymmetric quiver gauge theories of this type has been investigated in \cite{Balasubramanian:2008qf} and we shall recover some of their results within the context of quadrilateral brane tilings in section~\ref{sec:quadbranetilings}. 

Our main result is a structure theorem in section~\ref{sec:quadtile} which provides a method for deconstructing and reconstructing any quadrilateral tiling via iterating combinations of certain graph-theoretic moves applied to the tiling. We will also describe the refinement of this procedure for the class of \lq smooth' quadrilateral tilings which contain no bivalent vertices. The exclusion of bivalent vertices in the tiling is desirable since they encode mass terms in the superpotential which imply an inconsistency in the superconformal quiver gauge theory. We will see that the structure theory for quadrilateral brane tilings is closely related to the structure theory for the class of eulerian directed graphs they encode, and we often make use of this connection. The later sections in the paper will be concerned with providing a physical interpretation for the generating moves in the associated class of supersymmetric quiver gauge theories. In particular, we investigate the effect of each generating move on the superpotential and vacuum moduli space for the associated field theory and determine whether it maps within the class of admissible quadrilateral brane tilings. We will also identify a couple of composite moves with particular manifestations of Seiberg duality within the class of quadrilateral brane tilings. A simple characterisation of a class of consistent quadrilateral brane tilings with \lq maximal' central charge will be described before going on to discuss the general structure of the class of admissible quadrilateral brane tilings.

Although we have chosen to focus on brane tilings in the better understood context of type IIB string theory, it is important to stress that they can also be used to encode M2-brane configurations in M-theory \cite{Hanany:2008cd,Hanany:2008fj,Franco:2008um,Hanany:2008gx,Davey:2009sr,Davey:2009qx,Davey:2009bp,Davey:2009et}. We will see that some of the generating moves for quadrilateral brane tilings produce inconsistent superconformal quiver gauge theories in four dimensions. However, these tilings may still be relevant in the context of M2-brane effective field theories in three dimensions where the consistency conditions are milder although not so clearly understood. That said, there has been significant progress recently and a general framework for determining the exact superconformal symmetry via \lq F-maximisation' is emerging \cite{Kapustin:2009kz,Drukker:2010nc,Herzog:2010hf,Jafferis:2010un,Kapustin:2010mh,Martelli:2011qj,Jafferis:2011zi} so it may also be of interest to consider quadrilateral brane tilings in that context. 

This paper is organised as follows. We begin in section~\ref{sec:graphtheory} with a short review of some basic concepts in graph theory that will be essential in the forthcoming analysis. Section~\ref{sec:graphs} defines the jargon we will use to describe properties of (directed) graphs and recalls the characterisation of eulerian (directed) graphs. Section~\ref{sec:moves} defines several important moves within the class of eulerian directed graphs and also the operation of mutation which underlies the definition of Seiberg duality for quivers. Section~\ref{sec:generatingeuleriandigraphs} summarises how some of the aforementioned moves can be used to generate any element in the class $\vec{\fG}$ of eulerian directed graphs and the structurally significant r\^{o}le played by the subclass $\vec{\fF}_2$ of loopless eulerian directed graphs with exactly two incoming and two outgoing arrows at each vertex. Some remarks on mutations within the class $\vec{\fF}_2$ will also be made. In section~\ref{sec:bipartitetilingsT2}, we define a bipartite tiling of the torus and how it encodes an eulerian directed graph. This will be followed by a description of how the generating moves for $\vec{\fG}$ are encoded by certain moves within the class of bipartite tilings of the torus. In section~\ref{sec:quadtile}, we focus attention on the subclass of quadrilateral tilings which encode elements in $\vec{\fF}_2$. Section~\ref{sec:facetypes} defines the different types of faces it is useful to distinguish in a quadrilateral tiling. Section~\ref{sec:facerecog} classifies the ways in which a given configuration of arrows at a vertex in an element in $\vec{\fF}_2$ can be encoded by a quadrilateral face. Section~\ref{sec:facecollapse} characterises which types of faces in a quadrilateral tiling can be \lq collapsed' to define a new quadrilateral tiling. Section~\ref{sec:removebivalentpinched} describes a procedure for removing all bivalent vertices and pinched faces from a given quadrilateral tiling. In section~\ref{sec:smoothquadtile}, the generating moves are derived for the class $\eQ$ of \lq smooth' quadrilateral tilings (whose elements contain no bivalent vertices and no pinched faces). Section~\ref{sec:mapQtoF} highlights some generic instances where two different tilings in $\eQ$ encode the same directed graph in $\vec{\fF}_2$ and also several types of directed graphs in $\vec{\fF}_2$ which cannot be encoded by a quadrilateral tiling. Section~\ref{sec:branetilings} reviews how bipartite tilings of the torus can be realised as brane tilings in IIB string theory. Section~\ref{sec:branepicture} describes the relevant configuration of intersecting $5$-branes which gives rise to a brane tiling. Section~\ref{sec:quiverandsuperpotential} details how the quiver representation and superpotential for the low-energy effective field theory are defined by the brane tiling. Section~\ref{sec:superconformalinvariance} discusses the necessary conditions which admissible brane tilings must obey in order to flow to a unitary interacting superconformal fixed point in the infrared. Section~\ref{sec:vacuummodulispace} describes the relevant branch of the superconformal vacuum moduli space in the quiver gauge theory to be identified with the toric Calabi-Yau cone in the dual geometry. Section~\ref{sec:seibergduality} discusses the relevant aspects of Seiberg duality for brane tilings. Finally, in section~\ref{sec:quadbranetilings}, we consider the physical aspects discussed in section~\ref{sec:branetilings} in the context of quadrilateral brane tilings and investigate the physical implications of the structure theorem in section~\ref{sec:quadtile}. Sections~\ref{sec:bivalentandpinched} and \ref{sec:gridsandchains} describe the physical inconsistency of having bivalent vertices and certain types of faces in a quadrilateral brane tiling. Sections~\ref{sec:duplag}, \ref{sec:compmoveI}, \ref{sec:compmoveIIandIII} and \ref{sec:smoothreconstruction} describe the physical effects of each of the moves used to generate quadrilateral brane tilings. The moves which do not map between admissible quadrilateral brane tilings are noted in the process. Section~\ref{sec:seiberg} describes a case of Seiberg duality mapping within the class of quadrilateral brane tilings and its interpretation as a particular composite move. Section~\ref{sec:examples} reconciles some of our generic results with properties of quadrilateral brane tilings in the extensive catalogue obtained in \cite{Davey:2009bp}. Section~\ref{sec:maxquad} describes the characterisation of \lq maximal' quadrilateral brane tilings. Section~\ref{sec:admissiblequadbranetilings} concludes by discussing the general structure of admissible quadrilateral brane tilings.

%%%%%%%%%%%%%%    
\section{Graph theory}
\label{sec:graphtheory}
This section contains a brief review of some essential concepts in the theory of graphs, focussing on certain structural aspects for the class of eulerian directed graphs that will be pertinent in our analysis of brane tilings in later sections. Standard graph theory textbooks such as \cite{BonMur,BJGut,Diestel} provide a more comprehensive introduction to most of this material and a less succinct review can be found in sections 2 and 3 of \cite{deMedeiros:2010pr}, which is summarised here in the interests of keeping the exposition self-contained. The discussion of mutations follows section 3 of \cite{Keller}.
  
%%%%%%%%%%  
\subsection{Graphs and digraphs}
\label{sec:graphs}
A {\emph{graph}} $G$ consists of a set of {\emph{vertices}} $V$ and a set of {\emph{edges}} $E$. (Both sets are taken to be finite with $|V| = n$ and $|E| =e$.) To each edge in $E$, one must also assign a pair of vertices in $V$ that it connects. An edge connecting a vertex to itself is called a {\emph{loop}}. The {\emph{degree}} ${\mathrm{deg}}(v)$ of a vertex $v \in V$ is the number of edges in $E$ that end on $v$ (with each loop attached to $v$ counting twice in ${\mathrm{deg}}(v)$). The so-called handshaking lemma provides the relation $\sum_{v \in V} {\mathrm{deg}}(v) = 2e$. A graph is called $k${\emph{-regular}} if ${\mathrm{deg}}(v) = k$ for all $v \in V$, in which case $kn=2e$. The {\emph{cycle graph}} $C_n$ with $n$ vertices is defined such that its edges form the sides of an $n$-sided polygon, whence it is $2$-regular. The {\emph{complete graph}} $K_n$ with $n$ vertices is defined such that each pair of distinct vertices in it are connected by a single edge, whence it is $(n-1)$-regular.    

An edge is called {\emph{simple}} if it is the only one which connects the pair of vertices assigned to it. A graph is then said to be simple if it contains only simple edges (if loops are included here then a simple graph must be loopless). A {\emph{walk}} in $G$ consists of a sequence of vertices in $V$ such that each pair of consecutive vertices in the sequence are connected by an edge in $E$. A walk with no repeated vertices is called a {\emph{path}} and a closed path is called a {\emph{cycle}}. A walk with no repeated edges is called a {\emph{trail}} and a closed trail is called a {\emph{circuit}}. A pair of vertices in $V$ are said to be {\emph{connected}} if there exists a path between them in $G$. A graph is connected if all its vertices are connected. A path (cycle) is called {\emph{hamiltonian}} if it contains every vertex in $V$ exactly once and a graph is called hamiltonian if it admits a hamiltonian cycle. A trail (circuit) is called {\emph{eulerian}} if it traverses every edge in $E$ exactly once and a graph is called eulerian if it admits an eulerian circuit. The following equivalent statements provide a characterisation of eulerian graphs:
\begin{itemize}
 \item {\emph{$G$ is eulerian.}}
 \item {\emph{$G$ is connected and contains no vertices of odd degree.}}
 \item {\emph{$G$ is connected and its edge set can be partitioned into subsets which define edge-disjoint cycle graphs on the vertices of $G$.}}
\end{itemize} 
The equivalence of the first two statements is Euler's theorem. A corollary of the third statement is that any eulerian graph with $n$ vertices and $e$ edges can be obtained by identifying precisely $e-n$ appropriately chosen vertices in the cycle graph $C_e$. The statements above imply that a connected $k$-regular graph is eulerian when $k$ is even (e.g. every connected $2$-regular graph is a cycle graph). 

An {\emph{orientation}} on a graph is defined by making each edge into an {\emph{arrow}} connecting the same pair of vertices but now with a specified direction pointing from one vertex to the other (a loop based at a vertex in the graph being made into an arrow pointing from the base vertex to itself). A graph $G$ equipped with an orientation is called a {\emph{directed graph}} or {\emph{digraph}}, written $\vec{G}$. For example, a digraph obtained by equipping the complete graph $K_n$ with an orientation is called a {\emph{tournament}}. Many of the concepts and results described above for graphs can be extended in an obvious way for digraphs. We shall therefore discuss only those properties which are refined in a particular way for digraphs. 

One such refinement is that the degree of any vertex $v$ in a digraph can be written ${\mathrm{deg}}(v) = {\mathrm{deg}}^+(v) + {\mathrm{deg}}^-(v)$, where ${\mathrm{deg}}^\pm (v)$ denote the number of arrows pointing from/to $v$ and are respectively referred to as the {\emph{out-/in-degree}} of $v$ (with each loop arrow based at $v$ in the digraph contributing one to both ${\mathrm{deg}}^+(v)$ and ${\mathrm{deg}}^-(v)$). The handshaking lemma is then refined such that $\sum_{v \in V} {\mathrm{deg}}^+(v) = \sum_{v \in V} {\mathrm{deg}}^-(v) = e$ for any digraph with $e$ arrows. A digraph is said to be {\emph{balanced}} if ${\mathrm{deg}}^+(v) = {\mathrm{deg}}^-(v)$ for all $v \in V$. A balanced digraph $\vec{G}$ will be called $k$-regular if ${\mathrm{deg}}^+(v) = k$ for all vertices $v$ (this means that the underlying unoriented graph $G$ is $2k$-regular). Thus $kn =e$ for a $k$-regular balanced digraph with $n$ vertices and $e$ arrows.    

We define an arrow in a digraph $\vec{G}$ to be {\emph{undirected simple}} if the corresponding undirected edge in $G$ is simple. An arrow in $\vec{G}$ pointing from one vertex $v$ to another vertex $w$ is called {\emph{simple}} if it is the only one pointing from $v$ to $w$ (i.e. the arrow is simple even if there is another arrow pointing from $w$ to $v$). Thus any undirected simple arrow must be simple but the converse need not be true. A digraph is then said to be (undirected) simple if it contains only (undirected) simple arrows. Thus not every simple digraph is obtained by defining an orientation on a simple graph. 

All the different kinds of walks that were defined in a graph $G$ extend in the obvious way to directed walks in a digraph $\vec{G}$ (i.e. one can only proceed in directions defined by the orientation of the arrows). A digraph $\vec{G}$ is called {\emph{weakly connected}} if its underlying undirected graph $G$ is connected. A digraph $\vec{G}$ is called {\emph{strongly connected}} if it contains a directed path from $v$ to $w$ and a directed path from $w$ to $v$ for all pairs of vertices $(v,w)$. The characterisation of eulerian digraphs is provided by the following equivalent statements:
\begin{itemize}
 \item {\emph{$\vec{G}$ is eulerian.}} 
 \item {\emph{$\vec{G}$ is weakly connected and balanced (which implies it is also strongly connected).}} 
 \item {\emph{$\vec{G}$ is strongly connected and its arrow set can be partitioned into arrow-disjoint directed cycles on the vertices of $\vec{G}$.}}  
\end{itemize}
Similar to the undirected case, a corollary of the third statement above is that any eulerian digraph with $n$ vertices and $e$ arrows can be obtained by identifying precisely $e-n$ appropriately chosen vertices in the directed cycle graph $\vec{C}_e$. Note that the cycle digraph is eulerian with the in-degree and out-degree for each of its vertices equal to one (it is the only possible type of $1$-regular eulerian digraph). It is the less trivial structure of (loopless) $2$-regular eulerian digraphs that will be relevant for most of the results in this paper. 

%%%%%%%%%%  
\subsection{Moves}
\label{sec:moves}

Let us now define some canonical operations or {\emph{moves}} which act on a digraph. 

\begin{figure}[h!]
\includegraphics[scale=1.2]{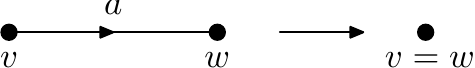} \hspace*{.6in}
\includegraphics[scale=1.2]{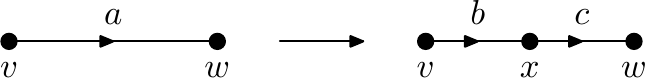}
\caption{Contraction and subdivision of an arrow.}
\label{fig1}
\end{figure}

The {\emph{contraction}} of an arrow $a$ pointing from a vertex $v$ to a different vertex $w$ in a digraph $\vec{G}$ is defined by deleting $a$ then identifying the vertices $v$ and $w$ to create a new digraph. The {\emph{subdivision}} of an arrow $a$ pointing from a vertex $v$ to a vertex $w$ in $\vec{G}$ is defined by deleting $a$ then adding a new vertex $x$ and two new arrows $b$ and $c$ such that $b$ points from $v$ to $x$ and $c$ points from $x$ to $w$. Whence $x$ is balanced with out-degree $1$ in the subdivided digraph whilst the out- and in-degrees of both $v$ and $w$ remain as they were in $\vec{G}$. This can also be thought of as placing an extra vertex on $a$ in between vertices $v$ and $w$. (The definition of subdivision is taken to also apply to loops when $v=w$.) The reverse operation of removing from a digraph a balanced vertex with out-degree $1$ (i.e. contracting an arrow pointing either to or from a balanced vertex with out-degree $1$) is called {\emph{smoothing}} and a digraph is {\emph{smooth}} if it contains no balanced vertices with out-degree $1$. The contraction and subdivision of $a$ are depicted in figure~\ref{fig1}.
         
\noindent Performing either of these two operations on any arrow in an eulerian digraph will produce a new digraph that must also be eulerian. The only modification of vertex degrees are such that ${\mathrm{deg}}^+(v) + {\mathrm{deg}}^+(w) -1 = {\mathrm{deg}}^+(v=w)$ for contraction and ${\mathrm{deg}}^+(x)=1$ for subdivision. Only the contraction of an undirected simple arrow will not create a loop in the new digraph while subdivision of an arrow can never create a loop. (Subdivision of a loop based at vertex $v$ creates a $\vec{C}_2$ traversing $v$ and the new vertex $x$.) Contracting an arrow can never create a subdivision. Thus contraction of (undirected simple) arrows is an operation within the class of (loopless) smooth eulerian digraphs and subdivision of arrows is an operation within the class of loopless eulerian digraphs. It is worth remarking that, if the initial digraph $\vec{G}$ has $n$ vertices and $e$ arrows, then contraction/subdivision of an arrow reduces/increases both of these parameters by one in the new digraph so that their difference is left invariant by both these operations. 

\begin{figure}[h!]
\includegraphics{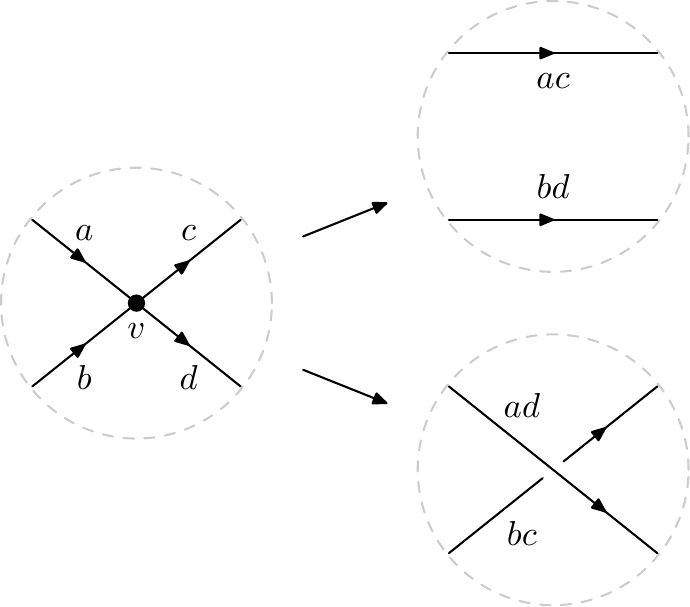}
\caption{Splitting an out-degree $2$ vertex.}
\label{fig2}
\end{figure}

Consider an eulerian digraph $\vec{G}$ that contains at least one vertex with out-degree $2$. Select one such vertex $v$ in $\vec{G}$ and label its two incoming and outgoing arrows $(a,b)$ and $(c,d)$. The {\emph{splitting}} of vertex $v$ is defined by first deleting $v$ and then connecting the head of $a$ to the tail of either $c$ or $d$ (forming a new arrow $ac$ or $ad$) and the head of $b$ to the tail of either $d$ or $c$ (forming a new arrow $bd$ or $bc$). The two possible outcomes of this operation are shown in figure~\ref{fig2}.
  
\noindent Let us denote by $\vec{H}_=$ and $\vec{H}_\times$ the two digraphs which result from the two possible splittings of $v$. Generally $\vec{H}_=$ and $\vec{H}_\times$ will not be isomorphic though they must both have all their vertices balanced. Moreover, $\vec{H}_=$ and $\vec{H}_\times$ will also be weakly connected whenever the respective arrow pairs $ac$ and $ad$ are adjacent in an eulerian circuit in $\vec{G}$. Clearly this must always be the case for either $\vec{H}_=$ or $\vec{H}_\times$. When it is the case for only one, it is this option which is chosen for the splitting so that in all cases the resulting digraph is eulerian. It is worth noting that, if $\vec{G}$ has $n$ vertices and $e$ arrows, then splitting $v$ in either way reduces $n$ by one and $e$ by two, thus reducing $e-n$ by one. 

\begin{figure}[h!]
\centering
\begin{tabular}{cccc}
\includegraphics{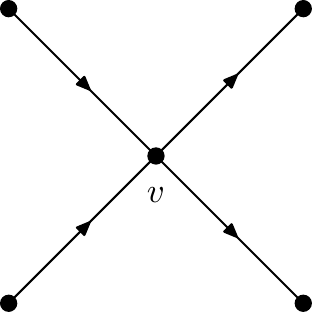} & \includegraphics{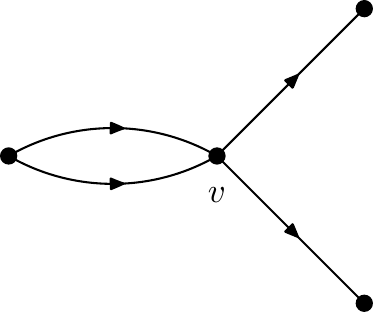} & \includegraphics{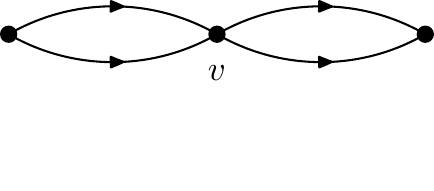} & \\ [.1in]
\includegraphics{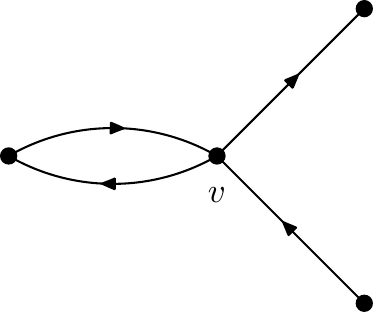} & \includegraphics{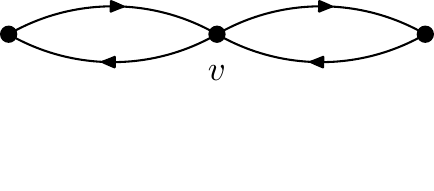} & \includegraphics{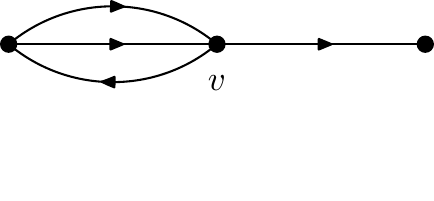} & \includegraphics{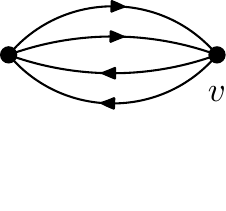} \\ 
\end{tabular}
\caption{Inequivalent arrow connections for a balanced out-degree $2$ vertex $v$.}  
\label{fig3}
\end{figure}

If $\vec{G}$ is smooth then so are both $\vec{H}_=$ and $\vec{H}_\times$. If $\vec{G}$ is smooth and loopless, the conditions under which $\vec{H}_=$ and $\vec{H}_\times$ will also be loopless and eulerian depend on the nature of the connections with other vertices that are made by the four arrows attached to the vertex $v$ to be split. There are seven distinct scenarios that are drawn in figure~\ref{fig3}. Only the four arrows attached to $v$ and the other vertices they are connected to are shown. It will be unnecessary to distinguish the other connections obtained by reversing the directions of all four arrows attached to $v$ in figure~\ref{fig3}. 

\noindent A splitting of $v$ in a smooth loopless eulerian digraph $\vec{G}$ is defined to be loopless if the corresponding digraph $\vec{H}_=$ or $\vec{H}_\times$ is also smooth, loopless and eulerian. Thus, by definition, loopless splitting of out-degree $2$ vertices is an operation within the class of smooth loopless eulerian digraphs. In the three cases shown in the first row of figure~\ref{fig3}, $\vec{H}_=$ and $\vec{H}_\times$ are both loopless. In the first two cases in the second row, either $\vec{H}_=$ or $\vec{H}_\times$ is loopless (but not both). In the last two cases in the second row, neither $\vec{H}_=$ nor $\vec{H}_\times$ is loopless and so a loopless splitting of $v$ is impossible here. In each of the other five cases, there always exists at least one loopless splitting of $v$. More precisely, there are two possible loopless splittings of $v$ in the second and third cases and only one possible loopless splitting of $v$ in the fourth and fifth cases. In the first case, despite both $\vec{H}_=$ and $\vec{H}_\times$ being loopless, one of them may not be eulerian if it is not weakly connected. However, it is easily seen that this scenario can only occur when $\vec{G}$ is built from two disjoint eulerian trails such that the start and end points of each trail are connected to a pair vertices attached to an outgoing and incoming arrow at $v$ in the first case in figure~\ref{fig3}. For example, given eulerian trails $\gamma_1$ and $\gamma_2$, connecting up $\gamma_1$ to the vertices on which arrows $a$ and $c$ end and $\gamma_2$ to the vertices on which arrows $b$ and $d$ end defines an eulerian digraph which has an eulerian circuit $( ad \gamma_2 bc \gamma_1 ) $ but does not have an eulerian circuit containing $ac$ and indeed $\vec{H}_=$ is not weakly connected. 

\begin{figure}[h!]
\includegraphics{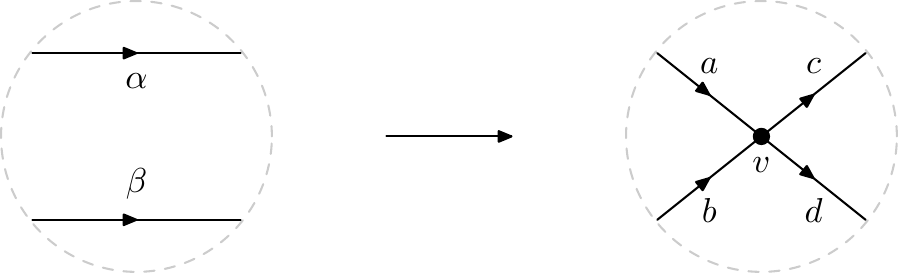}
\caption{Simple immersion of $\vec{H}$ in $\vec{G}$.}
\label{fig4}
\end{figure}

Reversing the loopless splitting of a out-degree $2$ vertex in a smooth loopless eulerian digraph $\vec{G}$ defines an operation that was referred to as {\emph{simple immersion}} in \cite{deMedeiros:2010pr}. (It corresponds to the simplest non-trivial case of the more general notion of an immersion that was introduced in \cite{Johnson} as the natural containment relation in the context of eulerian digraphs.) This reverse operation is depicted in figure~\ref{fig4} and proceeds by isolating any pair of arrows $\alpha$ and $\beta$ in a smooth loopless eulerian digraph $\vec{H}$ and joining them to form a new balanced out-degree $2$ vertex $v$ in a smooth loopless eulerian digraph $\vec{G}$ (which is the same as $\vec{H}$ everywhere except within the dashed circle). The tails of the new arrows $a$ and $b$ in $\vec{G}$ touch the same vertices as the tails of  $\alpha$ and $\beta$ in $\vec{H}$  while the heads of the new arrows $c$ and $d$ in $\vec{G}$ touch the same vertices as the heads of  $\alpha$ and $\beta$ in $\vec{H}$.
 
\noindent By construction, $\vec{H}$ therefore follows from a loopless splitting of $v$ in $\vec{G}$. Relative to the labelling of arrows in figure~\ref{fig2}, $\alpha$ and $\beta$ in $\vec{H}$ in figure~\ref{fig4} would be identified with $ac$ and $bd$ in $\vec{H}_=$, in cases where it could be obtained from a loopless splitting of $v$. Relabelling $c \leftrightarrow d$ in figure~\ref{fig4} and making the same comparison with figure~\ref{fig2} would identify $\alpha$ and $\beta$ in $\vec{H}$ with $ad$ and $bc$ in $\vec{H}_\times$.

The final move we shall make use of in the forthcoming analysis is defined as follows. Consider a loopless digraph $\vec{G}$ which has no circuits containing exactly two arrows (i.e. if $\vec{G}$ contains at least one arrow pointing from a vertex $v$ to another vertex $w$ then it does not contain an arrow pointing from $w$ to $v$). The move proceeds by first selecting a vertex $v$ in $\vec{G}$. Now consider the set of ordered pairs of arrows in $\vec{G}$ defined such that each pair $(a,b)$ in the set has arrow $a$ pointing to $v$ from another vertex $w$ while arrow $b$ points from $v$ to another vertex $x$. For a given pair $(a,b)$, let $r$ denote the number of arrows connecting $w$ and $x$, such that $r \geq 0$ means $r$ arrows pointing from $w$ to $x$ while $r<0$ means $|r|$ arrows  pointing from $x$ to $w$. For each pair $(a,b)$, the rule is to add one new arrow pointing from $w$ to $x$ when $r\geq 0$ and delete one arrow pointing from $x$ to $w$ when $r<0$. Having applied this rule to all arrow pairs in the set above, the last step is to reverse the orientation of each arrow connected to $v$. This operation defines a new loopless digraph $\vec{G}^\prime$ which also has no circuits containing exactly two arrows and the map $\mu_v : \vec{G} \mapsto \vec{G}^\prime$ is called an elementary {\emph{mutation}} of $\vec{G}$ with respect to $v$. It is straightforward to check that this definition implies $\mu_v$ is an involution. A general mutation consists of a composition of elementary mutations. The utility of mutations in the study of cluster algebras was pioneered in \cite{FomZelI,FomZelII,FomZelIII,FomZelIV} and they have many physical applications in the context of BPS states and dualities in string theory \cite{Feng:2001bn,Feng:2002kk,Berenstein:2002fi,Herzog:2004qw,Cecotti:2011rv}. In particular, the connection between Seiberg duality in supersymmetric quiver gauge theories in four dimensions and mutations of the quiver will be discussed in more detail in section~\ref{sec:branetilings}.

Taking $\vec{G}$ to be loopless with no two-arrow circuits avoids potential ambiguities in the prescription above. However, it is worth emphasising that $\mu_v$ is also well-defined (and involutive) when applied to a vertex $v$ in any loopless digraph $\vec{G}$ provided each arrow in any two-arrow circuit in $\vec{G}$ does not point to or from $v$ and is also not contained in any three-arrow circuit containing $v$ in $\vec{G}$ (i.e. such that any two-arrow circuits in $\vec{G}$ are unaffected by $\mu_v$). As opposed to the other moves described in this section, mutation is not an operation within the class of eulerian digraphs. For example, mutating the cycle digraph $\vec{C}_{n>3}$ with respect to any vertex $v$ defines a new weakly connected digraph with $n+1$ arrows and $n$ vertices, of which $n-2$ are balanced (with out-degree $1$) while the remaining two to either side of $v$ are not.    

%%%%%%%%%%%%%%%%%%%
\subsection{Generating eulerian digraphs}
\label{sec:generatingeuleriandigraphs} 

Let us denote the set of all eulerian digraphs by $\vec{\fG}$ and the subset of all smooth loopless eulerian digraphs by $\vec{\fF}$. To any element in $\vec{\fF}$, one can add a loop based at any vertex and subdivide any arrow (or new loop that is introduced) to produce an element in the complement of $\vec{\fF} \subset \vec{\fG}$. Indeed any element in $\vec{\fG}$ can be obtained from an element in $\vec{\fF}$ (or the trivial graph) by applying some combination of loop additions and subdivisions. 

By definition, any $\vec{G} \in \vec{\fF}$ with $n$ vertices and $e$ arrows must contain no subdivisions, so that ${\mathrm{deg}}^+(v) >1$ for every vertex $v$ in $\vec{G}$. Consequently, the handshaking lemma $\sum_{v \in V} {\mathrm{deg}}^+(v) = e$ implies that $e \geq 2n$ with $e=2n$ occurring only if $\vec{G}$ is $2$-regular. Let us denote the set of all $2$-regular eulerian digraphs by $\vec{\fG}_2 \subset \vec{\fG}$ and the subset of all loopless $2$-regular eulerian digraphs by $\vec{\fF}_2 \subset \vec{\fF}$. Elements in $\vec{{\mathfrak G}_2}$ are necessarily smooth and there cannot be more than two loops based at any of its vertices. The only possibility of having two loops based at any one vertex is if that is the only vertex in the eulerian digraph. All other elements in the complement of $\vec{\fF}_2 \subset \vec{\fG}_2$ must contain a single loop on some number of their vertices. Furthermore, any such element in $\vec{\fG}_2$ can be obtained from an element in $\vec{\fF}_2$ (or the digraph with two loops based at a single vertex) by applying some number of times the single composite move formed by subdividing an arrow and then adding a loop based at the new vertex created by the subdivision. 

\begin{figure}[h!]
\centering
\begin{tabular}{ccccc}
\includegraphics{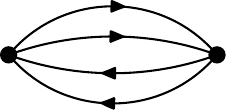} & & & & \\
\includegraphics{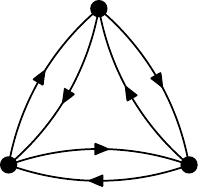} & \includegraphics{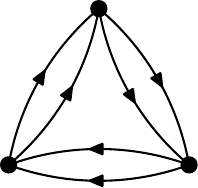}  & & & \\ 
\includegraphics{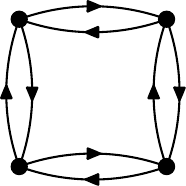} & \includegraphics{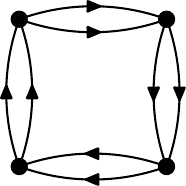}  & \includegraphics{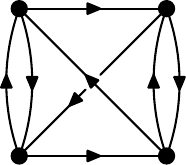} & \includegraphics{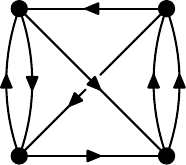} & \includegraphics{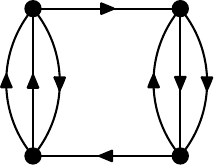} \\ 
\end{tabular}
\caption{Elements in $\vec{\mathfrak F}_2^{[t]}$ are drawn in row $t-1$ for $t=2,3,4$.}  
\label{fig5}
\end{figure}

For each set of eulerian digraphs above, let us define a subset labelled by a superscript $[t]$ such that it contains all the elements in the corresponding set which have the same value of $e-n=:t$. As already noted, contracting or subdividing an arrow in a digraph does not change the value of $t$ so these operations map between different members of the same family $\vec{\fG}^{[t]}$. Since any element in $\vec{\fF}$ has $e \geq 2n$ and $n >1$ then members of the family $\vec{\mathfrak F}^{[t]}$ must have $2 \leq n \leq t$ vertices and $e=n+t$ arrows. The parents in family $\vec{\fF}^{[t]}$ correspond to elements with the maximum number of vertices $n=t$, which are precisely the $2$-regular members comprising $\vec{\fF}_2^{[t]}$. For example, the elements in $\vec{\fF}_2^{[2]}$, $\vec{\fF}_2^{[3]}$ and $\vec{\fF}_2^{[4]}$ are shown in figure~\ref{fig5}. Any child in family $\vec{\fF}^{[t]}$, corresponding to an element in the complement of  $ \vec{\fF}_2^{[t]} \subset \vec{\fF}^{[t]}$ with $2 \leq n<t$, can be obtained from a parent via the contraction of some number of undirected simple arrows. 

\begin{figure}[h!]
\includegraphics{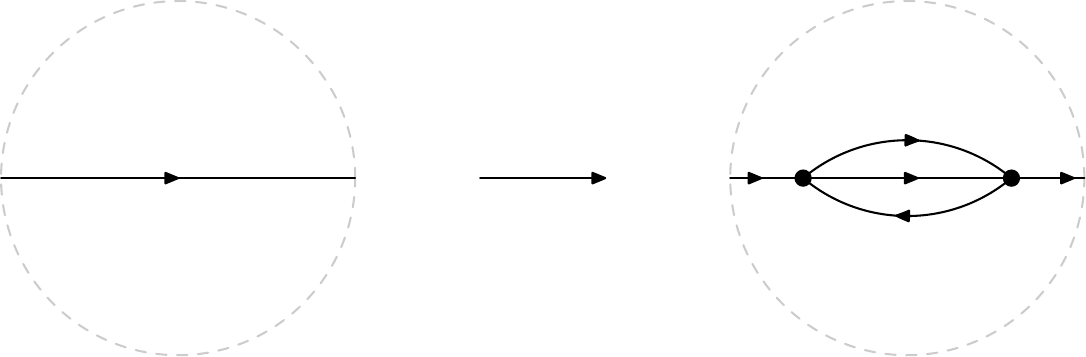}
\caption{A useful composite move creating two balanced out-degree $2$ vertices.}
\label{fig6}
\end{figure}

In this way, it was shown in \cite{deMedeiros:2010pr} that one can obtain any eulerian digraph in $\vec{\fG}$ from some loopless $2$-regular eulerian digraph in $\vec{\fF}_2$ (or the trivial graph) by applying some combination of loop additions, subdivisions and contractions. Moreover, any element in $\vec{\fF}_2$ can be obtained by applying to the unique element $\vec{G}_+$ in $\vec{\fF}_2^{[2]}$ some number of iterations of the move depicted in figure~\ref{fig6} (mapping $\vec{\fF}_2^{[t]} \rightarrow \vec{\fF}_2^{[t+2]}$), followed by some combination of simple immersions (as defined in figure~\ref{fig4} mapping $\vec{\fF}_2^{[t]} \rightarrow \vec{\fF}_2^{[t+1]}$). The connections made by the head and tail of the arrow intersecting the dashed circle on the right and left sides are the same before and after the move in figure~\ref{fig6}.

\noindent The operation in figure~\ref{fig6} can be thought of as the composite move formed by first subdividing any arrow in an element in $\vec{\fF}_2^{[t]}$, then attaching a loop based at this new vertex before finally applying a simple immersion as in figure~\ref{fig4} but with $\alpha$ identified with the aforementioned loop and $\beta$ identified with the arrow whose tail is attached to its base to produce a new element in $\vec{\fF}_2^{[t+2]}$. Despite being expressible in terms of existing moves, it is useful to distinguish it so that each of the two moves above is an operation within the class $\vec{\fF}_2$. 
 
Let us conclude with some remarks concerning elementary mutations of elements in $\vec{\fF}_2$ that will be useful to recall in later sections. We have already noted that the elementary mutation of a loopless eulerian digraph produces a loopless digraph that is not necessarily eulerian. However, the elementary mutation of a loopless $2$-regular eulerian digraph must always produce a loopless eulerian digraph (though it is not necessarily $2$-regular). With respect to a vertex $v$ in any $\vec{G} \in \vec{\fF}_2$, the mutation $\mu_v$ is only defined provided the criteria described at the end of section~\ref{sec:moves} are met. This means that $v$ must look like one of three scenarios in first row of figure~\ref{fig3}. Furthermore, for each pair of arrows $(a,b)$ with $a$ pointing from $w$ to $v$ and $b$ pointing from $v$ to $x$ as before, since $\vec{G}$ is $2$-regular, the number of arrows connecting $w$ and $x$ must be either $r=1,0,-1,-2$. Moreover, there are precisely four different ways to pair up an incoming arrow $a$ with an outgoing arrow $b$ at $v$ in $\vec{G}$. Going through all the permitted combinations is therefore straightforward but tedious. The upshot is that $\vec{G}^\prime = \mu_v ( \vec{G} )$ is always weakly connected with all vertices balanced, whence eulerian. Only the out-degrees of the balanced vertices connected to $v$ by an arrow in $\vec{G}$ can be modified by $\mu_v$ and the out-degree of any one of these vertices can change by either $+2,+1,0,-1$, depending on the nature of the connections with $v$ in $\vec{G}$. It is not worth going through the different cases in any more detail but it will be useful to note the following two special situations which will be found in later sections to arise naturally in the context of Seiberg dualities for the class of tilings to be considered in section~\ref{sec:quadtile}.. 

\begin{figure}[h!]
\includegraphics{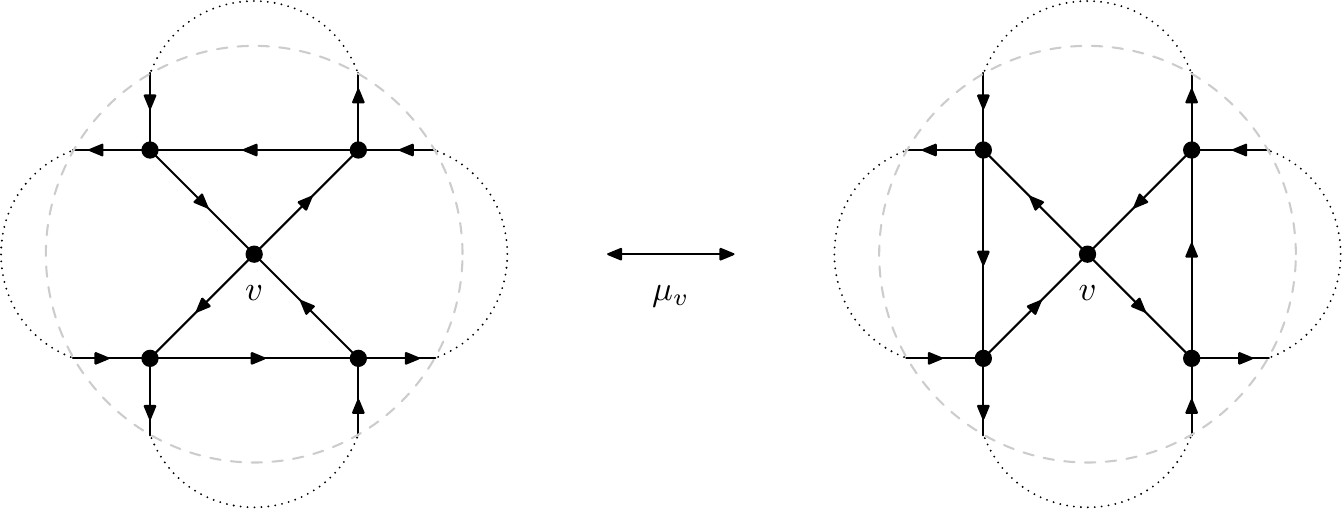}
\caption{Mutation $\mu_v$ mapping between $\vec{G}$ and $\vec{G}^\prime$ in $\vec{\fF}_2^{[t]}$.}
\label{fig6a}
\end{figure}

\begin{figure}[h!]
\includegraphics{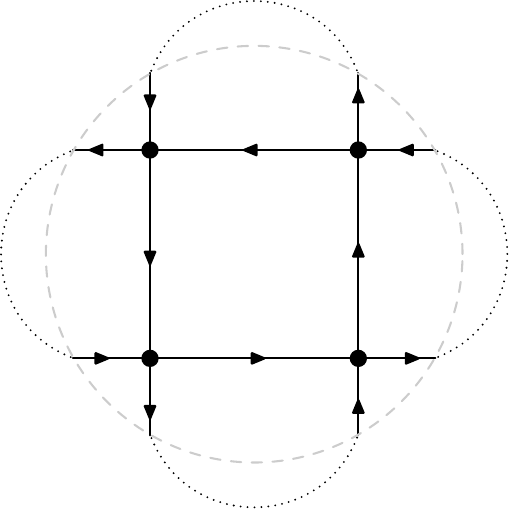}
\caption{The element $\vec{H} \in \vec{\fF}_2^{[t-1]}$ which can be simply immersed in both $\vec{G}$ and $\vec{G}^\prime$.}
\label{fig6b}
\end{figure}

With respect to a vertex $v$ in any $\vec{G} \in \vec{\fF}_2^{[t]}$, if the mutation $\mu_v$ is defined and $\vec{G}^\prime = \mu_v ( \vec{G} ) \in \vec{\fF}_2^{[t]}$ then, provided $\mu_v$ is not an isomorphism of $\vec{G}$, $v$ must take the form shown in figure~\ref{fig6a}. Each dotted arc outside the dashed circle in figure~\ref{fig6a} denotes a permitted identification of the pair of arrows it connects in $\vec{G}$ (and $\vec{G}^\prime$). 
Generically $\mu_v$ is not an isomorphism when there are either zero, one, two or three such identifications in $\vec{G}$  (and $\vec{G}^\prime$). The action of $\mu_v$ here can be thought of as bisecting the arrows which cross the dashed circle in figure~\ref{fig6a} then rotating the interior by $\tfrac{\pi}{2}$ before gluing it back together. Whence $\mu_v$ is indeed an involution since a $\pi$ rotation gives an isomorphism. Notice that there must always exist a loopless splitting of vertex $v$ in both $\vec{G}$ and $\vec{G}^\prime$ which produces the same $\vec{H} \in \vec{\fF}_2^{[t-1]}$, as shown in figure~\ref{fig6b}. 
Conversely, from the circuit in $\vec{H}$ formed by the four arrows connecting the four vertices in figure~\ref{fig6b}, pinching together either the left and right or top and bottom arrows defines the simple immersion of $\vec{H}$ in either $\vec{G}$ or $\vec{G}^\prime$. It is worth noting that although this is the most general elementary mutation within the class $\vec{\fF}_2$, there exist many other mutations within $\vec{\fF}_2$ which contain some number of elementary mutations that are not within $\vec{\fF}_2$. 

\begin{figure}[h!]
\includegraphics{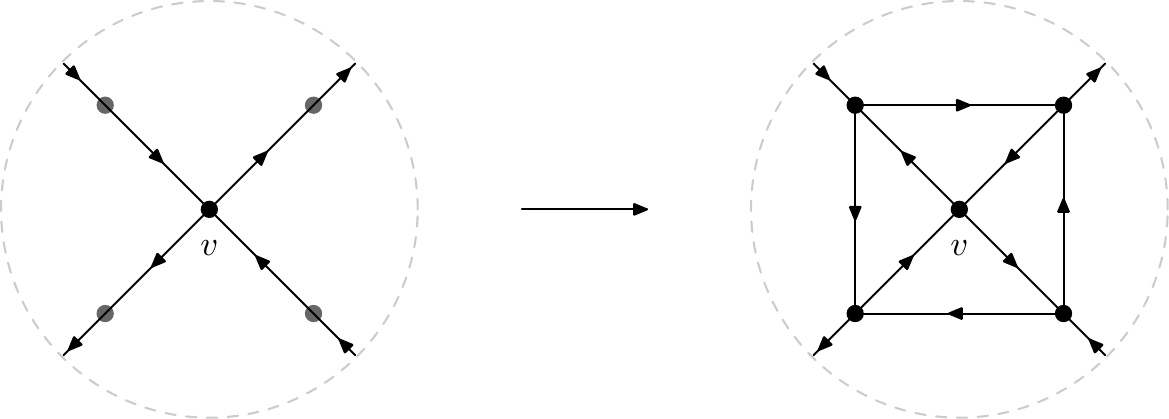}
\caption{The composite move acting on $\vec{H} \in \vec{\fF}_2^{[t]}$ with respect to $v$ formed by subdividing each arrow attached to $v$ then mutating to give $\vec{G} \in \vec{\fF}_2^{[t+4]}$.}
\label{fig6c}
\end{figure}

The other special case we shall consider involves the elementary mutation of an element in $\vec{\fF}_2$ such that the loopless eulerian digraph produced by the mutation has the maximum number of four balanced vertices with out-degree one (i.e. all its remaining vertices are balanced with out-degree $2$). Unlike the one described above, this mutation therefore does not act within the class $\vec{\fF}_2$. However, when combined with the subdivision move defined in section~\ref{sec:moves}, it can be used to define a generic operation within $\vec{\fF}_2$ in the following way. Take any $\vec{H} \in \vec{\fF}_2^{[t]}$ and select a vertex $v$ in $\vec{H}$. Next subdivide each of the four arrows attached to $v$ which defines $\vec{H}^\prime \in \vec{\fG}^{[t]}$ containing $t+4$ vertices. The criteria described at the end of section~\ref{sec:moves} are guaranteed to be met by $\vec{H}^\prime$ with respect to $v$, for any choice of $\vec{H}$. The four arrows attached to $v$ in $\vec{H}^\prime$ end on the four different vertices created in the aforementioned subdivision of $\vec{H}$ and no two of these four vertices are connected by an arrow in $\vec{H}^\prime$ since $\vec{H}$ is loopless. By mutating $\vec{H}^\prime$ with respect to $v$, one can therefore obtain $\vec{G} = \mu_v ( \vec{H}^\prime )  \in \vec{\fF}_2^{[t+4]}$. The effect of this operation, mapping any $\vec{H} \in \vec{\fF}_2^{[t]}$ to $\vec{G} \in \vec{\fF}_2^{[t+4]}$, is shown in figure~\ref{fig6c} (grey dots indicate the subdivisions of arrows in $\vec{H}$).
Conversely, given an elementary mutation $\mu_v$ that is defined with respect to a vertex $v$ in some $\vec{G} \in \vec{\fF}_2$ such that $\mu_v ( \vec{G} )$ contains exactly four balanced vertices with out-degree one, $v$ must take the form shown on the right hand side in figure~\ref{fig6c}.  The move in figure~\ref{fig6c} arises naturally in the context of Seiberg duality in supersymmetric quiver gauge theories in four dimensions (e.g. it corresponds to a special case of figure 12 in \cite{Franco:2005rj}). We will revisit this connection in later sections.

%%%%%%%%%%%%%%    
\section{Bipartite tilings of the torus}
\label{sec:bipartitetilingsT2}

This section contains a review of the particular class of tilings which can be used to encode supersymmetric quiver gauge theories in four dimensions. In particular, recalling how each such tiling encodes an eulerian digraph on which the quiver representation is based (the physical data encoded by the tiling will be reviewed in section~\ref{sec:branetilings}). We then note how some of the structural aspects for eulerian digraphs that were discussed in the previous section are interpreted in terms of tilings. This will prepare the way for a more detailed analysis of the class of quadrilateral tilings which encode eulerian digraphs in $\vec{\fF}_2$ in section~\ref{sec:quadtile}.

Any element in the class of tilings of interest can be thought of as a particular kind of graph drawn on the surface of the torus without any edges crossing. An equivalent representation in the plane with two directions periodically identified is often more convenient, such that all the vertices and edges in the aforementioned graph are contained within a minimal non-repeating region or {\emph{fundamental domain}}. From this perspective, the faces enclosed by the edges are labelled and form the tiles. Each vertex in the tiling must have degree greater than one (i.e. each vertex is at least {\emph{bivalent}}). Furthermore, the tiling must be {\emph{bipartite}} meaning that each of its vertices can be coloured either white or black in such a way that every edge in the tiling connects a pair of adjacent vertices with opposite colour. Consequently, each face in the tiling must be bounded by an even number of edges. (It is possible that some of the boundary edges for a given face are repeated once, in which case the replicas do not reside within a fundamental domain of the tiling.) The bipartition of the vertices in the tiling defines an orientation to each of the edges. This follows by assigning to each white/black vertex in the tiling a clockwise/anticlockwise circulation of the edges connected to it. Since each edge forms part of the boundary of adjacent faces in the tiling, this circulation provides a well-defined notion of each edge pointing from one face to another.      

Any such bipartite tiling of the torus (henceforth just referred to as a tiling) encodes a digraph such that each face in the tiling represents a labelled vertex in the digraph and each oriented edge in the tiling represents an arrow in the digraph pointing from one vertex to another. The circulation of edges connected to each vertex in the tiling represents a circuit in the digraph. Since each face in the tiling is bounded by an even number of edges which connect adjacent vertices with alternating white/black colour going around the boundary then each vertex in the associated digraph must be balanced (a repeated boundary edge around a face in the tiling describes a loop based at the corresponding vertex in the digraph). Moreover, since any pair of faces in the tiling must be connected by some sequence of intermediate adjacent faces (i.e. there can exist no disjoint partition of the set of all faces in the tiling)  then there must exist a path between any pair of vertices in the associated digraph. The digraph encoded by the tiling must therefore be both balanced and weakly connected, whence eulerian. If a tiling $\tau_{\vec{G}}$ encodes an eulerian digraph $\vec{G} \in \vec{\fG}^{[t]}$ with $n$ vertices and $e$ arrows then $\tau_{\vec{G}}$ must have precisely $n$ faces, $e$ edges and $t$ vertices contained within a fundamental domain, due to the formula $n-e+t=0$ for the euler character of a torus. However, it is important to stress that this map from tilings to eulerian digraphs is not bijective. There can sometimes exist different tilings which encode the same eulerian digraph, so the map is not injective. There also exist eulerian digraphs which cannot be encoded by any tiling, so the map is not surjective either. 

Any tiling $\tau_{\vec{G}}$ which encodes an eulerian digraph $\vec{G}$ containing $p>0$ loops based at a vertex $v$ must contain a face $v$ described by a $2\, {\mathrm{deg}}^+(v)$-sided polygon with precisely $2\, {\mathrm{deg}}^+(v) -p$ different boundary edges contained within a fundamental domain of $\tau_{\vec{G}}$ (with face $v$ being adjacent to itself on precisely $p$ of those edges). Moreover, since each pair of repeated edges on the boundary of face $v$ must be identified periodically across a fundamental domain, clearly they must sit on opposite sides of the polygon, separated by ${\mathrm{deg}}^+(v)-1$ other edges on both sides of the boundary. Thus, due to the bipartition of vertices in the tiling, the identification of repeated edges requires ${\mathrm{deg}}^+(v)$ to be odd. Note that this condition implies there exists no tiling which can encode any $2$-regular eulerian digraph that is contained in the complement of $\vec{\fF}_2 \subset \vec{\fG}_2$ (i.e. if a given $2$-regular eulerian digraph can be encoded by a tiling then it is necessarily loopless). Some obstructions to the existence of tilings which encode the elements in $\vec{\fF}_2$ will also be discussed in section~\ref{sec:mapQtoF}. It is also worth noting that, given a tiling $\tau_{\vec{G}}$ which encodes an eulerian digraph $\vec{G}$, there can be no tiling which encodes the eulerian digraph obtained by adding a loop based at any vertex in $\vec{G}$ with odd out-degree. 

Consider now any pair of faces $v$ and $w$ that are adjacent on an edge $a$ in a tiling $\tau_{\vec{G}}$, such that $a$ points from $v$ to $w$ in $\vec{G} \in \vec{\fG}^{[t]}$. By adding a new edge to $\tau_{\vec{G}}$ that connects the same pair of vertices as $a$, one obtains a new tiling which encodes an eulerian digraph in $\vec{\fG}^{[t]}$ that is obtained by subdividing arrow $a$ in $\vec{G}$ (just as in figure~\ref{fig1}). No new vertices are introduced in $\tau_{\vec{G}}$ by this operation and one new face is created, bounded by just the new edge and $a$. As opposed to loop addition, subdivision therefore has a clear interpretation as the operation described above within the class of bipartite tilings of the torus. This move of duplicating an edge in a tiling is well-known in the literature \cite{Hanany:2008fj,Davey:2009sr,Davey:2009bp}, and is sometimes referred to as \lq edge doubling'. Note that any tiling which encodes a smooth eulerian digraph must therefore contain no faces that are bounded by just two edges. Since each face is bounded by an even number of edges, the minimal number of bounding edges for any face in a tiling which encodes a smooth eulerian digraph is four. Any such quadrilateral face in a tiling $\tau_{\vec{G}}$ must be bounded by four different edges since it represents a balanced out-degree $2$ vertex in $\vec{G}$ (i.e. this vertex is necessarily loopless). Furthermore, every face in a tiling $\tau_{\vec{G}}$ is quadrilateral only if $\vec{G} \in \vec{\fF}_2^{[t]}$.  
   
For any two different faces $v$ and $w$ that are adjacent on an edge $a$ in a tiling $\tau_{\vec{G}}$, such that arrow $a$ points from $v$ to $w$ in $\vec{G} \in \vec{\fG}^{[t]}$, contracting arrow $a$ in $\vec{G}$ (as in figure~\ref{fig1}) corresponds to deleting the bounding edge $a$ in $\tau_{\vec{G}}$ and identifying faces $v=w$. However, the removal of a bounding edge in this way does not always define a new tiling encoding the eulerian digraph in $\vec{\fG}^{[t]}$ obtained via contraction of the corresponding arrow. For instance, if arrow $a$ is not the only arrow connecting vertices $v$ and $w$ in $\vec{G}$ then its contraction must create at least one loop based at the identified vertex $v=w$. Since ${\mathrm{deg}}^+(v=w) = {\mathrm{deg}}^+(v) + {\mathrm{deg}}^+(w) -1$ then clearly there can exist no tiling which encodes the resulting eulerian digraph whenever the out-degree of one of the two vertices $v$ and $w$ is even and that of the other one is odd. On the other hand, if arrow $a$ is the only arrow connecting vertices $v$ and $w$ in $\vec{G}$ (i.e. it is undirected simple) then the removal of edge $a$ in tiling $\tau_{\vec{G}}$, corresponding to contracting $a$ in $\vec{G}$, always defines a new tiling. More generally, in cases where a new tiling is obtained by removing an edge, this move is also well-known in the brane tiling literature \cite{Douglas:1997de,Morrison:1998cs,Feng:2000mi,GarciaEtxebarria:2006aq} and has a clear physical interpretation in terms of \lq Higgsing' the matter field associated with arrow $a$ in the corresponding supersymmetric quiver gauge theory.

Given a tiling $\tau_{\vec{G}}$ which contains a quadrilateral face $v$ bounded by edges labelled $a$, $b$, $c$ and $d$, such that the pair of arrows $(a,b)$/$(c,d)$ point to/from the out-degree $2$ vertex $v$ in $\vec{G} \in \vec{\fG}^{[t]}$, the cyclic ordering of the boundary edges in a given direction around face $v$ must be either $(acbd)$ or $(adbc)$, since $\tau_{\vec{G}}$ is bipartite. Splitting vertex $v$ in $\vec{G}$ (as in figure~\ref{fig2}) to produce either digraph $\vec{H}_=$ or $\vec{H}_\times$ requires the identification in $\tau_{\vec{G}}$ of either adjacent edges $a=c$ and $b=d$ or $a=d$ and $b=c$, in both cases such that the face $v$ they enclosed in $\tau_{\vec{G}}$ is collapsed. Note that the vertices in $\tau_{\vec{G}}$ which form the corners of face $v$ where the four pairs $(a,c)$, $(a,d)$, $(b,c)$, $(b,d)$ of boundary edges meet must be such that the $(a,c)$ and $(b,d)$ corners have one colour while the $(a,d)$ and $(b,c)$ corners have the opposite colour. However, these corner vertices need not all be different. For the case of $\vec{H}_= \in \vec{\fG}^{[t-1]}$, in order to identify adjacent edges $a=c$ and $b=d$ and collapse face $v$ in $\tau_{\vec{G}}$, it is necessary that the $(a,d)$ and $(b,c)$ corner vertices are different in $\tau_{\vec{G}}$ and must also be identified to obtain a tiling $\tau_{\vec{H}_=}$ provided neither of the other corner vertices $(a,c)$ and $(b,d)$ is bivalent. Similarly, for the case of $\vec{H}_\times \in \vec{\fG}^{[t-1]}$, in order to identify adjacent edges $a=d$ and $b=c$ and collapse face $v$ in $\tau_{\vec{G}}$, it is necessary that the $(a,c)$ and $(b,d)$ corner vertices are different in $\tau_{\vec{G}}$ and must also be identified to obtain a tiling $\tau_{\vec{H}_\times}$ provided neither of the other corner vertices $(a,d)$ and $(b,c)$ is bivalent. More details on the implementation of this move will follow in the next section for the class of quadrilateral tilings which encode elements in $\vec{\fF}_2$. Sufficed to say that an immediate consequence of the preceding discussion is that, given a tiling $\tau_{\vec{H}}$ encoding an eulerian digraph $\vec{H}$ containing a pair of arrows $\alpha$ and $\beta$, the simple immersion shown in figure~\ref{fig4} has no canonical implementation on $\tau_{\vec{H}}$ unless the corresponding edges $\alpha$ and $\beta$ are adjacent in $\tau_{\vec{H}}$.

%%%%%%%%%%%%%%    
\section{Quadrilateral tilings}
\label{sec:quadtile}
 
In principle, continuing the structural analysis above would lead to a set of moves within the class of bipartite tilings of the torus (corresponding to certain composite moves within the class of eulerian digraphs they encode) that could be used to generate them all. In practice, this task is beyond the scope of the present work and henceforth we shall restrict attention to the subclass of quadrilateral tilings which encode loopless $2$-regular eulerian digraphs. Just as the subclass $\vec{\fF}_2 \subset \vec{\fG}$ plays a structurally significant r\^{o}le within the class of eulerian digraphs, so too the subclass of quadrilateral tilings will be found to display many of the important structural features within the class of tilings as a whole, but within a less complicated framework. The physical interpretation of these structural results in the context of supersymmetric quiver gauge theories based on quadrilateral tilings will be discussed in section~\ref{sec:quadbranetilings}.  

%%%%%%%%%%%%%%    
\subsection{Face types}
\label{sec:facetypes}

Any face $v$ in a tiling $\tau_{\vec{G}}$ which encodes $\vec{G} \in \vec{\fF}_2^{[t]}$ is bounded by four different edges which, for definiteness, we shall label $a$, $b$, $c$ and $d$ such that they appear in order $(a \circ d \bullet b \circ c \, \bullet )$ around the boundary of face $v$ (so the pair of arrows $(a,b)$/$(c,d)$ must point to/from vertex $v$ in $\vec{G}$). The $\circ$/$\bullet$ symbols denote the white/black vertices forming the corners of face $v$, which may correspond to either $2$, $3$ or $4$ different vertices in $\tau_{\vec{G}}$ (respectively describing either $2$, $3$ or $4$ different circuits in $\vec{G}$). It will be convenient to distinguish between the different types of quadrilateral faces which can occur in $\tau_{\vec{G}}$, to which end we introduce the following nomenclature. 

Face $v$ will be referred to as being {\emph{isolated}} if its corners correspond to four different vertices in $\tau_{\vec{G}}$. These vertices describe circuits in $\vec{G}$ of the form $(ad \gamma_\circ )$, $(bd \gamma_\bullet )$, $(bc \gamma^\circ )$ and $(ac \gamma^\bullet )$, where none of the trails $\gamma_\circ$, $\gamma_\bullet$, $\gamma^\circ$ and $\gamma^\bullet$ in $\vec{G}$ contain arrows $a$, $b$, $c$ and $d$.    

If the corners of face $v$ correspond to three different vertices in $\tau_{\vec{G}}$ and two of these vertices are white/black then the three different vertices describe circuits in $\vec{G}$ of the form $(ac \gamma^\bullet bd \gamma_\bullet )$/$(ad \gamma_\circ bc \gamma^\circ )$, $(ad \gamma_\circ )$/$(bd \gamma_\bullet )$ and $(bc \gamma^\circ )$/$(ac \gamma^\bullet )$. Thinking of this situation in terms of the identification of a fourth black/white vertex in an isolated face drawn on the torus, there are two topologically distinct ways in which this can occur. That is, any curve drawn inside the face which connects the two same colour vertices to be identified becomes closed upon their identification such that it is either contractible or not contractible to a point on the torus. If it is not contractible then face $v$ will be called a {\emph{chain}} in $\tau_{\vec{G}}$ and the identified vertex will be referred to as the {\emph{link}} of $v$. If it is contractible then face $v$ will be referred to as being {\emph{pinched}} and the identified vertex will be referred to as the {\emph{pinched vertex}} of $v$.  

If the corners of face $v$ correspond to two different vertices in $\tau_{\vec{G}}$ then these vertices describe circuits in $\vec{G}$ of the form $(ad \gamma_\circ bc \gamma^\circ )$ and $(ac \gamma^\bullet bd \gamma_\bullet )$. In terms of the identification of both a white and black vertex in an isolated face, this situation occurs such that any two curves drawn inside the face which connect both the white and black pairs of vertices to be identified must, upon the identification of these two vertices, define a basis for the first homology group of the torus. (Other situations like either one or both of the closed curves being contractible cannot occur.) In this case, face $v$ will be called a {\emph{grid}} in $\tau_{\vec{G}}$ and both its vertices will be referred to as links. 

%%%%%%%%%%%%%%%%%%%%    
\subsection{Recognising faces}
\label{sec:facerecog}  

Let us now consider the characteristic properties that each type of face $v$ in $\tau_{\vec{G}}$ must possess in order for it to encode a given scenario in figure~\ref{fig3} for vertex $v$ in $\vec{G}$. The impatient reader may wish to skip to section~\ref{sec:summary} for a summary.  

%%%%%%%%%%%%%%%%%%%%    
\subsubsection{Isolated faces}
\label{sec:isolatedfaces}  

If $v$ is an isolated face in $\tau_{\vec{G}}$, no more than two of its four different corner vertices can be bivalent (implying no more than two of the four trails $\gamma_\circ$, $\gamma_\bullet$, $\gamma^\circ$ and $\gamma^\bullet$ in $\vec{G}$ can be empty). If $v$ contained either three or four bivalent corner vertices, it would need to be adjacent to another quadrilateral face $w$ on all four of its boundary edges, but clearly no such $w$ can exist. If $v$ has no bivalent corner vertices then it cannot be adjacent to a face $w$ on any pair of adjacent boundary edges. If it could then $w$ must be a chain whose link is the corner vertex of $v$ on which the pair of boundary edges are adjacent. However, this is only possible if $v$ is also a chain (whose link is the corner vertex with the opposite colour to the link of $w$), contradicting the assumption that it is an isolated face. Thus, if $v$ has no bivalent corner vertices, it can only be adjacent to the same face on a pair of boundary edges on opposite sides of $v$. Consequently, any isolated face $v$ with no bivalent corner vertices is adjacent to either four, three or two different faces in $\tau_{\vec{G}}$, corresponding respectively to the first, second and third scenarios in the first row of figure~\ref{fig3}. If $v$ has one bivalent corner vertex then it must be adjacent to a face $w$ on the pair of boundary edges connected to the bivalent vertex. The bivalent corner vertex of $v$ must therefore also be a bivalent corner vertex of $w$. This corresponds to the first scenario in the second row of figure~\ref{fig3}. If $v$ has two bivalent corner vertices with the same colour then it must be adjacent to a face $w$ on a pair of boundary edges connected to one bivalent vertex and adjacent to another face $x$ on the other two boundary edges connected to the other bivalent vertex. Therefore one bivalent corner vertex of $v$ must be a bivalent corner vertex of $w$ and the other bivalent corner vertex of $v$ must be a bivalent corner vertex of $x$. This corresponds to the second scenario in the second row of figure~\ref{fig3}. If $v$ has two bivalent corner vertices with opposite colours then it must be adjacent to a face $w$ on the three boundary edges connected to these two bivalent vertices. Therefore both bivalent corner vertices of $v$ must also be bivalent corner vertices of $w$. This corresponds to the third scenario in the second row of figure~\ref{fig3}. By a process of elimination, there exists no isolated face $v$ which can represent the fourth scenario in the second row of figure~\ref{fig3}.

%%%%%%%%%%%%%%%%%%%%    
\subsubsection{Chains and pinched faces}
\label{sec:chainsandpinchedfaces}

If $v$ is either a chain or a pinched face in $\tau_{\vec{G}}$ with two of its three corner vertices white/black then its remaining black/white corner vertex cannot be bivalent. By definition, all the four boundary edges of $v$ must end on this black/white corner vertex so it must be at least $4$-valent. 

If $v$ is a chain in $\tau_{\vec{G}}$, its link defines a partition of the four boundary edges into two pairs, with each pair lying to either side of the non-contractible closed curve running through $v$ and its link on the torus. If a face $w$ in $\tau_{\vec{G}}$ is adjacent to $v$ on exactly two of its four boundary edges, these two edges cannot lie on either side of the non-contractible curve such that they are adjacent around the boundary of $v$. If they could then $w$ would need to be a chain (or a grid) sharing the same white/black link as $v$. However, this would require the identification of a black/white corner vertex in $v$ as another link, implying $v$ must be a grid and contradicting the assumption that it is a chain. Thus, any given face in $\tau_{\vec{G}}$ must be adjacent to $v$ on no more than two of its four boundary edges since it is inevitable that the situation above would be encountered whenever $v$ is adjacent to another face on at least three boundary edges. Consequently, there exists no chain $v$ which can represent the third and fourth scenarios in the second row of figure~\ref{fig3}. The remaining scenarios are best explained in terms of the composition of the four trails $\gamma_\circ$, $\gamma_\bullet$, $\gamma^\circ$ and $\gamma^\bullet$ in $\vec{G}$ that are encoded by the three corner vertices of chain $v$. If none of these trails is empty (implying no corner vertices of $v$ are bivalent), $v$ can be adjacent to either four, three or two different faces in $\tau_{\vec{G}}$, corresponding respectively to the first, second and third scenarios in the first row of figure~\ref{fig3}. When a face $w$ in $\tau_{\vec{G}}$ is adjacent to $v$ on exactly two boundary edges here, $w$ is also a chain with the same link as $v$ and the two boundary edges lie on either side of the non-contractible curve but are not adjacent around the boundary of $v$ (i.e. they are on opposite sides of $v$). Since the white/black link of $v$ is never bivalent, $v$ will also contain no bivalent corner vertices when exactly one or both of the trails $\gamma_\circ / \gamma_\bullet$ and $\gamma^\circ / \gamma^\bullet$ is empty (both being empty means that the link of $v$ is exactly $4$-valent). For each such empty trail, there must exist a face $w$ in $\tau_{\vec{G}}$ that is adjacent to $v$ on exactly two boundary edges which both lie to one side of the non-contractible curve (i.e. the side on which there are no edges associated with the empty trail). Moreover, the face $w$ must also be a chain whose link is one of the black/white corner vertices of $v$ on the corresponding side of the non-contractible curve. The first or second scenario in the second row of figure~\ref{fig3} is represented respectively by one or both of these trails being empty. However, this is not the only way in which the last two scenarios can be represented by a chain in $\tau_{\vec{G}}$. If either $\gamma_\bullet / \gamma_\circ$ or $\gamma^\bullet / \gamma^\circ$ is empty then $v$ contains one bivalent black/white corner vertex and must be adjacent to a face $w$ in $\tau_{\vec{G}}$ on the two boundary edges connected to the bivalent vertex on the corresponding side of the non-contractible curve. The face $w$ must therefore also be a chain sharing the same white/black link and black/white bivalent corner vertex with $v$. This situation is only possible provided the other trail $\gamma_\circ / \gamma_\bullet$ or $\gamma^\circ / \gamma^\bullet$ on the same side of the non-contractible curve as the empty one is not also empty. On the other hand, the remaining trail $\gamma^\circ / \gamma^\bullet$ or $\gamma_\circ / \gamma_\bullet$ encoded by the white/black link of $v$ on the other side of the non-contractible curve may or may not be empty. If it is not empty, $v$ represents the first scenario in the second row of figure~\ref{fig3}. If it is empty, $v$ represents the second scenario in the second row of figure~\ref{fig3}. The only remaining possibility is that $\gamma_\bullet / \gamma_\circ$ and $\gamma^\bullet / \gamma^\circ$ are both empty (with $\gamma_\circ / \gamma_\bullet$ and $\gamma^\circ / \gamma^\bullet$ not empty), in which case $v$ contains two bivalent black/white corner vertices and is adjacent to two different faces in $\tau_{\vec{G}}$; one on each pair of boundary edges connected to the two bivalent vertices on either side of the non-contractible curve. Both adjacent faces are chains sharing the same white/black link with $v$ and each face shares a different black/white bivalent corner vertex with $v$. In this case, $v$ again represents the second scenario in the second row of figure~\ref{fig3}.

If $v$ is a pinched face in $\tau_{\vec{G}}$, it defines a partition of $\tau_{\vec{G}}$ such that any vertices and edges not contained in the boundary of $v$ must be contained in either the {\emph{interior}} or {\emph{exterior}} of $v$ in $\tau_{\vec{G}}$ (with each region adjacent to $v$ on precisely two of its four boundary edges). Consequently, copies of a given face in $\tau_{\vec{G}}$ can never appear in both the interior and exterior regions of any pinched face in $\tau_{\vec{G}}$ (and any face contained in the interior region of a pinched face must be either isolated or pinched). Thus, any given face in $\tau_{\vec{G}}$ must be adjacent to $v$ on no more than two of its four boundary edges. Moreover, for any face $w$ that is adjacent to $v$ on exactly two of its boundary edges, these two edges must correspond to the boundary of either the interior or exterior of $v$ in $\tau_{\vec{G}}$. Either way, $w$ must also share with $v$ the two corner vertices with opposite colour on which the two boundary edges end. Therefore, if $w$ is in the interior of $v$ in $\tau_{\vec{G}}$, it must also be a pinched face, {\emph{nested}} inside $v$. If $w$ is in the exterior of $v$ in $\tau_{\vec{G}}$, it must again also be a pinched face, with $v$ nested inside it. In either case, the two boundary edges must represent a pair of arrows in $\vec{G}$, with one arrow pointing from $v$ to $w$ and the other arrow pointing from $w$ to $v$.  Consequently, there exists no pinched face $v$ which can represent any of the four scenarios in figure~\ref{fig3} involving a pair of arrows which connect the same two vertices with the same orientation. The three remaining scenarios correspond to the different nesting options for $v$. The first scenario in the first row of figure~\ref{fig3} is represented by any $v$ which neither nests nor is nested in another pinched face in $\tau_{\vec{G}}$. The first scenario in the second row of figure~\ref{fig3} is represented by any $v$ which either nests or is nested in another pinched face in $\tau_{\vec{G}}$. The second scenario in the second row of figure~\ref{fig3} is represented by any $v$ which nests one pinched face and is nested in another pinched face in $\tau_{\vec{G}}$. 

%%%%%%%%%%%%%%%%%%%%    
\subsubsection{Grids}
\label{sec:grids}

If $v$ is a grid in $\tau_{\vec{G}}$, by definition, all four boundary edges of $v$ must end on both its links. Therefore both vertices in any grid must be at least $4$-valent. Furthermore, the four boundary edges of $v$ enclose a region in $\tau_{\vec{G}}$ whose interior is contained within a fundamental domain of the tiling. Consequently, no vertex contained within this region can be the link for a face in $\tau_{\vec{G}}$. If each face contained within this region is either isolated or pinched then $v$ represents the first scenario in the first row of figure~\ref{fig3}. The only way a face $w$ in $\tau_{\vec{G}}$ can be a chain or a grid is if it shares one or both links with $v$. If $w$ is a chain in $\tau_{\vec{G}}$, sharing one link with $v$ implies that $v$ must be the only grid in $\tau_{\vec{G}}$ and that every other chain in $\tau_{\vec{G}}$ must have the same link as $w$. Aside from the link, a chain $w$ in $\tau_{\vec{G}}$ must contain either one or both of its other corner vertices (with the opposite colour to the link) in the interior of the region enclosed by $v$, otherwise $w$ would also be a grid. If it contains both these corner vertices in the interior region then $w$ cannot be adjacent to $v$ since there exist no common boundary edges. If it contains just one corner vertex in the interior region then $w$ can be adjacent to $v$ on up to two adjacent boundary edges (and no more than two chains in $\tau_{\vec{G}}$ can be adjacent to $v$ on the maximum number of two boundary edges). If $\tau_{\vec{G}}$ contains any number of chains, but with none of them adjacent to $v$ on two boundary edges, then $v$ again represents the first scenario in the first row of figure~\ref{fig3} (any chain that is adjacent to $v$ on a single boundary edge must correspond to one of the four vertices connected to $v$ by an arrow in $\vec{G}$). If one of the chains in $\tau_{\vec{G}}$ is adjacent to $v$ on two boundary edges, $v$ represents the first scenario in the second row of figure~\ref{fig3}. If two of the chains in $\tau_{\vec{G}}$ are both adjacent to $v$ on two boundary edges, $v$ represents the second scenario in the second row of figure~\ref{fig3}. If $w$ is a grid in $\tau_{\vec{G}}$, sharing both links with $v$ implies that every other face in $\tau_{\vec{G}}$ must be either isolated or pinched. Thus, any tiling can contain no more than two grids and any tiling with exactly two grids must contain no chains. A grid $w$ in $\tau_{\vec{G}}$ may be adjacent to $v$ on some number of boundary edges they have in common. If they share no more than one boundary edge, $v$ represents the first scenario in the first row of figure~\ref{fig3}. If they share two boundary edges on opposite sides, $v$ represents the second scenario in the first row of figure~\ref{fig3}. If they share two boundary edges on adjacent sides, $v$ represents the first scenario in the second row of figure~\ref{fig3}. If they share three boundary edges, $v$ represents the third scenario in the second row of figure~\ref{fig3}. Finally, if they share all four boundary edges, $v$ represents the fourth scenario in the second row of figure~\ref{fig3}. In this last case, the pair of grids $v$ and $w$ are the only faces in the tiling we shall henceforth refer to as $\tau_+$ (since it encodes the unique element $\vec{G}_+ \in \vec{\fF}_2^{[2]}$ depicted in figure~\ref{fig5}). By a process of elimination, there exists no grid $v$ which can represent the third scenario in the first row of figure~\ref{fig3}.
 
%%%%%%%%%%%%%%%%%%%%    
\subsubsection{Summary}
\label{sec:summary} 

Before moving on to look at how the different types of faces may be collapsed, let us end this subsection by collecting a few important conclusions drawn from the analysis above.
\begin{itemize}

\item The first scenario in the first row of figure~\ref{fig3} can be represented by a face $v$ of any type, provided $v$ is adjacent to four different faces in $\tau_{\vec{G}}$. For each type of face, this implies none of the trails $\gamma_\circ$, $\gamma_\bullet$, $\gamma^\circ$ and $\gamma^\bullet$ in $\vec{G}$ is empty and therefore $v$ contains no bivalent corner vertices.   

\item The second scenario in the first row of figure~\ref{fig3} can be represented by a face $v$ of any type except pinched, provided $v$ is adjacent to three different faces in $\tau_{\vec{G}}$ with one face adjacent on two boundary edges on opposite sides of $v$. This again implies none of the trails $\gamma_\circ$, $\gamma_\bullet$, $\gamma^\circ$ and $\gamma^\bullet$ in $\vec{G}$ is empty and $v$ contains no bivalent corner vertices.   

\item The third scenario in the first row of figure~\ref{fig3} can only be represented by a face $v$ that is either isolated or a chain, provided $v$ is adjacent to two different faces in $\tau_{\vec{G}}$ with each face adjacent on two boundary edges on opposite sides of $v$. Once again, this implies none of the trails $\gamma_\circ$, $\gamma_\bullet$, $\gamma^\circ$ and $\gamma^\bullet$ in $\vec{G}$ is empty and $v$ contains no bivalent corner vertices.

\item The first scenario in the second row of figure~\ref{fig3} can be represented by a face $v$ of any type, provided $v$ is adjacent to three different faces in $\tau_{\vec{G}}$ with one face adjacent on two boundary edges on adjacent sides of $v$. For each type of face, this implies exactly one of the trails $\gamma_\circ$, $\gamma_\bullet$, $\gamma^\circ$ and $\gamma^\bullet$ in $\vec{G}$ must be empty. If $v$ is isolated, this means it must contain a single bivalent corner vertex. If $v$ is a chain, the empty trail must be associated with either the link of $v$ or one of its other two corner vertices. Only if it is not associated with the link will $v$ contain a single bivalent corner vertex (corresponding to the other corner vertex that is associated with the empty trail). If $v$ is pinched, it must be either nested in or have nested within it another pinched face in $\tau_{\vec{G}}$. Again, only if the empty trail is not associated with the pinched vertex of $v$ will $v$ contain a single bivalent corner vertex (in which case the bivalent corner vertex is shared with the adjacent pinched face that either nests or is nested in $v$). If $v$ is a grid, it contains no bivalent corner vertices irrespective of which trail is empty. 

\item The second scenario in the second row of figure~\ref{fig3} can be represented by a face $v$ of any type, provided $v$ is adjacent to two different faces in $\tau_{\vec{G}}$ with each face adjacent on two boundary edges on adjacent sides of $v$. For each type of face, this implies exactly two of the trails $\gamma_\circ$, $\gamma_\bullet$, $\gamma^\circ$ and $\gamma^\bullet$ in $\vec{G}$ must be empty. If $v$ is isolated, this means it must contain two bivalent corner vertices with the same colour. If $v$ is a chain, each empty trail must be associated with either the link of $v$ or one of its other two corner vertices. If both empty trails are associated with the link then $v$ will contain no bivalent corner vertices. If just one empty trail is associated with the link then the corner vertex of $v$ associated with the other empty trail is bivalent. If neither empty trail is associated with the link then the other two corner vertices of $v$ with the same colour are both bivalent. If $v$ is pinched, it must be nested in one pinched face and have nested in it another pinched face in $\tau_{\vec{G}}$. Again, if both empty trails are associated with the pinched vertex of $v$ then $v$ will contain no bivalent corner vertices. If just one empty trail is associated with the pinched vertex then the corner vertex of $v$ associated with the other empty trail is bivalent (and is shared with an adjacent pinched face that either nests or is nested in $v$). If neither empty trail is associated with the pinched vertex then the other two corner vertices of $v$ with the same colour are both bivalent (one being shared with the face that nests $v$ and the other being shared with the face that is nested in $v$). If $v$ is a grid, it contains no bivalent corner vertices though both empty trails must be associated with the same link and both faces adjacent to $v$ must be chains.

\item The third scenario in the second row of figure~\ref{fig3} can only be represented by a face $v$ that is either isolated or a grid, provided $v$ is adjacent to two different faces in $\tau_{\vec{G}}$ with one face adjacent to $v$ on three boundary edges. If $v$ is isolated, this means it must contain two bivalent corner vertices with the opposite colour that are connected by a boundary edge. Moreover, the face adjacent to $v$ on three boundary edges must also be isolated and share all four corner vertices with $v$. Indeed any pair of bivalent vertices that are connected by an edge in $\tau_{\vec{G}}$ must be enclosed by two isolated faces of this form, as shown in figure~\ref{fig7}. 
\begin{figure}[h!]
\includegraphics[scale=1.2]{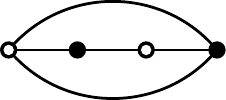} 
\caption{Two bivalent vertices connected by an edge in a quadrilateral tiling.}
\label{fig7}
\end{figure}
If $v$ is a grid, it contains no bivalent corner vertices and the face adjacent to $v$ on three boundary edges must also be a grid. In this case, each remaining face in $\tau_{\vec{G}}$ must be either isolated or pinched and $\vec{G}$ must follow by replacing one arrow in $\vec{G}_+$ with some eulerian trail (containing vertices and arrows encoded by the isolated and pinched faces in  $\tau_{\vec{G}}$).  

\item The fourth scenario in the second row of figure~\ref{fig3} must be represented by $\tau_+$ (i.e. this tiling with two grids and no other data is the only one which can encode $\vec{G}_+$).

\item No more than two grids can be contained in any $\tau_{\vec{G}}$. If $\tau_{\vec{G}}$ contains two grids then it must contain no chains and both links for each grid must be the same. If the two grids are adjacent on $0 \leq p \leq 4$ boundary edges in $\tau_{\vec{G}}$ then $\vec{G}$ must follow by replacing $4-p$ arrows in $\vec{G}_+$ with $4-p$ eulerian trails. The $4-p$ boundary edges not shared by each grid pair up to enclose $4-p$ regions, with each region containing only isolated and pinched faces encoding the corresponding eulerian trail in $\vec{G}$.   

\end{itemize}

%%%%%%%%%%%%%%    
\subsection{Collapsing faces}
\label{sec:facecollapse}

Depending on which type of face $v$ is and how it is contained in $\tau_{\vec{G}}$, it may be possible to identify edges $a=c$, $b=d$ and two different white corner vertices or edges $a=d$, $b=c$ and two different black corner vertices to produce a new quadrilateral tiling in which $v$ has been {\emph{collapsed}}. Clearly this is not possible if $v$ is a grid but may be possible for the remaining face types. Since $\tau_{\vec{G}}$ encodes $\vec{G} \in \vec{\fF}_2^{[t]}$, whenever face $v$ can be collapsed it must correspond to a loopless splitting of vertex $v$ in $\vec{G}$, defining a tiling $\tau_{\vec{H}_=}$ or $\tau_{\vec{H}_\times}$ which encodes $\vec{H}_=$ or $\vec{H}_\times$ in $\vec{\fF}_2^{[t-1]}$. However, this need not always imply that for each loopless splitting of a vertex $v$ in $\vec{G}$, it is possible to collapse the corresponding face $v$ in $\tau_{\vec{G}}$ which encodes it. Let us now describe precisely which options are available for collapsing the different face types. It will be convenient to label the faces which are adjacent to face $v$ on edges $a$, $b$, $c$ and $d$ respectively by $A$, $B$, $C$ and $D$ (with the proviso that these four adjacent faces need not all be different -- the different possible identifications corresponding to the inequivalent arrow connections for vertex $v$ in $\vec{G}$ in figure~\ref{fig3}). 

%%%%%%%%%%%%%%%%%%%%    
\subsubsection{Isolated faces}
\label{sec:isolatedfaces2}  

If $v$ is an isolated face in $\tau_{\vec{G}}$ then it can be collapsed to define $\tau_{\vec{H}_=}$ provided $\vec{H}_= \in \vec{\fF}_2^{[t-1]}$ (whence $A \neq C$ and $B \neq D$ implying trails $\gamma^\bullet$ and $\gamma_\bullet$ are not empty so neither black corner vertex of $v$ is bivalent) or it can be collapsed to define $\tau_{\vec{H}_\times}$ provided $\vec{H}_\times \in \vec{\fF}_2^{[t-1]}$ (whence $A \neq D$ and $B \neq C$ implying trails $\gamma_\circ$ and $\gamma^\circ$ are not empty so neither white corner vertex of $v$ is bivalent). 

If $A$, $B$, $C$ and $D$ are all different, corresponding to the first scenario in the first row of figure~\ref{fig3}, $v$ can be collapsed in both ways so long as $\vec{H}_=$ and $\vec{H}_\times$ are both weakly connected. Recall that this must always be the case except when $\vec{G}$ can be obtained by attaching two disjoint eulerian trails to the four vertices $A$, $B$, $C$ and $D$ adjacent to $v$ such that $\vec{G} \in \vec{\fF}_2^{[t]}$ is based on an eulerian circuit of the form $(ac \gamma_1 bd \gamma_2 )$ or $(ad \gamma_1 bc \gamma_2 )$, for some pair of disjoint loopless $2$-regular eulerian trails $\gamma_1$ and $\gamma_2$. In that case, $\vec{H}_=$ or $\vec{H}_\times$ will be weakly connected but $\vec{H}_\times$ or $\vec{H}_=$ will not be. However, since face $v$ is isolated in $\tau_{\vec{G}}$, there must exist four circuits in $\vec{G}$ of the form $(ad \gamma_\circ )$, $(bd \gamma_\bullet )$, $(bc \gamma^\circ )$ and $(ac \gamma^\bullet )$ with $a$, $b$, $c$ and $d$ absent from each trail $\gamma_\circ$, $\gamma_\bullet$, $\gamma^\circ$ and $\gamma^\bullet$. Clearly it is impossible to realise all four circuits of this kind in the type eulerian digraph described above -- for $\vec{G}$ based on $(ac \gamma_1 bd \gamma_2 )$ or $(ad \gamma_1 bc \gamma_2 )$, there exist no circuits of the form $(bd \gamma_\bullet )$, $(ac \gamma^\bullet )$ or $(ad \gamma_\circ )$, $(bc \gamma^\circ )$. Consequently, any face $v$ with $A$, $B$, $C$ and $D$ all different in any $\tau_{\vec{G}}$ for which only one loopless splitting of vertex $v$ in $\vec{G}$ is possible cannot be isolated. Thus any isolated face $v$ with $A$, $B$, $C$ and $D$ all different can always be collapsed in both ways. If $A=B$ or $C=D$ with no further identifications, corresponding to the second scenario in the first row of figure~\ref{fig3}, $v$ can be collapsed in both ways. If $A=B$ and $C=D$ with no further identification, corresponding to the third scenario in the first row of figure~\ref{fig3}, $v$ can again be collapsed in both ways. 

If $A=C$ or $B=D$ ($A=D$ or $B=C$) with no further identifications, corresponding to the first scenario in the second row of figure~\ref{fig3}, $v$ can be collapsed in only one way to define $\tau_{\vec{H}_\times}$ ($\tau_{\vec{H}_=}$). If $A=C$ and $B=D$ ($A=D$ and $B=C$) with no further identification, corresponding to the second scenario in the second row of figure~\ref{fig3}, $v$ can again be collapsed in only one way to define $\tau_{\vec{H}_\times}$ ($\tau_{\vec{H}_=}$). 

If exactly three of $A$, $B$, $C$ and $D$ are the same, corresponding to the third scenario in the second row of figure~\ref{fig3}, face $v$ in $\tau_{\vec{G}}$ cannot be collapsed since there exists no loopless splitting of vertex $v$ in $\vec{G}$. Finally, $A=B=C=D$ cannot occur when $v$ is isolated.  

In summary, the ways in which an isolated face $v$ in $\tau_{\vec{G}}$ can be collapsed are precisely the ways in which the corresponding vertex $v$ in $\vec{G}$ can be split without creating a loop. This bijective correspondence will cease to exist for the other types of faces. In both the second and third cases above, up to a relabelling of arrows, $\vec{H}_=$ and $\vec{H}_\times$ are necessarily isomorphic though $\tau_{\vec{H}_=}$ and $\tau_{\vec{H}_\times}$ need not be. 

%%%%%%%%%%%%%%%%%%%%    
\subsubsection{Chains and pinched faces}
\label{sec:chainsandpinchedfaces2}

If $v$ is either a chain or a pinched face in $\tau_{\vec{G}}$ then the way it may be collapsed is dictated by the colour of its link or pinched vertex. If this colour is white then $v$ can only be collapsed to define $\tau_{\vec{H}_\times}$ provided $\vec{H}_\times \in \vec{\fF}_2^{[t-1]}$. If this colour is black then $v$ can only be collapsed to define $\tau_{\vec{H}_=}$ provided $\vec{H}_= \in \vec{\fF}_2^{[t-1]}$. 

If $A$, $B$, $C$ and $D$ are all different, whether a chain or a pinched face, $v$ can be collapsed in exactly one way to define $\tau_{\vec{H}_=}$/$\tau_{\vec{H}_\times}$ when the link or pinched vertex of $v$ is black/white so long as $\vec{H}_=$/$\vec{H}_\times$ is weakly connected. This could only fail to occur when $\vec{G}$ is based on an eulerian circuit of the form $(ad \gamma_1 bc \gamma_2 )$/$(ac \gamma_1 bd \gamma_2 )$, for some pair of disjoint loopless $2$-regular eulerian trails $\gamma_1$ and $\gamma_2$. However, since the link or pinched vertex of $v$ is black/white, there must exist three circuits in $\vec{G}$ of the form $(ac \gamma^\bullet bd \gamma_\bullet )$/$(ad \gamma_\circ bc \gamma^\circ )$, $(ad \gamma_\circ )$/$(bd \gamma_\bullet )$ and $(bc \gamma^\circ )$/$(ac \gamma^\bullet )$ with $a$, $b$, $c$ and $d$ absent from each trail $\gamma_\circ$, $\gamma_\bullet$, $\gamma^\circ$ and $\gamma^\bullet$. Clearly though it is impossible to realise any one of these three kinds of circuits in the type of eulerian digraph described above. Whence, any chain or pinched face $v$ with $A$, $B$, $C$ and $D$ all different can always be collapsed in the unique manner described above. If $A=B$ or $C=D$ with no further identifications, $v$ must be a chain and can be collapsed in exactly one way. If $A=B$ and $C=D$ with no further identification, again $v$ must be a chain and can be collapsed in exactly one way. Note that in each of the three previous cases, even when both $\vec{H}_=$ and $\vec{H}_\times$ are in $\vec{\fF}_2^{[t-1]}$, there is always just one way to collapse $v$ defining $\tau_{\vec{H}_=}$/$\tau_{\vec{H}_\times}$ when the link or pinched vertex of $v$ is black/white.

If either $A=C$, $B=D$, $A=D$ or $B=C$ with no further identifications, whether a chain or a pinched face, $v$ can be collapsed in exactly one way to define $\tau_{\vec{H}_=}$/$\tau_{\vec{H}_\times}$ when the link or pinched vertex of $v$ is black/white so long as $\vec{H}_=$/$\vec{H}_\times$ is loopless. If the link or pinched vertex of $v$ is black then $\vec{H}_= \in \vec{\fF}_2^{[t-1]}$ in the $A=D$ and $B=C$ cases while the other $A=C$ and $B=D$ cases cannot occur. If the link or pinched vertex of $v$ is white then $\vec{H}_\times \in \vec{\fF}_2^{[t-1]}$ in the $A=C$ and $B=D$ cases while the other $A=D$ and $B=C$ cases cannot occur. The reason why two of the four possible identifications are always excluded follows from the analysis in section~\ref{sec:chainsandpinchedfaces}. If $v$ is a chain, the argument in the second paragraph of section~\ref{sec:chainsandpinchedfaces} shows that the adjacency with $v$ in the two excluded cases would require $v$ to be a grid. If $v$ is a pinched face, the argument in the third paragraph of section~\ref{sec:chainsandpinchedfaces} shows that the adjacency with $v$ in the two excluded cases would require a prohibited identification of two faces in the interior and exterior of $v$. Whence, any chain or pinched face $v$ with $A=C$, $B=D$, $A=D$ or $B=C$ and no further identifications can always be collapsed in the unique manner described above. If both $A=C$ and $B=D$ (or both $A=D$ and $B=C$) with no further identification, the permitted adjacencies imply that the link or pinched vertex of $v$ must be white (or black) and $v$ can be collapsed in exactly one way to define $\tau_{\vec{H}_\times}$ (or $\tau_{\vec{H}_=}$).  

Finally, there exists no chain or pinched face $v$ which can represent the third and fourth scenarios in the second row of figure~\ref{fig3}, such that at least three of $A$, $B$, $C$ and $D$ are the same.

In summary, any chain or pinched face $v$ in $\tau_{\vec{G}}$ with a white/black link or pinched vertex can be collapsed in exactly one way to define a tiling $\tau_{\vec{H}_\times}$/$\tau_{\vec{H}_=}$. In each case, this corresponds to the loopless splitting of vertex $v$ in $\vec{G} \in \vec{\fF}_2^{[t]}$ which yields $\vec{H}_\times$/$\vec{H}_= \in \vec{\fF}_2^{[t-1]}$. In the first case where $A$, $B$, $C$ and $D$ are all different, there could exist another loopless splitting of $v$ in $\vec{G}$ yielding $\vec{H}_=$/$\vec{H}_\times \in \vec{\fF}_2^{[t-1]}$ but there is no way to represent this by collapsing the corresponding chain or pinched face $v$ in $\tau_{\vec{G}}$. If $v$ is a chain, the same applies in the second (third) case where $A=B$ or (and) $C=D$ with no further identification. Recall that $\vec{H}_=$ and $\vec{H}_\times$ are isomorphic (up to relabelling arrows) in these two cases but the white/black colour of the link of $v$ dictates the tiling $\tau_{\vec{H}_\times}$/$\tau_{\vec{H}_=}$ which can be obtained from collapsing $v$. In all remaining cases, the unique way in which $v$ can be collapsed in $\tau_{\vec{G}}$ corresponds to the unique loopless splitting of vertex $v$ in $\vec{G}$. 
 
%%%%%%%%%%%%%%    
\subsection{Removing bivalent vertices and pinched faces}
\label{sec:removebivalentpinched}

Let us now consider how to deconstruct a given quadrilateral tiling $\tau_{\vec{G}}$. It is convenient to begin this task by first removing from $\tau_{\vec{G}}$ those substructures which do not play a structurally significant r\^{o}le in the tiling. In section~\ref{sec:branetilings}, it will be seen that the innocuous substructures we consider here actually indicate a physical inconsistency in the associated superconformal quiver gauge theories. Thus their removal is both structurally and physically desirable. 

The first substructure corresponds to any face $v$ in $\tau_{\vec{G}}$ which contains either one bivalent corner vertex or two bivalent corner vertices with the same colour. Assuming $\tau_{\vec{G}}$ contains at least one such face then $v$ cannot be a grid and it must represent the first or second scenario in the second row of figure~\ref{fig3} when it contains respectively one or two bivalent corner vertices. If $v$ is a chain or a pinched face, the link or pinched vertex cannot be bivalent. Thus, whether the face $v$ is isolated, pinched or a chain, it can always be collapsed such that a bivalent corner vertex is identified with the other corner vertex with the same colour (which is also bivalent when $v$ contains two bivalent corner vertices). If $v$ is isolated, the two corner vertices with opposite colour to the bivalent one(s) are both at least trivalent. The aforementioned collapse of $v$ will produce a new tiling with a bivalent corner vertex of $v$ in $\tau_{\vec{G}}$ removed and with either one of the two other corner vertices with opposite colour made into a bivalent vertex in the new tiling only if it is exactly trivalent in $\tau_{\vec{G}}$. If $v$ is a chain or a pinched face, the link or pinched vertex must be at least $4$-valent and the unique collapse of $v$ will produce a new tiling, again with a bivalent corner vertex of $v$ in $\tau_{\vec{G}}$ removed, and with the link or pinched vertex made bivalent in the new tiling only if it is exactly $4$-valent in $\tau_{\vec{G}}$. Repeating this procedure must eventually result in a tiling which contains no faces with either one bivalent corner vertex or two bivalent corner vertices with the same colour (though it may contain some number of faces with two different coloured bivalent corner vertices).          

The next substructure corresponds to any pinched face $v$ in $\tau_{\vec{G}}$. Let us assume that the procedure described in the paragraph above has already been performed so that $\tau_{\vec{G}}$ contains no faces with either one bivalent corner vertex or two bivalent corner vertices with the same colour but that it does contain at least one pinched face $v$ (with no bivalent corner vertices). There always exists a unique way of collapsing $v$ that identifies its two corner vertices with the opposite colour to that of the pinched vertex. This collapse creates a bivalent vertex only if the pinched vertex of $v$ is exactly $4$-valent. However, the pinched vertex of $v$ is exactly $4$-valent only if it nests one pinched face and is nested in another pinched face in $\tau_{\vec{G}}$ (such that the pinched vertices for these interior and exterior nested faces correspond to the pair of corner vertices of $v$ that are not pinched). The same applies to both these other pinched faces. The sequence of nestings must therefore continue in this way until a pinched face whose pinched vertex is at least $5$-valent (and with no bivalent corner vertices) is reached in both the interior and exterior regions of $v$ in $\tau_{\vec{G}}$. Collapsing either one of these pinched faces in the unique manner does not create a bivalent vertex. Furthermore, the nested pinched face it was adjacent to in $\tau_{\vec{G}}$ is also a pinched face in the new tiling whose pinched vertex must be at least $5$-valent (and with no bivalent corner vertices). Proceeding in this manner, one can systematically remove all pinched faces from $\tau_{\vec{G}}$ without creating any new bivalent vertices. The reason why pinched faces are prioritised amongst the other types of faces which can be collapsed in a unique manner without creating bivalent vertices is that the collapse of a pinched face never modifies the type of any of the remaining faces in the tiling. Their removal will therefore not modify the structure of $\tau_{\vec{G}}$ too radically. Recall that any face which is adjacent to $v$ on two boundary edges in $\tau_{\vec{G}}$ must also be pinched and such that one of them is nested in the other. The sequence of interior and exterior nestings for a given pinched face ends when the respective pairs of interior and exterior boundary edges both separate two different faces in $\tau_{\vec{G}}$. Consequently, after collapsing all the pinched faces in $\tau_{\vec{G}}$, any two edges which connect the same pair of vertices in the resulting tiling must enclose a region separating two different faces. 

The final substructure corresponds to any face $v$ in $\tau_{\vec{G}}$ which contains two different coloured bivalent corner vertices. From section~\ref{sec:summary}, we recognise that any such $v$ must be an isolated face that is adjacent to another isolated face $w$ in $\tau_{\vec{G}}$ on three boundary edges, as depicted in figure~\ref{fig7}. Let us assume that the procedures described in the two paragraphs above have already been performed so that $\tau_{\vec{G}}$ contains no pinched faces nor any faces with just one bivalent corner vertex or two bivalent corner vertices with the same colour. The two boundary edges not shared by any such pair of adjacent isolated faces $v$ and $w$ in $\tau_{\vec{G}}$ must connect the same pair of vertices (which are both at least $4$-valent since $\tau_{\vec{G}}$ contains no pinched faces) and enclose a region (containing just $v$ and $w$) separating two different faces in $\tau_{\vec{G}}$. Although it is impossible to collapse either $v$ or $w$, one can define a new tiling $\tau_{\vec{H}}$ by deleting the two bivalent vertices they enclose together with the three boundary edges on which they are adjacent plus any one of their two other boundary edges. That is, the region containing $v$ and $w$ which separated two different faces in $\tau_{\vec{G}}$ is just replaced by a single edge in  $\tau_{\vec{H}}$. Clearly this operation does not modify the type of any of the remaining faces in the tiling and therefore cannot create a pinched face. If one of the two non-bivalent corner vertices of $v$ and $w$ is exactly $4$-valent in $\tau_{\vec{G}}$, it will be made into a bivalent vertex in $\tau_{\vec{H}}$ and one must then subject $\tau_{\vec{H}}$ to the procedures described in the two paragraphs above. If both of these vertices are $4$-valent in $\tau_{\vec{G}}$ then they are made into a pair of bivalent vertices connected by a single edge in $\tau_{\vec{H}}$, and therefore necessarily form the corner vertices of a pair of isolated faces that are adjacent on three boundary edges in $\tau_{\vec{H}}$. In this case, one needs only to repeat the operation described above in order to remove them. 

In summary, we have have found that subjecting any tiling to a combination of the three procedures described above must result in a tiling with no bivalent vertices and no pinched faces. Henceforth we shall refer to any quadrilateral tiling containing no bivalent vertices and no pinched faces as being {\emph{smooth}}. The set of all smooth quadrilateral tilings will be denoted by $\eQ$ and $\eQ^{[t]} \subset \eQ$ denotes the set of all smooth quadrilateral tilings with $t$ vertices (whence $2t$ edges and $t$ faces), whose elements encode eulerian digraphs in a subset of $\vec{\fF}_2^{[t]}$.

\begin{figure}[h!]
\includegraphics[scale=1.2]{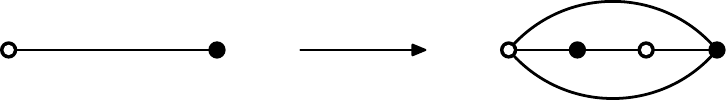} 
\caption{Lagging an edge in a tiling.}
\label{fig8}
\end{figure}

The new operation defined above which produces the tiling $\tau_{\vec{H}}$ from $\tau_{\vec{G}}$, via the removal of two bivalent vertices connected by an edge, encodes precisely the reverse of the composite move shown in figure~\ref{fig6} that is required to generate $\vec{\fF}_2$. In particular, if the single edge separating two different faces in $\tau_{\vec{H}}$ encodes the arrow in $\vec{H} \in \vec{\fF}_2^{[t-2]}$ depicted in the dashed circle on the left hand side of figure~\ref{fig6} then $\vec{G} \in \vec{\fF}_2^{[t]}$ which results from this move on the right hand side is exactly what $\tau_{\vec{G}}$ encodes (such that the two new balanced out-degree two vertices in $\vec{G}$ are identified with the adjacent isolated faces $v$ and $w$ in $\tau_{\vec{G}}$). For any tiling $\tau_{\vec{H}}$ which encodes some $\vec{H} \in \vec{\fF}_2^{[t]}$, let us define the {\emph{lagging}} of an edge $a$ in $\tau_{\vec{H}}$ to be the replacement depicted in figure~\ref{fig8}
of $a$ in $\tau_{\vec{H}}$ with the ensemble in figure~\ref{fig7} between the same pair of vertices to define a new tiling $\tau_{\vec{G}}$. The tiling $\tau_{\vec{G}}$ encodes $\vec{G} \in \vec{\fF}_2^{[t+2]}$ obtained by applying the composite move shown in figure~\ref{fig6} to arrow $a$ in $\vec{H}$.  

\begin{figure}[h!]
\includegraphics[scale=1.2]{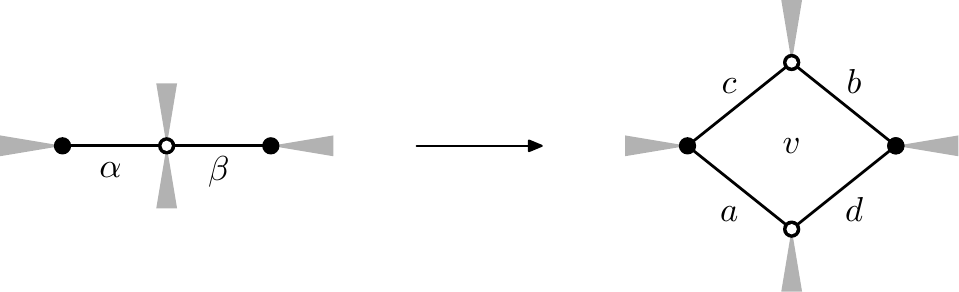} 
\caption{Reconstructing a face $v$ in $\tau_{\vec{G}}$.}
\label{fig9}
\end{figure}

Recall that collapsing a face in $\tau_{\vec{G}}$ is identified with performing a loopless splitting of the corresponding vertex in $\vec{G}$ and the reverse move, simple immersion, corresponds to the other move required to generate $\vec{\fF}_2$. If it is possible to collapse a face $v$ in a tiling $\tau_{\vec{G}}$ which encodes $\vec{G} \in \vec{\fF}_2^{[t]}$, to define a tiling $\tau_{\vec{H}}$ which encodes $\vec{H} \in \vec{\fF}_2^{[t-1]}$, the two pairs of adjacent boundary edges in $v$ that are identified in its collapse define two edges $\alpha$ and $\beta$ in $\tau_{\vec{H}}$ (either $\alpha =a=c$, $\beta =b=d$ for $\vec{H} = \vec{H}_=$ or $\alpha =a=d$, $\beta =b=c$ for $\vec{H} = \vec{H}_\times$). The edges $\alpha$ and $\beta$ must both end on at least one common vertex in $\tau_{\vec{H}}$ (i.e. the one identified in the collapse of $v$ in $\tau_{\vec{G}}$) but need not bound the same face in $\tau_{\vec{H}}$. If $v$ is an isolated face in $\tau_{\vec{G}}$, $\alpha$ and $\beta$ must end on only one common vertex in $\tau_{\vec{H}}$ (i.e. their other two endpoints must correspond to two different vertices in $\tau_{\vec{H}}$). If $v$ is a chain in $\tau_{\vec{G}}$, $\alpha$ and $\beta$ must end on the same two vertices in $\tau_{\vec{H}}$, such that they describe a non-contractible cycle connecting these two vertices on the torus. If $v$ is a pinched face in $\tau_{\vec{G}}$, $\alpha$ and $\beta$ must end on the same two vertices in $\tau_{\vec{H}}$, such that they describe a contractible cycle connecting these two vertices on the torus. In all cases, the reverse move mapping $\tau_{\vec{H}}$ to $\tau_{\vec{G}}$ will be referred to as {\emph{reconstructing}} a face $v$ in $\tau_{\vec{G}}$ and encodes precisely the simple immersion depicted in figure~\ref{fig4}, acting on the corresponding pair of arrows $\alpha$ and $\beta$ in $\vec{H}$ to produce $\vec{G}$ containing the new balanced out-degree $2$ vertex $v$. An example of the reconstruction of a face $v$ is shown in figure~\ref{fig9}. 
The shaded wedges indicate any other edges that are connected to the endpoints of edges $\alpha$ and $\beta$ in $\tau_{\vec{H}}$. (Their appearance in circulations around the corner vertices of $v$ in $\tau_{\vec{G}}$ encode the trails $\gamma_\circ$, $\gamma_\bullet$, $\gamma^\circ$ and $\gamma^\bullet$, giving three circuits of the form $(\alpha \gamma^\bullet )$, $(\beta \gamma_\bullet )$ and $(\alpha \gamma^\circ \beta \gamma_\circ )$ in $\vec{H}$ encoded by circulations around the endpoints of $\alpha$ and $\beta$ in $\tau_{\vec{H}}$.) This example shows $v$ as an isolated face in $\tau_{\vec{G}}$ with $\tau_{\vec{H}}$ obtained by collapsing $v$ such that the white corner vertices are identified. If $v$ were a chain or a pinched face in $\tau_{\vec{G}}$ then there would be only one other black corner vertex of $v$ in $\tau_{\vec{G}}$ such that $\alpha$ and $\beta$  in $\tau_{\vec{H}}$ would describe a non-contractible or contractible cycle on the torus. The general picture is therefore obtained in the obvious way, taking the negative of figure~\ref{fig9} when $\tau_{\vec{H}}$ is obtained by the collapse of $v$ which identifies its black corner vertices. Thus for any tiling $\tau_{\vec{H}}$ which encodes some $\vec{H} \in \vec{\fF}_2^{[t]}$, by selecting any vertex together with a pair of edges $\alpha$ and $\beta$ which both end on this vertex in $\tau_{\vec{H}}$, one can reconstruct a face $v$ to define a new tiling $\tau_{\vec{G}}$ in the manner described above. The tiling $\tau_{\vec{G}}$ obtained in figure~\ref{fig9} encodes precisely $\vec{G} \in \vec{\fF}_2^{[t+1]}$ obtained in figure~\ref{fig4} via the simple immersion of $\vec{H}$ in $\vec{G}$.  

The reversal of the three procedures used to eliminate all bivalent vertices and pinched faces from a given tiling are generated by special cases of the two operations described above. Any pair of bivalent vertices that are connected by an edge in a tiling can be obtained by lagging an edge in another tiling according to figure~\ref{fig8}. Any pinched face in a tiling can be obtained as a reconstructed face from another tiling in terms of some pair of edges $\alpha$ and $\beta$ in that tiling which form a contractible cycle on the torus connecting two vertices with opposite colour. Any face which contains a single bivalent corner vertex in a tiling can be obtained as a reconstructed face from another tiling in terms of some pair of edges $\alpha$ and $\beta$ which form adjacent boundary edges of a face $w$ in that tiling. The reconstructed face $v$ is adjacent to $w$ on exactly two common boundary edges connected to the bivalent vertex created in the new tiling. Any face which contains two bivalent corner vertices with the same colour can be obtained as a special case of this move such that the common vertex shared by $\alpha$ and $\beta$ is bivalent. In the example shown in figure~\ref{fig9}, this would correspond to either one or both of the shaded wedges attached to the white vertex containing no edges. Consequently we have now established that any quadrilateral tiling can be obtained from some smooth tiling in $\eQ$ by applying to it some combined iteration of the three moves described above.

%%%%%%%%%%%%%%    
\subsection{Smooth quadrilateral tilings}
\label{sec:smoothquadtile}

The preceding analysis has shown that any quadrilateral tiling can be deconstructed via collapsing faces and de-lagging edges. The deconstruction process must end only when a tiling is obtained containing no collapsible faces and no lagged edges. Such a tiling must contain only grids and is therefore isomorphic to the unique tiling $\tau_+ \in \eQ^{[2]}$ encoding $\vec{G}_+ \in \vec{\fF}_2^{[2]}$. Conversely, any quadrilateral tiling can be obtained via reconstructing faces and lagging edges in $\tau_+$. Naturally these two generating moves for quadrilateral tilings in figures~\ref{fig9} and \ref{fig8} encode precisely the two generating moves for loopless $2$-regular eulerian digraphs in figures~\ref{fig4} and \ref{fig6}. However, given a tiling $\tau_{\vec{H}}$ which encodes some $\vec{H} \in \vec{\fF}_2^{[t]}$, typically there will exist simple immersions of $\vec{H}$ in some $\vec{G} \in \vec{\fF}_2^{[t+1]}$ which involve pairs of arrows $\alpha$ and $\beta$ in $\vec{H}$ that are encoded by pairs of edges which do not end on a common vertex in $\tau_{\vec{H}}$. Consequently, there exists no way to represent a simple immersion of this type by the reconstruction of a face in the associated tiling. There are two possible explanations. The first is that there exists no tiling $\tau_{\vec{G}}$ encoding $\vec{G}$ which follows from such a simple immersion. The second is that a $\tau_{\vec{G}}$ exists but the collapse of face $v$ in it which would encode the loopless splitting of vertex $v$ in $\vec{G}$ from which $\vec{H}$ is recovered is impossible. 

Let us now consider how to deconstruct a given quadrilateral tiling $\tau_{\vec{G}}$ within the class $\eQ$. To accomplish this objective, we must identify the options which are available for collapsing faces in $\tau_{\vec{G}}$ in such a way that no bivalent vertices and no pinched faces are produced, ensuring a new tiling in $\eQ$ is obtained. Any such collapse of a face in $\tau_{\vec{G}}$ will be referred to as a {\emph{smooth collapse}}. The first step is to parse the results in section~\ref{sec:summary} concerning the recognition of characteristic features of the different face types which can encode one of the scenarios in figure~\ref{fig3}, assuming now that $\tau_{\vec{G}} \in \eQ^{[t]}$ and restricting attention to only the collapsible face types which may be contained in $\tau_{\vec{G}}$ (i.e. ignoring pinched faces and grids). 

If $v$ is an isolated face in $\tau_{\vec{G}}$ then none of the trails $\gamma_\circ$, $\gamma_\bullet$, $\gamma^\circ$ and $\gamma^\bullet$ in $\vec{G}$ is empty and $v$ must encode one of the three scenarios in the first row of figure~\ref{fig3}. From the results of section~\ref{sec:isolatedfaces2}, $v$ can therefore always be collapsed in both ways. The new tiling obtained by collapsing $v$ will not contain a pinched face but will contain a bivalent vertex whenever either of the two corner vertices of $v$ with the opposite colour to the pair which are identified in collapsing $v$ is trivalent. For example, the collapse of $v$ obtained by reversing the arrow in figure~\ref{fig9} would produce a tiling $\tau_{\vec{H}} \in \eQ^{[t-1]}$ provided neither of the black corner vertices of $v$ in $\tau_{\vec{G}}$ is trivalent (i.e. provided the shaded wedges attached to the black corner vertices each contain at least two edges, encoding at least two arrows in both $\gamma_\bullet$ and $\gamma^\bullet$). Hence, there exists no smooth collapse of an isolated face $v$ in $\tau_{\vec{G}}$ if $v$ contains at least two trivalent corner vertices with opposite colours. There exists only one smooth collapse if $v$ contains exactly one trivalent corner vertex (or two trivalent corner vertices with the same colour), namely the collapse involving the identification of the trivalent vertex. Both ways of collapsing $v$ are smooth only when $v$ contains no trivalent corner vertices.  

If $v$ is a chain in $\tau_{\vec{G}}$ with a white (black) link vertex then neither of the two trails $\gamma_\bullet$, $\gamma^\bullet$  ($\gamma_\circ$, $\gamma^\circ$) in $\vec{G}$ is empty. If neither of the two remaining trails $\gamma_\circ$, $\gamma^\circ$ ($\gamma_\bullet$, $\gamma^\bullet$) encoded by the link of $v$ is empty, $v$ must encode one of the three scenarios in the first row of figure~\ref{fig3}. If either $\gamma_\circ$ or $\gamma^\circ$ ($\gamma_\bullet$ or $\gamma^\bullet$) is empty, $v$ must encode the first scenario in the second row of figure~\ref{fig3}. If both $\gamma_\circ$ and $\gamma^\circ$ ($\gamma_\bullet$ and $\gamma^\bullet$) are empty, $v$ must encode the second scenario in the second row of figure~\ref{fig3}. From the results of section~\ref{sec:chainsandpinchedfaces2}, $v$ can always be collapsed in exactly one way to define a tiling $\tau_{\vec{H}_\times}$ ($\tau_{\vec{H}_=}$) encoding $\vec{H}_\times$ ($\vec{H}_=$) in $\vec{\fF}_2^{[t-1]}$. The new tiling obtained by collapsing $v$ will not contain a pinched face but will contain a bivalent vertex whenever the link of $v$ is exactly $4$-valent. Clearly this can only occur in the last situation described above when both $\gamma_\circ$ and $\gamma^\circ$ ($\gamma_\bullet$ and $\gamma^\bullet$) are empty. Hence, the unique collapse of a chain $v$ in $\tau_{\vec{G}}$ is always smooth except when it encodes the second scenario in the second row of figure~\ref{fig3}.

Reversing the smooth collapse of $v$ corresponds to reconstructing a face $v$ in $\tau_{\vec{G}}$ such that no bivalent vertices and no pinched faces are produced. This just corresponds to the remaining type of face reconstruction that was discussed in section~\ref{sec:removebivalentpinched}, which shall henceforth be referred to as a {\emph{smooth reconstruction}} of $v$ in $\tau_{\vec{G}}$. If $v$ is an isolated face in $\tau_{\vec{G}}$ which does not contain two trivalent corner vertices with opposite colours then it can be obtained as a reconstructed face from another smooth tiling $\tau_{\vec{H}} \in \eQ^{[t-1]}$ in terms of some pair of edges $\alpha$ and $\beta$ which both end on exactly one common vertex and which do not form adjacent boundary edges for any face in $\tau_{\vec{H}}$. For example, the reconstruction of isolated face $v$ in $\tau_{\vec{G}}$ shown in figure~\ref{fig9} is smooth whenever $\tau_{\vec{H}}$ is smooth (thus $\gamma_\bullet$ and $\gamma^\bullet$ both contain at least two arrows) and neither $\gamma_\circ$ or $\gamma^\circ$ is empty. If $v$ is a chain in $\tau_{\vec{G}}$ which does not encode the second scenario in the second row of figure~\ref{fig3} then it can be obtained as a reconstructed face from another smooth tiling $\tau_{\vec{H}} \in \eQ^{[t-1]}$ in terms of some pair of edges $\alpha$ and $\beta$ which form a non-contractible cycle on the torus connecting two vertices with opposite colours in $\tau_{\vec{H}}$ encoding circuits of the form $( \alpha \gamma_\circ \beta \gamma^\circ )$ and $( \alpha \gamma_\bullet \beta \gamma^\bullet )$ in $\vec{H}$ such that at most one of the trails $\gamma_\circ$, $\gamma_\bullet$, $\gamma^\circ$ and $\gamma^\bullet$ can be empty. 

\begin{figure}[h!]
\includegraphics[scale=1.1]{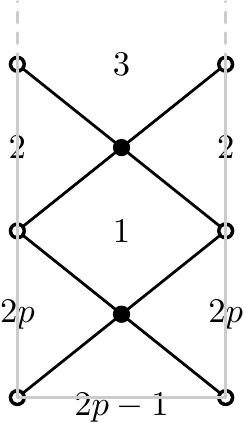} 
\caption{The unique smooth tiling $\tau_{\vec{A}_{2p}} \in \eQ^{[2p]}$.}
\label{fig10}
\end{figure}

\begin{figure}[h!]
\includegraphics{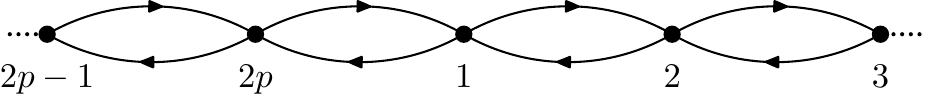} 
\caption{The necklace digraph $\vec{A}_{2p} \in \vec{\fF}_2^{[2p]}$.}
\label{fig11}
\end{figure}

The deconstruction of a given tiling $\tau_{\vec{G}} \in \eQ^{[t]}$ within the class $\eQ$ begins by continuing to smoothly collapse isolated faces and chains in $\tau_{\vec{G}}$ until a tiling $\tau_{\vec{H}} \in \eQ^{[t-s]}$ is obtained (after $s \geq 0$ iterations) which contains no faces that can be smoothly collapsed. By definition, any face in $\tau_{\vec{H}}$ must therefore be either a grid, a chain which encodes the second scenario in the second row of figure~\ref{fig3} or an isolated face containing at least two trivalent corner vertices with opposite colours. Let us now consider the possible composition of such faces in $\tau_{\vec{H}}$, utilising the results of section~\ref{sec:facerecog}. If $\tau_{\vec{H}}$ contains two grids then it must contain no chains and so any remaining faces must be isolated of the type described above. If $\tau_{\vec{H}}$ contains one grid then each chain in $\tau_{\vec{H}}$ must share the same link as one of the link vertices of the grid. However, each chain in $\tau_{\vec{H}}$ must also contain no bivalent corner vertices and encode the second scenario in the second row of figure~\ref{fig3}. Any such chain $v$ must therefore be adjacent to one chain $w$ on two boundary edges to one side of the link of $v$ and to another chain $x$ on the two boundary edges to the other side of the link of $v$. Since $v$ contains no bivalent vertices, the two corner vertices of $v$ to either side of its link must provide the links for $w$ and $x$ respectively. Thus $v$, $w$ and $x$ cannot share the same link, contradicting the assumption that $\tau_{\vec{H}}$ contains a grid and so $\tau_{\vec{H}}$ cannot contain both a grid and a chain. That is, if $\tau_{\vec{H}}$ contains one grid then any remaining faces must be isolated of the type described above. If $\tau_{\vec{H}}$ contains no grids then each chain must again encode the second scenario in the second row of figure~\ref{fig3}. As we just saw, any such chain $v$ must be adjacent to a pair of chains $w$ and $x$ running parallel with $v$ on either side of its link in $\tau_{\vec{H}}$. Since $w$ and $x$ are also chains in $\tau_{\vec{H}}$, they too must encode the second scenario in the second row of figure~\ref{fig3}. Iterating this logic therefore shows that the only way $\tau_{\vec{H}}$ may contain a chain is if it contains no other face types and consists solely of a number of adjacent chains running parallel to each other in $\tau_{\vec{H}}$. The necessary identification for $\tau_{\vec{H}}$ to define a bipartite tiling of the torus here requires the number of chains to be even, with $t-s=2p$ for some $p>1$. A picture of $\tau_{\vec{H}}$ is shown in figure~\ref{fig10}, with the chains labelled $i=1,...,2p$ such that chain $i$ is adjacent to chain $i-1$ below and chain $i+1$ above its link (modulo $2p$) and with links for odd/even values of $i$ coloured white/black (grey lines represent the periodic identifications to be made in the horizontal and vertical directions).
The tiling in figure~\ref{fig10} for $\tau_{\vec{H}}$ encodes $\vec{H}$ isomorphic to the \lq necklace' digraph $\vec{A}_{2p}$ depicted in figure~\ref{fig11}.       
(The case $p=1$ is excluded since $\tau_{\vec{H}}$ would be isomorphic to the unique tiling $\tau_+$ which encodes $\vec{G}_+ = \vec{A}_2$ and consists of just two grids.) It is easily seen that, modulo relabelling, figure~\ref{fig10} represents the unique smooth tiling which encodes the necklace digraph (whence, there exists no smooth tiling which encodes $\vec{A}_{2p+1}$ for any $p>0$). The only remaining possibility is that $\tau_{\vec{H}}$ contains only isolated faces with at least two trivalent corner vertices with opposite colours. In summary, $\tau_{\vec{H}}$ is either isomorphic to the tiling $\tau_{\vec{A}_{2p}}$ in figure~\ref{fig10} (for some $p>1$) or else it must contain no chains and each isolated face must contain at least two trivalent corner vertices with opposite colours. 

\begin{figure}[h!]
\includegraphics[scale=1.2]{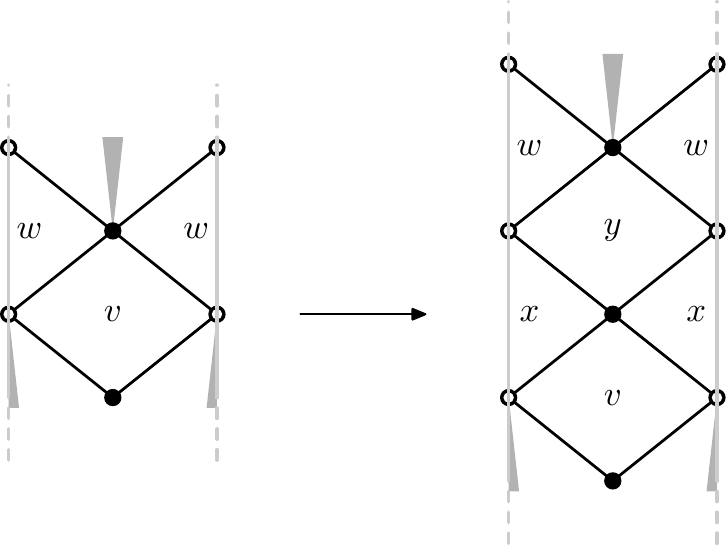} 
\caption{Composite move I mapping $\tau_{\vec{H}} \in \eQ^{[t]}$ to $\tau_{\vec{G}} \in \eQ^{[t+2]}$.}
\label{fig12}
\end{figure}

The deconstruction of tiling $\tau_{\vec{A}_{2p}}$ in figure~\ref{fig10} within the class $\eQ$ is straightforward. Collapsing any chain in $\tau_{\vec{A}_{2p}}$ will create a single bivalent vertex and thus defines a tiling $\tau_{\vec{A}_{2p-1}}$ which is not smooth. However, by collapsing one of the two chains in $\tau_{\vec{A}_{2p-1}}$ which contains this bivalent vertex as a corner, one obtains a smooth tiling isomorphic to $\tau_{\vec{A}_{2(p-1)}} \in \eQ^{[2(p-1)]}$ when $p>2$ or $\tau_+ \in \eQ^{[2]}$ when $p=2$. This composite collapse therefore defines a map within the class $\eQ$ which just reduces by one the value of $p$ in $\tau_{\vec{A}_{2p}}$. Whence, $\tau_+$ is always recovered after precisely $p-1$ iterations. Reversing this composite collapse describes a special case of a more general composite move mapping $\eQ^{[t]} \rightarrow \eQ^{[t+2]}$ that is defined as follows. Consider any smooth tiling $\tau_{\vec{H}} \in \eQ^{[t]}$ containing a pair of chains $v$ and $w$ which are adjacent on exactly two boundary edges but do not share the same link vertex (i.e. each shared boundary edge ends on the link of $v$ and the link of $w$). This means they must each encode either the first or second scenario in the second row of figure~\ref{fig3}, such that the corresponding vertices $v$ and $w$ in $\vec{H}$ are connected by a pair of arrows with opposite orientations which are encoded by the two common boundary edges of chains $v$ and $w$ in $\tau_{\vec{H}}$. By selecting the link of $w$ and the two common boundary edges of $v$ and $w$ which end on it, one can reconstruct a face $y$ in a new tiling. This encodes the simple immersion involving the pair of arrows connecting $v$ and $w$ in $\vec{H}$. The face $y$ is necessarily a chain sharing the same link as $v$ and is adjacent to $v$ on exactly two boundary edges connected to the single bivalent vertex in the new tiling (which forms a corner vertex for both $y$ and $v$). This tiling is therefore not smooth. The next step is to select in this tiling the link of $v$ and the two common boundary edges of $v$ and $y$ which end on it (their opposite endpoint being the bivalent vertex). This selection defines the reconstruction of another face $x$ to give a smooth tiling $\tau_{\vec{G}} \in \eQ^{[t+2]}$. It encodes the simple immersion in $\vec{G}$ involving the pair of arrows connecting vertices $v$ and $y$ in the digraph encoded by the aforementioned intermediate tiling which is not smooth. The face $x$ is necessarily a chain in $\tau_{\vec{G}}$ and it is adjacent to chain $y$ in $\tau_{\vec{G}}$ on exactly two boundary edges such that they do not share the same link vertex (i.e. in the same manner as chains $v$ and $w$ in $\tau_{\vec{H}}$). Thus the net effect of this operation is to sandwich these two new adjacent chains $x$ and $y$ in between the chains $v$ and $w$ in $\tau_{\vec{H}} \in \eQ^{[t]}$ to define $\tau_{\vec{G}} \in \eQ^{[t+2]}$. This operation will be referred to as {\emph{composite move I}} and is depicted in figure~\ref{fig12} (wherein the links of $v$ and $w$ are coloured white and black). 
It is straightforward to check that performing the two constituent face reconstructions in the opposite order, by selecting first the link of $v$ and then the link of $w$, defines a smooth tiling which is isomorphic to $\tau_{\vec{G}}$.   

Let us now consider how to deconstruct within the class $\eQ$ a smooth tiling $\tau_{\vec{G}} \in \eQ^{[t]}$ containing no chains and whose isolated faces all have at least two trivalent corner vertices with opposite colours. The first step is to understand the possible composition of such faces in $\tau_{\vec{G}}$. If $\tau_{\vec{G}}$ contains no grids then all its faces are isolated. If $\tau_{\vec{G}}$ contains one grid then all its other faces are isolated and must be contained within a single region enclosed by the four boundary edges of the grid (the interior of this region is contained within a fundamental domain of $\tau_{\vec{G}}$). If $\tau_{\vec{G}}$ contains two grids then all its other faces are isolated and must be contained within one of four possible regions enclosed by pairs of boundary edges of the two grids, in the manner described in the final item of section~\ref{sec:summary}. 

\begin{figure}[h!]
\includegraphics[scale=1.2]{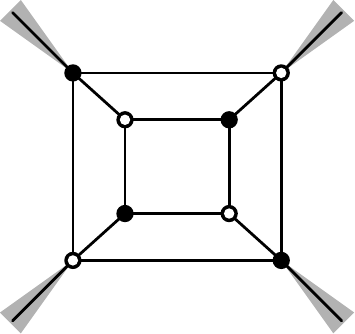} \hspace*{.8in}
\includegraphics[scale=1.2]{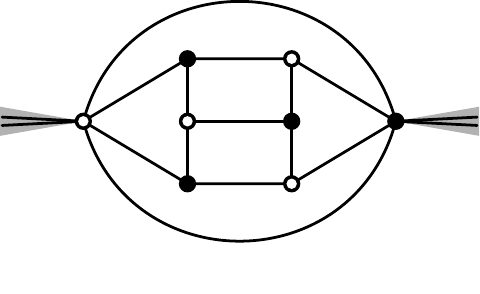}
\caption{The two possible neighbourhoods of a $4$-face in $\tau_{\vec{G}}$.}
\label{fig13}
\end{figure}

It will be convenient to refer to an isolated face as a $p$-face when it contains exactly $p\geq 2$ trivalent corner vertices with at least two having opposite colours (so any isolated face in $\tau_{\vec{G}}$ is a $p$-face with $p$ equal to either $2$, $3$ or $4$). Since the link vertices of a grid are both at least $4$-valent, any grid in $\tau_{\vec{G}}$ must be adjacent to either another grid or a $2$-face on each of its boundary edges (and if a $2$-face then the two trivalent corner vertices must be contained in the interior of one of the aforementioned regions). Consequently, any $3$- or $4$-face in $\tau_{\vec{G}}$ must be adjacent to another $p$-face in $\tau_{\vec{G}}$ on each of its four boundary edges. These adjacent isolated faces need not all have the same value of $p$ though the values are restricted by the assumed structure of $\tau_{\vec{G}}$. In particular, any $4$-face in $\tau_{\vec{G}}$ must be adjacent to either four $2$-faces or one $4$-face, two $3$-faces and one $2$-face (and with the two $3$-faces adjacent on opposite sides). Indeed it is straightforward to check that the neighbourhood of faces adjacent to any $4$-face in $\tau_{\vec{G}}$ must look like one of the two scenarios in figure~\ref{fig13}. 
Each of the four shaded wedges in the first scenario must contain at least one edge while each of the two shaded wedges in the second scenario must contain at least two edges. 

\begin{figure}[h!]
\includegraphics[scale=1.2]{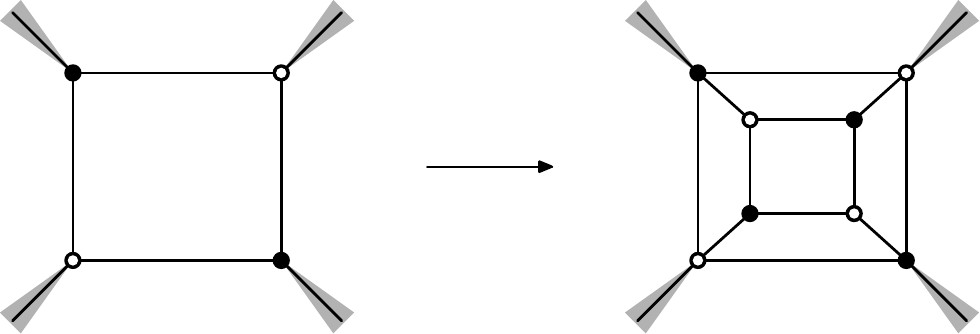} 
\caption{Composite move II mapping $\tau_{\vec{H}} \in \eQ^{[t]}$ to $\tau_{\vec{G}} \in \eQ^{[t+4]}$.}
\label{fig14}
\end{figure}

In the first scenario in figure~\ref{fig13}, one can obtain a new tiling $\tau_{\vec{H}} \in \eQ^{[t-4]}$ (when $t \geq 6$) by deleting the four trivalent corner vertices of the $4$-face together with the eight edges connected to them in $\tau_{\vec{G}}$. This can also be thought of as the composite collapse of each of the four $2$-faces adjacent to the $4$-face, though each collapse on its own is not smooth and produces a single bivalent vertex. The net effect of this composite collapse is the replacement of the $4$-face and its four adjacent $2$-faces in $\tau_{\vec{G}}$ with a single face $v$ in $\tau_{\vec{H}}$. The four corner vertices of $v$ in $\tau_{\vec{H}}$ each have degree one less than the corresponding vertices in $\tau_{\vec{G}}$, while the degrees of all the other vertices contained in both $\tau_{\vec{H}}$ and $\tau_{\vec{G}}$ are unchanged. Consequently, $\tau_{\vec{H}}$ must be smooth since each corner vertex of $v$ is at least trivalent (i.e. no bivalent vertices and no pinched faces are produced by this operation). Each face in $\tau_{\vec{H}}$ must be of the same type as in $\tau_{\vec{G}}$ but, depending on the number of grids contained in $\tau_{\vec{G}}$, $v$ could be either a grid, a chain or an isolated face which need not contain at least two trivalent corner vertices with opposite colours. However, in any eventuality, one can continue the deconstruction procedure on $\tau_{\vec{H}}$ within the class $\eQ$ that was described above, via smoothly collapsing faces in all possible ways until a simpler smooth tiling is inevitably recovered which contains only grids and $p$-faces. Reversing the composite collapse described in this paragraph defines what we shall refer to as {\emph{composite move II}} and consists of inserting a single $4$-face inside a face in any smooth tiling $\tau_{\vec{H}} \in \eQ^{[t]}$, as shown in figure~\ref{fig14}, to define a new smooth tiling $\tau_{\vec{G}} \in \eQ^{[t+4]}$.
Identifying the chosen face in $\tau_{\vec{H}} \in \eQ^{[t]}$ above with a vertex $v$ in $\vec{H} \in \vec{\fF}_2^{[t]}$, notice that composite move II encodes precisely the operation shown in figure~\ref{fig6c} involving subdivision and mutation with respect to $v$. As mentioned in section~\ref{sec:bipartitetilingsT2}, subdividing an arrow $a$ in $\vec{H}$ corresponds to inserting a new edge in $\tau_{\vec{H}}$ connecting the same pair of vertices as edge $a$. Whence, the intermediate subdivision of the four arrows connected to $v$ in $\vec{H}$ (denoted by grey dots in figure~\ref{fig6c}) corresponds to duplicating each of the four boundary edges of the corresponding face in $\tau_{\vec{H}}$ above. The identification above has been observed already in \cite{Franco:2005rj} and our composite move II corresponds a special case of their figure 13 (i.e. they are not working within the class of smooth quadrilateral tilings).   

\begin{figure}[h!]
\includegraphics[scale=1.2]{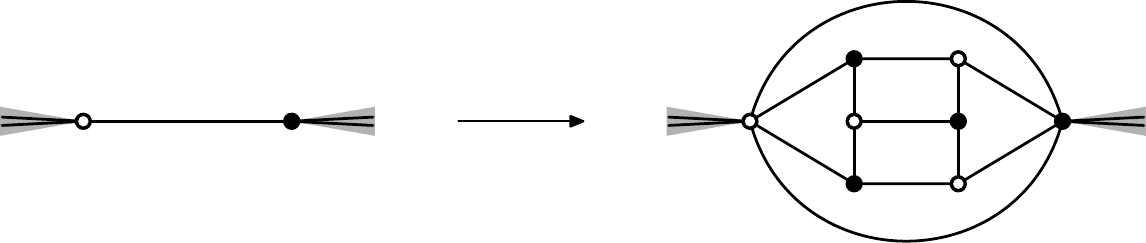} 
\caption{Composite move III mapping $\tau_{\vec{H}} \in \eQ^{[t]}$ to $\tau_{\vec{G}} \in \eQ^{[t+6]}$.}
\label{fig15}
\end{figure}

In the second scenario in figure~\ref{fig13}, one can again obtain a new tiling by deleting from either one of the two $4$-faces its four trivalent corner vertices and the eight edges connected to them. This tiling is not smooth since the aforementioned deletion necessarily produces a lagged edge (see figure~\ref{fig8}) containing two bivalent vertices connected by an edge. However, by de-lagging this edge, a tiling $\tau_{\vec{H}} \in \eQ^{[t-6]}$ (when $t \geq 8$) is obtained. The net effect of this composite collapse and de-lagging is to replace all the vertices and edges in $\tau_{\vec{G}}$ depicted between the two vertices that the shaded wedges are connected to in the second scenario in figure~\ref{fig13} with a single edge in $\tau_{\vec{H}}$. Only the degrees of the two vertices connected to the shaded wedges are modified by this operation, with each one having degree three less in $\tau_{\vec{H}}$ than it had in $\tau_{\vec{G}}$. Consequently, $\tau_{\vec{H}}$ must be smooth since each of these vertices is at least trivalent (i.e. no bivalent vertices and no pinched faces are produced by this operation). Furthermore, not only is each face in $\tau_{\vec{H}}$ of the same type as in $\tau_{\vec{G}}$ but each isolated face in $\tau_{\vec{H}}$ must be a $p$-face. The subsequent deconstruction of $\tau_{\vec{H}}$ within $\eQ$ will therefore not involve the smooth collapse of any of its faces. If $\tau_{\vec{H}}$ contains a $4$-face then either the procedure described in this paragraph or the one above defines its subsequent deconstruction within $\eQ$, depending respectively on whether the neighbourhood of the $4$-face looks like the second or first scenario in figure~\ref{fig13}. Reversing the composite collapse and de-lagging described in this paragraph defines what we shall refer to as {\emph{composite move III}} and consists of replacing an edge in any smooth tiling $\tau_{\vec{H}} \in \eQ^{[t]}$ with the ensemble in the second scenario in figure~\ref{fig13} to define a new smooth tiling $\tau_{\vec{G}} \in \eQ^{[t+6]}$. This operation is shown in figure~\ref{fig15}.

The analysis above has shown how the reversal of composite moves II and III define the deconstruction within $\eQ$ of any $4$-face in a smooth tiling containing only grids and $p$-faces. Thus, when combined with the smooth collapse of faces, the deconstruction can only be obstructed by the existence of a smooth tiling $\tau_{\vec{G}}$ containing only grids and $p$-faces with $p=2,3$. It therefore remains to show that no such $\tau_{\vec{G}}$ (except $\tau_+$ with two grids and no other faces) exists. By definition, a $2$-face contains exactly one boundary edge connecting its two trivalent corner vertices. If $\tau_{\vec{G}}$ contains a $2$-face $v$ then the face $w$ adjacent to it on the aforementioned boundary edge must be a $p$-face (i.e. the existence of a boundary edge connecting two trivalent corner vertices of $w$ means it cannot be a grid). However, $w$ must have $p=4$ or else one boundary edge of $v$ and one boundary edge of $w$ that both end on the same trivalent vertex must form part of the boundary of another isolated face that is not a $p$-face (i.e. it must contain two non-trivalent corner vertices with the same colour). Whence, either option contradicts the assumed structure of $\tau_{\vec{G}}$ and so $\tau_{\vec{G}}$ must contain no $2$-faces. There now remains only the possibility that $\tau_{\vec{G}}$ contains only grids and $3$-faces. Clearly this is impossible if $\tau_{\vec{G}}$ contains one grid since any $p$-face adjacent to the grid must have $p=2$. If $\tau_{\vec{G}}$ contains two grids then it must be isomorphic to $\tau_+$. We therefore now have only to exclude the possibility that $\tau_{\vec{G}}$ contains only $3$-faces. To do so, consider the non-trivalent corner vertex of any $3$-face in $\tau_{\vec{G}}$, which must have degree $4+q$ for some $q \geq 0$. This $(4+q)$-valent vertex must define the non-trivalent corner vertex for precisely $4+q$ $3$-faces in $\tau_{\vec{G}}$, such that one $3$-face is adjacent to the next on successive edges circulating around this vertex. By definition, all three other corner vertices for each of these $3$-faces must be trivalent. The union of all these $3$-faces can therefore be thought of as forming a $2(4+q)$-sided polygon on $2(4+q)$ trivalent vertices in $\tau_{\vec{G}}$ with precisely $4+q$ edges connecting the single interior $(4+q)$-valent vertex to each of the $4+q$ trivalent vertices with the opposite colour on the corners of the polygon. The remaining $4+q$ trivalent vertices (with the same colour as the interior one) on the corners of the polygon each have one remaining edge and each of these edges must be connected to a single exterior vertex in order that they enclose quadrilateral faces. This final exterior vertex must therefore also be $(4+q)$-valent and have opposite colour to the $(4+q)$-valent vertex in the interior of the polygon. Tallying up, $\tau_{\vec{G}}$ must therefore contain $t= 2+2(4+q)$ vertices, $e= 4(4+q)$ edges and $n= 2(4+q)$ $3$-faces. However, this implies $t-e+n =2$ which describes the euler character of a sphere rather than a torus! We can therefore conclude that any smooth tiling containing only grids and $p$-faces must contain at least one $4$-face or else it is isomorphic to $\tau_+$. 

In summary, we have shown that any smooth tiling can be generated within the class $\eQ$ by applying to the unique tiling $\tau_+ \in \eQ^{[2]}$ some combined iteration of the four moves: \\ [-.3cm]

\begin{itemize}
\item Smooth reconstruction (see figure~\ref{fig9}), mapping $\eQ^{[t]} \rightarrow \eQ^{[t+1]}$. \\ [-.3cm]

\item Composite move I (see figure~\ref{fig12}), mapping $\eQ^{[t]} \rightarrow \eQ^{[t+2]}$. \\ [-.3cm]

\item Composite move II (see figure~\ref{fig14}), mapping $\eQ^{[t]} \rightarrow \eQ^{[t+4]}$. \\ [-.3cm]

\item Composite move III (see figure~\ref{fig15}), mapping $\eQ^{[t]} \rightarrow \eQ^{[t+6]}$. 
\end{itemize}

%%%%%%%%%%%%%%    
\subsection{The tiling function}
\label{sec:mapQtoF}

As was noted in section~\ref{sec:bipartitetilingsT2}, although every bipartite tiling of the torus encodes an eulerian digraph, the coding function which maps tilings to eulerian digraphs is not bijective. Using the preceding results in this section, we can now describe a few generic instances where bijectivity fails within the domain of smooth quadrilateral tilings in $\eQ$ which encode loopless $2$-regular eulerian digraphs in $\vec{\fF}_2$.    

%%%%%%%%%%%%%%    
\subsubsection{Lack of injectivity}
\label{sec:injectivity}

A situation where there must generically exist two non-isomorphic smooth tilings which encode the same eulerian digraph is as follows. Consider a tiling $\tau_{\vec{H}} \in \eQ^{[t]}$ which contains a pair of edges $\alpha$ and $\beta$ that form a non-contractible cycle on the torus connecting two vertices with opposite colours, such that these vertices encode circuits of the form $( \alpha \gamma_\circ \beta \gamma^\circ )$ and $( \alpha \gamma_\bullet \beta \gamma^\bullet )$ in $\vec{H}$ with none of the trails $\gamma_\circ$, $\gamma_\bullet$, $\gamma^\circ$ and $\gamma^\bullet$ being empty. Selecting the white vertex in $\tau_{\vec{H}}$ together with edges $\alpha$ and $\beta$ which end on it specifies the smooth reconstruction of a face $v$ which defines a chain with a black link in a tiling $\tau_{\vec{G}} \in \eQ^{[t+1]}$. On the other hand, selecting the black vertex in $\tau_{\vec{H}}$ together with edges $\alpha$ and $\beta$ which end on it specifies the smooth reconstruction of a face $v$ which defines a chain with a white link in a tiling $\tau^\prime_{\vec{G}} \in \eQ^{[t+1]}$. In general, the tilings $\tau_{\vec{G}}$ and $\tau^\prime_{\vec{G}}$ will not be isomorphic but they must both encode the same $\vec{G} \in \vec{\fF}_2^{[t+1]}$ which follows from $\vec{H} \in \vec{\fF}_2^{[t]}$ via the simple immersion shown in figure~\ref{fig4}. In both $\tau_{\vec{G}}$ and $\tau^\prime_{\vec{G}}$, relative to the labelling of arrows in $\vec{G}$ in figure~\ref{fig4}, the chain $v$ must have boundary edges $a$ and $d$ to one side of the link with $b$ and $c$ to the other side. 

One can take this process a stage further by selecting the link vertex of $v$ together with edges $b$ and $c$ which end on it in both $\tau_{\vec{G}}$ and $\tau^\prime_{\vec{G}}$. Doing so in $\tau_{\vec{G}}$ specifies the smooth reconstruction of a face $w$ which defines a chain with a white link in a tiling $\tau_{\vec{F}} \in \eQ^{[t+2]}$ (in which $v$ is also contained as a chain with a black link). On the other hand, doing this in $\tau^\prime_{\vec{G}}$ specifies the smooth reconstruction of a face $w$ which defines a chain with a black link in a tiling $\tau^\prime_{\vec{F}} \in \eQ^{[t+2]}$ (in which $v$ is also contained as a chain with a white link). In general, the tilings $\tau_{\vec{F}}$ and $\tau^\prime_{\vec{F}}$ will not be isomorphic but they must both encode the same $\vec{F} \in \vec{\fF}_2^{[t+2]}$ which follows from $\vec{G} \in \vec{\fF}_2^{[t+1]}$ by pinching together arrows $b$ and $c$ to form vertex $w$. Up to relabelling, selecting the link vertex of $v$ together with any other pair of edges from $a$, $b$, $c$ and $d$ in $\tau_{\vec{G}}$ and $\tau^\prime_{\vec{G}}$ which specifies the smooth reconstruction of a face will give rise to tilings that are always isomorphic.     

%%%%%%%%%%%%%%    
\subsubsection{Lack of surjectivity}
\label{sec:surjectivity}

We saw in section~\ref{sec:smoothquadtile} that the tiling $\tau_{\vec{A}_t}$ which encodes the necklace digraph $\vec{A}_t \in \vec{\fF}_2^{[t]}$ is unique, up to relabelling, and that $\tau_{\vec{A}_t} \in \eQ^{[t]}$ only if $t$ is even (see figure~\ref{fig10}). If $t$ is odd then $\tau_{\vec{A}_t}$ must contain exactly one bivalent vertex and so there is no smooth tiling which can encode the necklace digraph on an odd number of vertices.

From the sixth item in section~\ref{sec:summary}, any face $v$ in a smooth tiling which encodes the third scenario in the second row of figure~\ref{fig3} must be a grid and, from the eighth item, any tiling must contain no more than two grids. Whence there can exist no smooth tiling which encodes any digraph in $\vec{\fF}_2$ with more than one vertex of the form shown in the third scenario in the second row of figure~\ref{fig3}. 

\begin{figure}[h!]
\includegraphics{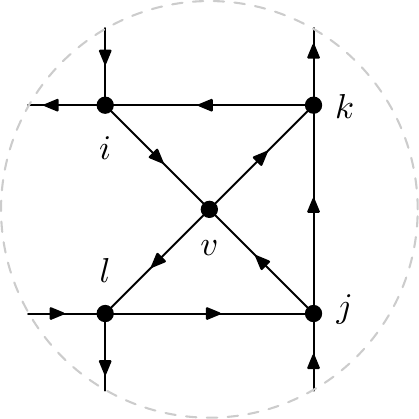}
\caption{No tiling can encode this configuration.}
\label{fig15a}
\end{figure}

It is worth remarking that some digraphs in $\vec{\fF}_2$ cannot be encoded by any quadrilateral tiling, whether smooth or not. For example, consider the existence of a tiling $\tau_{\vec{G}}$ such that $\vec{G}$ contains the arrangement of vertices and arrows shown in figure~\ref{fig15a}. 
Face $v$ must be adjacent to four different faces $i$, $j$, $k$ and $l$ in $\tau_{\vec{G}}$ on its four boundary edges. Since $\tau_{\vec{G}}$ is bipartite, face $v$ must be adjacent to $i$ and $j$ on opposite sides (and to $k$ and $l$ on its other two sides). Face $j$ must be adjacent to $v$ and $k$ on opposite sides (and to face $l$ on one other side). Face $k$ must be adjacent to $v$ and $j$ on opposite sides (and to face $i$ on one other side). However, the last two sentences imply that $v$ must be adjacent to $j$ and $k$ on opposite sides which contradicts the sentence preceding them. Whence, $\tau_{\vec{G}}$ does not exist.

%%%%%%%%%%%%%%    
\section{Brane tilings}
\label{sec:branetilings}
 
This section begins with a brief review of how bipartite tilings of the torus can be engineered via certain supersymmetric brane configurations in type IIB string theory, wherein they are referred to as {\emph{brane tilings}} \cite{Hanany:2005ve,Franco:2005rj,Franco:2005sm,Hanany:2005ss}. We will describe how the tiling data encodes both the gauge-matter couplings and superpotential in supersymmetric quiver gauge theories in four dimensions which provide an effective description of such brane configurations at low energies. The necessary consistency conditions for any such unitary theory to realise an exact superconformal symmetry will also be noted. We will then discuss the vacuum moduli spaces for this class of theories before concluding with some remarks on Seiberg duality. We will only sketch the salient features that will aid our analysis and a more detailed account of the construction can be found in either the original references above or reviews such as \cite{Kennaway:2007tq,Yamazaki:2008bt} and references therein. Summarising the relevant aspects of brane tilings here will prepare the way for a more detailed physical interpretation in section~\ref{sec:quadbranetilings} of the structural results obtained in previous sections, in the context supersymmetric quiver gauge theories based on quadrilateral tilings.        

%%%%%%%%%%%%%%    
\subsection{Brane picture}
\label{sec:branepicture}

Perhaps the most straightforward way to engineer a tiling involves a configuration of NS5- and D5-branes in IIB string theory. In this setup, all the branes extend along a common $(3+1)$-dimensional worldvolume and each NS5-brane wraps a holomorphic curve embedded in a $\CC^2$ spanning four of the six transverse directions. The canonically embedded $T^2 \subset \CC^2$ is precisely the torus on which the tiling is defined and the edges are described by its intersection with the aforementioned holomorphic curve for each NS5-brane. Each stack of coincident D5-branes in the configuration corresponds to a face in the tiling such that the stack extends within a region of the torus bounded by the edges formed by the intersection with each NS5-brane. The orientation of each NS5-brane provides an orientation for the corresponding edge in the tiling, pointing from one of the faces it bounds to the other one, whence the tiling encodes a weakly connected digraph. Interactions between the stacks of D5-branes are described by fundamental strings stretching across the NS5-brane boundary edges. A vertex in the tiling therefore describes a junction at which the corresponding sets of stretched fundamental strings interact locally. In general, there may be any number $N_i$ of coincident D5-branes in the stack associated with face $i$ in the tiling but it is only when all $N_i$ are equal that the tiling is necessarily bipartite, whence encoding an eulerian digraph. Any such configuration of $5$-branes preserves at least $\tfrac{1}{8}$ the maximal amount of supersymmetry and is thought to have a dual description in terms of D3-branes probing a toric conical singularity. 

%%%%%%%%%%%%%%    
\subsection{Quiver and superpotential}
\label{sec:quiverandsuperpotential}

At low energies, an effective description of the brane configuration described above is provided by a superconformal gauge theory on $\RR^{3,1}$ based on the following data extracted from the tiling. If the tiling contains $n$ faces and $e$ edges, the field content consists of $n$ gauge superfields and $e$ chiral matter superfields which are valued in the following representations of the gauge group $\cG = \prod_{i=1}^n U( N_i )$. The $i$th gauge superfield takes values in the lie algebra $\fu ( N_i )$, which should be thought of as the adjoint representation of $U( N_i )$. Let us denote by ${\bf N}_i$ the fundamental representation of $U ( N_i )$ (and ${\overline {\bf N}_i}$ its complex conjugate representation). The chiral matter superfields transform in a {\emph{quiver}} representation of $\cG$ defined such that the $a$th superfield transforms in the  representation $( {\bf N}_i , {\overline {\bf N}_j} )$ of $U( N_i ) \times U( N_j )$ if edge $a$ encodes an arrow in the digraph pointing from vertex $i$ to vertex $j$, or in the adjoint representation of $U ( N_i )$ if $a$ encodes a loop based at vertex $i$. The connections in the quiver can be encoded by an $n {\times} n$ {\emph{adjacency matrix}} $A$ which is defined such that entry $A_{ij}$ equals the number of arrows pointing from $i$ to $j$. The gauge coupling for each $U( N_i )$ factor in the quiver gauge theory is anomalous at one-loop unless $\sum_{j=1}^n A_{ij} N_j = \sum_{j=1}^n A_{ji} N_j$. If all $N_i$ equal the same number $N$, this is precisely the condition that all the vertices in the weakly connected digraph must be balanced. Henceforth we shall assume that this is the case such that the brane tilings of interest correspond to bipartite tilings of the torus.          

The data above fixes all the gauge-matter couplings in the classical lagrangian for the theory. It only remains to now specify the F-term superpotential. Supersymmetry in four dimensions demands only that the superpotential is defined by some holomorphic, $\cG$-invariant function $W$ of the bosonic matter fields $\{ X_a \}$ in the theory. (Matter field $X_a$ corresponds to the lowest component of the $a$th chiral matter superfield and it will sometimes be convenient to write this as $X_{ij}$ if arrow $a$ points from $i$ to $j$.) Each matter field $X_a$ can be thought of as being valued in ${\mbox{Mat}}_N ( \CC )$ and the trace of a matrix product of any number of different matter fields of the form ${\mbox{tr}} \left( X_{a_1} ... X_{a_p} \right)$ is $\cG$-invariant only if the cyclically ordered sequence of arrows $( a_1 ... a_p )$ forms a circuit in the underlying digraph. Recall that the circulation of edges around each vertex in a bipartite tiling $\tau_{\vec{G}}$ encodes a circuit in $\vec{G}$ and therefore specifies a $\cG$-invariant product of the matter fields as above. One can also specify an overall $+$ or $-$ sign for the product associated with a white or black vertex in $\tau_{\vec{G}}$. The superpotential $W_{\tau_{\vec{G}}}$ for the brane tiling is obtained by adding all such terms associated with all the different vertices in $\tau_{\vec{G}}$. The total number of terms in $W_{\tau_{\vec{G}}}$ therefore equals $t=e-n$ (i.e. the number of vertices in $\tau_{\vec{G}}$). Since $\tau_{\vec{G}}$ is bipartite, each matter field $X_a$ must appear in exactly two different terms with opposite signs in $W_{\tau_{\vec{G}}}$. In order to avoid unnecessary complications with traces and matrix ordering, we will typically assume $N=1$ in forthcoming expressions for $W_{\tau_{\vec{G}}}$. 

%%%%%%%%%%%%%%    
\subsection{Superconformal invariance}
\label{sec:superconformalinvariance}

At the classical level, bosonic matter fields in four dimensions have conformal dimension one and so the lagrangian is only conformally invariant when $W$ is either a cubic function or zero. Typically this not the case for the brane tiling superpotentials defined above which should instead undergo a non-trivial renormalisation group flow to an interacting superconformal fixed point in the infrared. Necessary criteria for the existence of a non-trivial superconformal fixed point are that the exact NSVZ $\beta$-functions \cite{Novikov:1983uc,Leigh:1995ep} for the both the gauge and superpotential couplings must vanish. These $\beta$-functions for an arbitrary brane tiling were computed in \cite{Franco:2005rj,Hanany:2005ss} and their vanishing at a non-trivial fixed point was found to be equivalent to certain linear relations being obeyed amongst the R-charges for the matter fields in the superconformal algebra. Let $R_a$ denote the R-charge for matter field $X_a$. The $\beta$-function for the gauge coupling associated with face $i$ in the tiling vanishes provided 
\begin{equation}\label{eq:gaugebeta}
\sum_{a \, \in \, \fd (i)} ( 1 - R_a ) =2~,
\end{equation}
where $\fd (i)$ denotes the set of edges bounding face $i$ in the tiling. The $\beta$-function for the superpotential term associated with vertex $\alpha$ in the tiling vanishes provided 
\begin{equation}\label{eq:superpotentialbeta}
\sum_{a \, | \, \alpha \, \in \, \fd (a)} R_a =2~,
\end{equation}
where $\fd (a)$ denotes the pair of vertices on which edge $a$ ends in the tiling. By summing \eqref{eq:gaugebeta} over all $n$ faces and \eqref{eq:superpotentialbeta} over all $t$ vertices in the tiling, one recovers the relation $t = e-n \, (= \sum_{a=1}^e R_a )$. The condition \eqref{eq:superpotentialbeta} is simply imposing that each term in the superpotential is exactly marginal at the superconformal fixed point. This follows since any chiral primary operator in the superconformal algebra must have conformal dimension $\Delta$ and R-charge $R$ related such that $\Delta = \tfrac{3}{2} R$ in four dimensions. In addition to \eqref{eq:gaugebeta} and \eqref{eq:superpotentialbeta}, if the superconformal field theory at the fixed point is unitary then we must have each $R_a >0$. 

It is important to stress that these conditions are necessary but generally not sufficient to guarantee that the associated superconformal field theory is consistent. For instance, any gauge-invariant scalar operator in a unitarity conformal field theory in four dimensions must have dimension no less than one. Therefore any such operator which is chiral primary in a unitary superconformal field theory in four dimensions must have R-charge no less than $\tfrac{2}{3}$ and this is not obviously implied by the conditions above. Furthermore, holography implies that the number of faces in a consistent tiling should equal the euler character of the dual toric Calabi-Yau cone in order that it contains a sufficient number holomorphic cycles for the D-branes to wrap \cite{Butti:2005vn,Hanany:2005ss,Gulotta:2008ef}. Although these conditions yield important information about consistent brane tilings, a definitive set of conditions which are necessary and sufficient for consistency has yet to be established. We will therefore limit our consideration to the class of {\emph{admissible}} brane tilings which obey \eqref{eq:gaugebeta} and \eqref{eq:superpotentialbeta} with all R-charges positive.

%%%%%%%%%%%%%%    
\subsubsection{Bivalent vertices}
\label{sec:bivalentvertices}
 
A white/black bivalent vertex in any tiling $\tau_{\vec{G}}$ encodes a mass term $\pm X_a X_b$ in $W_{\tau_{\vec{G}}}$, where $a$ and $b$ label the edges ending on the bivalent vertex. Clearly the scale generated by any such mass term is incompatible with superconformal invariance and cannot solve \eqref{eq:gaugebeta} and \eqref{eq:superpotentialbeta} with all R-charges positive. The standard argument is therefore to integrate out any massive fields and flow to a superconformal fixed point at energies well below the smallest mass scale. If the opposite endpoints of $a$ and $b$ correspond to different vertices in $\tau_{\vec{G}}$, the only other appearance of $X_a$ and $X_b$ in $W_{\tau_{\vec{G}}}$ is in a term of the form $\mp ( X_a M + X_b N )$, where $M$ and $N$ are monomials in the other matter fields associated with the circulation of edges around the two black/white vertices on which $a$ and $b$ end. As explained in section 5 of \cite{Franco:2005rj}, integrating out either $X_a$ or $X_b$ replaces the two terms above in $W_{\tau_{\vec{G}}}$ with the term $\mp MN$. This new superpotential is encoded by the tiling obtained by deleting $a$, $b$ and the white/black bivalent vertex and then identifying the other two black/white vertices in $\tau_{\vec{G}}$ (see figure 11 in \cite{Franco:2005rj}). 

%%%%%%%%%%%%%%    
\subsubsection{Isoradial embeddings}
\label{sec:isoradial}

Even if a given tiling contains no bivalent vertices, it still may not be possible to solve both \eqref{eq:gaugebeta} and \eqref{eq:superpotentialbeta} with all $R_a$ positive. A nice identification was made in \cite{Hanany:2005ss} between a solution of \eqref{eq:gaugebeta} and \eqref{eq:superpotentialbeta} with all $R_a >0$ and an  {\emph{isoradial embedding}} of the tiling on a torus equipped with flat metric. In an isoradially embedded tiling, each face is represented by a cyclic polygon with each of its corner vertices sitting on a circle with unit radius. The isoradial embedding is called strictly convex if the centre of each circle is contained in the interior of the face it is associated with. The edges bounding each face are represented by chords connecting adjacent vertices around the corresponding circle. 

Given any bipartite tiling $\tau_{\vec{G}}$, one can define an auxiliary tiling $\tau_{\vec{G}}^\diamond$ such that each vertex in $\tau_{\vec{G}}^\diamond$ corresponds to either a vertex or a face in $\tau_{\vec{G}}$ and each edge in $\tau_{\vec{G}}^\diamond$ connects a pair of vertices corresponding to a face and one of its corner vertices in $\tau_{\vec{G}}$. Whence $\tau_{\vec{G}}^\diamond$ is not necessarily bipartite but each of its faces is quadrilateral. As observed in \cite{KenSchlenk}, each strictly convex isoradial embedding for $\tau_{\vec{G}}$ corresponds to a {\emph{rhombic embedding}} for $\tau_{\vec{G}}^\diamond$ (such that each face in $\tau_{\vec{G}}^\diamond$ is represented by a rhombus with each side of unit length). If an edge $a$ bounds a face $i$ in an isoradially embedded tiling $\tau_{\vec{G}}$, the angle subtended between the two radii connecting the origin of the circle for $i$ to the endpoints of $a$ is identified with $\pi (1- R_a )$ in \cite{Hanany:2005ss}. Together with the chord representing $a$, these two radii form an isosceles triangle in which both remaining angles equal $\tfrac{\pi}{2} R_a$ and the length of the chord for $a$ is $2\, {\mathrm{cos}} \left( \tfrac{\pi}{2} R_a \right)$. Thus, if edges $a$ and $b$ are adjacent and bound the same face in an isoradially embedded tiling, the angle between $a$ and $b$ inside the face is $\tfrac{\pi}{2} ( R_a + R_b )$. The pair of faces which are adjacent on an edge $a$ in the isoradially embedded tiling $\tau_{\vec{G}}$ describe a pair of isosceles triangles that are adjacent on the chord for $a$ and form a rhombus parameterised by angle $\pi (1- R_a )$, corresponding to a face in the rhombically embedded tiling $\tau_{\vec{G}}^\diamond$.

The characterisation of rhombic embeddings for planar graphs with only quadrilateral faces was obtained in \cite{KenSchlenk}. A vital concept is the notion of a {\emph{train track}} which is defined as a maximally-extended sequence of adjacent quadrilateral faces such that each face in the sequence is adjacent to its predecessor and successor on opposite sides. A train track in any $\tau_{\vec{G}}^\diamond$ defines a closed curve on the torus traversing each of the edges on which the faces in the sequence are adjacent. One endpoint of each of these edges in $\tau_{\vec{G}}^\diamond$ corresponds to a vertex in $\tau_{\vec{G}}$ and all such vertices to one side of the closed curve in the train track are white while all those to the other side are black (specifying an orientation for the train track fixes which colour vertices sit on which side of the curve). From theorem 5.1 in \cite{KenSchlenk}, any $\tau_{\vec{G}}^\diamond$ admits a rhombic embedding only if each of its train tracks defines a simple closed curve on the torus and the lift to the universal cover of any pair of train tracks defines two periodic curves which intersect at most once. Each train track in $\tau_{\vec{G}}^\diamond$ corresponds to a maximally-extended closed walk in $\tau_{\vec{G}}$ zigzagging along the edges in $\tau_{\vec{G}}$ which connect vertices with opposite colours on opposite corners of each face along the train track in $\tau_{\vec{G}}^\diamond$. If a train track in $\tau_{\vec{G}}^\diamond$ defines a simple closed curve on the torus, the corresponding zigzag walk is a cycle in $\tau_{\vec{G}}$ (i.e. the closed curve does not intersect itself so no edges or vertices are repeated and the zigzag walk forms a closed path). Take the closed curve to be oriented such that all the vertices in $\tau_{\vec{G}}$ to its right-hand side along the train track in $\tau_{\vec{G}}^\diamond$ are white. The structure of the corresponding zigzag walk in $\tau_{\vec{G}}$ is defined by applying the following rule at each vertex encountered after traversing an edge in $\tau_{\vec{G}}$: if the vertex is white/black, proceed along the edge which is clockwise/anti-clockwise adjacent to the last one (i.e. the edge which is furthest to the left/right). Thus, a tiling $\tau_{\vec{G}}$ admits a strictly convex isoradial embedding only if each maximally-extended zigzag walk in $\tau_{\vec{G}}$ is a cycle and no two zigzag cycles intersect on more than one edge inside a fundamental domain of $\tau_{\vec{G}}$.  

%%%%%%%%%%%%%%    
\subsubsection{${\mathrm{a}}$-maximisation}
\label{sec:amax}

In a generic superconformal field theory in four dimensions, the method of a-maximisation \cite{Intriligator:2003jj} determines which $\fu (1) < \fsu (2,2|1)$ describes the R-symmetry in the superconformal algebra at the fixed point. Any $\fu (1) < \fsu (2,2|1)$ global symmetry of the superconformal field theory is parameterised by a set of \lq trial' R-charges for the fields. Equation (1.7) in \cite{Intriligator:2003jj} defines a particular cubic function \lq a' on the space of all such charges. Their remarkable result is that it is precisely the exact R-charges at the superconformal fixed point which (locally) maximise this function. The local maximum is identified with one of the central charges in the superconformal field theory.

The a-function in \cite{Intriligator:2003jj} to be maximised for a given brane tiling was found in \cite{Hanany:2005ss}. Up to an unimportant additive constant, it is given by 
\begin{equation}\label{eq:amax}
{\mathrm{a}} (R) = \frac{9}{32} \sum_{a=1}^e ( R_a - 1 )^3~,
\end{equation}
for a tiling with $e$ edges, as a function of trial R-charges $R_a$ which are subject to \eqref{eq:gaugebeta} and \eqref{eq:superpotentialbeta} with all $R_a >0$. Furthermore, for strictly convex isoradial embeddings, each $R_a < 1$ since angle $\pi (1- R_a )$ must be positive and so the a-function in \eqref{eq:amax} is negative-definite. 

Before moving on, let us make an observation that will be explored in more detail in section~\ref{sec:quadbranetilings}. For each white vertex $\alpha_\circ$ in a brane tiling, let index $a_{\alpha_\circ} = 1,..., {\mathrm{deg}}( \alpha_\circ )$ label the edges ending on $\alpha_\circ$. Since each edge ends on precisely one white vertex then all the edges in the tiling are accounted for by this indexing. Now write trial R-charges $R_{a_{\alpha_\circ}} = \tfrac{2}{ {\mathrm{deg}}( \alpha_\circ )} + r_{a_{\alpha_\circ}}$, where $-\tfrac{2}{ {\mathrm{deg}}( \alpha_\circ )} < r_{a_{\alpha_\circ}} < 1 -\tfrac{2}{ {\mathrm{deg}}( \alpha_\circ )}$ since $0 < R_{a_{\alpha_\circ}} < 1$ for a strictly convex isoradially embedded tiling. Substituting this into the a-function \eqref{eq:amax} gives 
\begin{align}\label{eq:amax2}
{\mathrm{a}} ( r_\circ ) &= \frac{9}{32} \sum_{\alpha_\circ} \left[ \sum_{a_{\alpha_\circ} =1}^{{\mathrm{deg}}( \alpha_\circ )}  ( R_{a_{\alpha_\circ}} - 1 )^3 \right] \nonumber \\
&= \frac{9}{32} \sum_{\alpha_\circ} \left[ \sum_{a_{\alpha_\circ} =1}^{{\mathrm{deg}}( \alpha_\circ )}  \left( \left( \frac{2}{ {\mathrm{deg}}( \alpha_\circ )} - 1 \right)^3 + r_{a_{\alpha_\circ}}^2 \left( r_{a_{\alpha_\circ}} + 3 \left( \frac{2}{ {\mathrm{deg}}( \alpha_\circ )} - 1 \right) \right)  \right) \right]~,
\end{align}
where the second equality follows using $\sum_{a_{\alpha_\circ} =1}^{{\mathrm{deg}}( \alpha_\circ )}  r_{a_{\alpha_\circ}}  =0$ which is condition \eqref{eq:superpotentialbeta} for each white vertex. Let us define ${\mathrm{a}}_\circ := \tfrac{9}{32} \sum_{\alpha_\circ} {\mathrm{deg}}( \alpha_\circ )  \left( \tfrac{2}{ {\mathrm{deg}}( \alpha_\circ )} - 1 \right)^3$. Since an isoradially embedded brane tiling must contain no bivalent vertices, for each white vertex, $\left( \tfrac{2}{ {\mathrm{deg}}( \alpha_\circ )} - 1 \right) <0$. Furthermore, since $r_{a_{\alpha_\circ}} < 1 -\tfrac{2}{ {\mathrm{deg}}( \alpha_\circ )}$ then $r_{a_{\alpha_\circ}} + 3 \left( \tfrac{2}{ {\mathrm{deg}}( \alpha_\circ )} - 1 \right) < 2 \left( \tfrac{2}{ {\mathrm{deg}}( \alpha_\circ )} - 1 \right) < 0$. We therefore conclude that ${\mathrm{a}} ( r_\circ ) \leq {\mathrm{a}}_\circ$ for any strictly convex isoradial embedding, with ${\mathrm{a}} ( r_\circ ) = {\mathrm{a}}_\circ$ only if all $r_{a_{\alpha_\circ}} =0$. However, it is only when the R-charge assignment $R_{a_{\alpha_\circ}} = \tfrac{2}{ {\mathrm{deg}}( \alpha_\circ )}$ also solves \eqref{eq:gaugebeta} for each face and \eqref{eq:superpotentialbeta} for each black vertex that this bound can be saturated by a-maximisation. In general, this will not be the case and maximisation within the space of consistent trial R-charges yields a value less than ${\mathrm{a}}_\circ$ (or a value greater than ${\mathrm{a}}_\circ$ which corresponds to an isoradial embedding that is not strictly convex). Indexing in terms of each black vertex $\alpha_\bullet$ in the brane tiling follows the same procedure as above but with $\circ \leftrightarrow \bullet$. We shall refer to a brane tiling as being {\emph{maximal}} if it admits a strictly convex isoradial embedding defined by either R-charge assignment $R_{a_{\alpha_\circ}} = \tfrac{2}{ {\mathrm{deg}}( \alpha_\circ )}$ or $R_{a_{\alpha_\bullet}} = \tfrac{2}{ {\mathrm{deg}}( \alpha_\bullet )}$, such that the a-function equals either ${\mathrm{a}}_\circ$ or ${\mathrm{a}}_\bullet$ at the superconformal fixed point. If a maximal tiling admits both these R-charge assignments, all its vertices must have the same degree $p$ and all the R-charges must equal $\tfrac{2}{p}$. The angle between any pair of adjacent edges which bound the same face in the tiling must therefore equal $\tfrac{2\pi}{p}$. Each face must therefore describe a regular polygon with $\tfrac{2p}{p-2}$ sides and this number must be an even integer since the tiling is bipartite. Thus all faces must be either hexagons (when $p=3$) or squares (when $p=4$) and ${\mathrm{a}}_\circ = {\mathrm{a}}_\bullet$ in both cases. All maximal tilings that are not regular admit only one of the two possible R-charge assignments described above. In a maximal tiling, notice that having R-charge $\tfrac{2}{ {\mathrm{deg}}( \alpha )}$ assigned to all the edges ending on vertex $\alpha$ implies that the chords for these edges must all have the same length $2\, {\mathrm{cos}} ( \frac{\pi}{{\mathrm{deg}}( \alpha )} )$ and have equal angular separation $\tfrac{2\pi }{ {\mathrm{deg}}( \alpha )}$ around $\alpha$.  

%%%%%%%%%%%%%%    
\subsection{Vacuum moduli space}
\label{sec:vacuummodulispace}

Points in the classical moduli space of supersymmetric vacua in a supersymmetric gauge theory in four dimensions correspond to gauge-equivalence classes of fields which solve both the D- and F-term equations (i.e. the zero loci of the scalar potential). Understanding this vacuum moduli space often provides crucial insights into the full phase structure of the theory in the presence of quantum corrections. However, the vacuum moduli space of a given theory will typically have a very complicated structure and may consist of a number of disconnected or intersecting branches of different dimension. It is often more practical to therefore focus on a particular branch which characterises important physical properties of the theory. For the superconformal quiver gauge theories of interest, this is taken to be the Higgs branch $\cM$ parameterised by those gauge-inequivalent constant non-zero matter fields which solve the D- and F-term equations (whence all gauge supermultiplets become massive and the gauge group is completely broken on this branch). Solutions of the F-term equations correspond to critical points of the superpotential $W$. Since $W$ is both holomorphic and gauge-invariant, it is actually invariant under the complexified gauge group and therefore so are the F-term equations. This extra gauge symmetry is not shared by the rest of the lagrangian and it does not preserve the D-term equations. However, at least for the class of superconformal quiver gauge theories we shall consider here, there exists a standard procedure that recovers the D-term equations from fixing all but the original real form of the complexified gauge group. In this sense, one can think of $\cM$ as a finite-dimensional irreducible algebraic variety formed by taking the holomorphic quotient of the space of critical critical points of $W$ by the group of complexified gauge transformations which act effectively on the constant matter fields. It is a fundamental result in algebraic geometry \cite{FulToric,CoxToric} that one may therefore identify this branch with the affine scheme of prime ideals of the coordinate ring of $\cM$ itself. From this perspective, $\cM$ is parameterised by gauge-invariant products of the constant matter fields subject to relations imposed by the F-term equations. 

For generic superconformal quiver gauge theories describing the low-energy dynamics of D3-branes probing a toric conical singularity in IIB string theory, the structure of the corresponding vacuum moduli spaces have been studied in detail in \cite{Forcella:2008bb,Forcella:2008ng} in terms of both the forward algorithm developed in \cite{Feng:2000mi,Feng:2001xr} and subsequent brane tiling techniques from \cite{Hanany:2005ve,Franco:2005rj,Franco:2006gc,Hanany:2005ss}. All such analyses are made possible by the special properties possessed by the superpotential $W_{\tau_{\vec{G}}}$ for a brane tiling which ensure that $\cM_{\tau_{\vec{G}}}$ is an affine toric variety. Indeed, for a single D3-brane, the gauge group is abelian and $\cM_{\tau_{\vec{G}}}$ corresponds to the affine toric Calabi-Yau three-fold transverse to the D3-brane worldvolume, as expected from holography in the strong coupling limit. For $N$ coincident D3-branes, the corresponding branch of the vacuum moduli space should take the form of an $N$-fold symmetric product of the aforementioned toric Calabi-Yau three-fold. Geometric invariant theory (see \cite{GIT3} for a review) allows one to equivalently think of the affine toric variety $\cM_{\tau_{\vec{G}}}$ as a toric symplectic quotient. From this perspective, the D-term equations correspond to moment maps associated with the hamiltonian torus action on matter fields defined by the quiver representation in the abelian theory. The Fayet-Iliopoulos parameters in the abelian theory correspond to integration constants in the moment map equations and must all equal zero when $\cM_{\tau_{\vec{G}}}$ is a cone. The moment maps take points in $\cM_{\tau_{\vec{G}}}$ to points in a certain convex rational polyhedral cone $\Lambda_{\tau_{\vec{G}}} \subset \RR^3$ and $\cM_{\tau_{\vec{G}}}$ can be thought of as a torus bundle over $\Lambda_{\tau_{\vec{G}}}$ such that circle fibres degenerate in a particular way on faces in $\Lambda_{\tau_{\vec{G}}}$. The canonical Reeb Killing vector field associated with the Sasaki-Einstein link of Calabi-Yau cone $\cM_{\tau_{\vec{G}}}$ defines a characteristic plane in $\RR^3$ whose intersection with $\Lambda_{\tau_{\vec{G}}}$ defines a convex lattice polygon $\Delta_{\tau_{\vec{G}}}$ (i.e. the convex hull of a finite set of points in $\ZZ^2$). The data from $\Delta_{\tau_{\vec{G}}}$ is sufficient to reconstruct $\cM_{\tau_{\vec{G}}}$ and therefore provides a useful way of encoding the geometries of interest -- $\Delta_{\tau_{\vec{G}}}$ is often referred to as the {\emph{toric diagram}} of $\cM_{\tau_{\vec{G}}}$ in the physics literature. Any other convex lattice polygon that is related to $\Delta_{\tau_{\vec{G}}}$ by an affine unimodular transformation of the lattice will also reconstruct $\cM_{\tau_{\vec{G}}}$. The euler character of $\cM_{\tau_{\vec{G}}}$ equals the twice the area $A ( \Delta_{\tau_{\vec{G}}} )$ of $\Delta_{\tau_{\vec{G}}}$ and so, for any consistent brane tiling $\tau_{\vec{G}}$ with $n$ faces, holography implies $A ( \Delta_{\tau_{\vec{G}}} ) = \tfrac{n}{2}$. An important conjecture made in \cite{Hanany:2005ss} is that the homology class in $H_1 ( T^2 , \ZZ )$ of each zigzag cycle in a consistent brane tiling $\tau_{\vec{G}}$ corresponds to the normal vector of an edge bounding $\Delta_{\tau_{\vec{G}}}$. More precisely, a zigzag cycle in $\tau_{\vec{G}}$ representing class $(p,q) \in H_1 ( T^2 , \ZZ )$ is identified with the lattice vector $(p,q) \in \ZZ^2$ that is normal to an edge connecting a pair of lattice points $(x,y)$ and $(x+q,y-p)$ on the boundary of $\Delta_{\tau_{\vec{G}}}$. This identification has been verified in all known consistent examples and provides an efficient way of translating between $\tau_{\vec{G}}$ and $\cM_{\tau_{\vec{G}}}$. 

%%%%%%%%%%%%%%    
\subsection{Seiberg duality}
\label{sec:seibergduality}

Different phases of many supersymmetric gauge theories in four dimensions are related by Seiberg duality \cite{Seiberg:1994pq}. Seiberg duality for theories whose vacuum moduli space involves a toric variety has been investigated in \cite{Feng:2000mi,Beasley:2001zp,Feng:2001bn,Cachazo:2001sg} and also \cite{Feng:2002zw,Berenstein:2002fi,Franco:2003ja,Franco:2005rj} in the context of supersymmetric quiver gauge theories. Two supersymmetric quiver gauge theories based on different brane tilings which are related by Seiberg duality must have the same vacuum moduli space and be identified with the same holographically dual toric Calabi-Yau three-fold at their common superconformal fixed point in the infrared. 

For a quiver representation of $\prod_{i=1}^n U( N_i )$ obeying $\sum_{j=1}^n A_{ij} N_j = \sum_{j=1}^n A_{ji} N_j$ as above, Seiberg duality acts with respect to some fixed vertex $v$ in the underlying digraph to define a new quiver representation of $U \left( -N_v + \sum_{j=1}^n A_{vj} N_j \right) \times \prod_{i \neq v} U( N_i )$ whose underlying digraph is obtained by mutating the original digraph with respect to $v$ via the recipe given at the end of section~\ref{sec:moves}. It is straightforward to check that the adjacency matrix obtained from this elementary mutation still obeys the one-loop gauge anomaly cancellation condition with $N_v \mapsto -N_v + \sum_{i=1}^n A_{vi} N_i$ and all other $N_i$ the same. This is a duality in the representation-theoretic sense since the map $\Psi_v$ defined by Seiberg dualising with respect to $v$ is an involution of the quiver representation (a stronger property than mutation $\mu_v$ being an involution of the underlying digraph). However, notice that Seiberg duality generally does not act within the class of quiver representations of interest with all $N_i =N$. Indeed the only way in which this can occur is with respect to a balanced out-degree $2$ vertex $v$. This point has been made in the context of brane tilings in section 6.2 of \cite{Franco:2005rj} and the action of $\Psi_v$ with respect to the corresponding quadrilateral face $v$ in the tiling generically takes the form of an \lq urban renewal' map shown in their figure 13. We have already noted that our composite move II within the class smooth quadrilateral tilings, shown in figure~\ref{fig14}, corresponds to a special case. Indeed Seiberg duality is therefore a natural operation on faces in smooth quadrilateral brane tilings for superconformal quiver gauge theories in four dimensions, which may be thought of as further physical motivation for their consideration. Of course, $\Psi_v$ is generally not an operation within the class of smooth quadrilateral brane tilings unless the corresponding mutation $\mu_v$ is within the class of eulerian digraphs ${\vec{\fF}}_2$ discussed at the end of section~\ref{sec:generatingeuleriandigraphs}. 

Seiberg duality for the wider class of supersymmetric quiver gauge theories based on arbitrary $2$-regular eulerian digraphs has also been studied in \cite{Balasubramanian:2008qf}. The authors of this reference derive a nice characterisation of \lq forbidden sets' of superpotential terms composed of matter fields for which no R-charges can solve the vanishing $\beta$-function equations in a unitary manner. We will see in the next section that a forbidden set of superpotential terms is bound to exist in any smooth quadrilateral tiling containing two trivalent vertices connected by an edge, as was first noted in figure 15 of \cite{Hanany:2005ss}.  

%%%%%%%%%%%%%%    
\section{Quadrilateral brane tilings}
\label{sec:quadbranetilings}

In this final section, in the context of the superconformal quiver gauge theories discussed in section~\ref{sec:branetilings}, we shall examine some physical implications of the structural results for quadrilateral brane tilings obtained in previous sections. The physical inconsistency of bivalent vertices and pinched faces in a quadrilateral brane tiling will be noted first. The implications of the consistency conditions for grids and chains will then be described. In passing, it will be useful to mention the effect of duplicating (see section~\ref{sec:bipartitetilingsT2}) and lagging (see figure~\ref{fig8}) an edge in the tiling. For a given smooth quadrilateral tiling $\tau_{\vec{G}}$, we will then consider the effect of the composite moves I, II and III described in section~\ref{sec:smoothquadtile} on the superpotential $W_{\tau_{\vec{G}}}$, Higgs branch $\cM_{\tau_{\vec{G}}}$ and consistency conditions \eqref{eq:gaugebeta} and \eqref{eq:superpotentialbeta}. In the process, we will establish the physical inconsistency in any smooth quadrilateral tiling of a pair of trivalent vertices connected by an edge. We will then make some remarks concerning the effect on the superpotential and consistency conditions of the smooth reconstruction move described in section~\ref{sec:smoothquadtile}. This analysis will prove that any admissible quadrilateral brane tiling must contain an equal number of white and black vertices. The effect of the mutation shown in figure~\ref{fig6a} will be described in terms of Seiberg duality for a smooth quadrilateral tiling. Some of the generic structural features of quadrilateral brane tilings will then be reconciled with the comprehensive survey conducted in \cite{Davey:2009bp} of brane tilings with no more than six vertices. A simple characterisation of the maximal (in the sense defined in section~\ref{sec:amax}) consistent quadrilateral brane tilings will be then be given. Finally, we conclude with a discussion of the general structure of the class of admissible quadrilateral brane tilings.

Following the notation of section~\ref{sec:facetypes}, for each face $v$ in $\tau_{\vec{G}}$ bounded by four different edges $a$, $b$, $c$ and $d$, condition \eqref{eq:gaugebeta} takes the form
\begin{equation}\label{eq:gaugebetaquad}
R_a + R_b + R_c + R_d =2~.
\end{equation}       

%%%%%%%%%%%%%%    
\subsection{Bivalent vertices and pinched faces}
\label{sec:bivalentandpinched}

The inconsistency of bivalent vertices in any brane tiling was discussed in section~\ref{sec:bivalentvertices} and indeed, for any pair of edges $a$ and $b$ connected to a bivalent vertex in a quadrilateral tiling, \eqref{eq:superpotentialbeta} and \eqref{eq:gaugebetaquad} imply the contradiction $R_a + R_b =0$. It is worth pointing out the technique we used for removing bivalent vertices by collapsing faces in section~\ref{sec:removebivalentpinched} is not the same as the procedure of integrating out massive fields that was described in section~\ref{sec:bivalentvertices} (i.e. the latter defines an operation that is not within the class of quadrilateral tilings). This distinction will utilised in section~\ref{sec:smoothreconstruction}. Either way, restricting attention to tilings without bivalent vertices is clearly desirable.

Let $v$ be a pinched face in $\tau_{\vec{G}}$ such that edges $a$ and $d$ bound the exterior while edges $b$ and $c$ bound the interior of $v$ in $\tau_{\vec{G}}$. Up to an overall sign, the matter fields associated with these boundary edges must appear in the form $X_a X_d \, M +  X_b X_c \, N - X_a X_b X_c X_d \, P$ in $W_{\tau_{\vec{G}}}$, where $M$, $N$ and $P$ are some monomials in the other matter fields (with $P$ being associated with the pinched vertex). However, \eqref{eq:gaugebetaquad} for $v$ combined with \eqref{eq:superpotentialbeta} for the pinched vertex imply the sum of R-charges for all the matter fields in $P$ must vanish, whence a contradiction since $P$ cannot be constant. Of course, this is to be expected given that it is impossible to represent the boundary edges of a pinched face as straight lines in the tiling without causing the face to degenerate. Thus any tiling containing a pinched face cannot admit an isoradial embedding. We conclude that any admissible quadrilateral brane tiling is necessarily smooth.

%%%%%%%%%%%%%%    
\subsection{Grids and chains}
\label{sec:gridsandchains}

If $v$ is a grid in $\tau_{\vec{G}}$, the matter fields associated with boundary edges of $v$ must appear in the form $X_a X_b X_c X_d \, ( M - N )$ in $W_{\tau_{\vec{G}}}$, where $M$ and $N$ are monomials in the other matter fields, such that $M$ is associated with the white link and $N$ is associated with the black link of $v$. Condition \eqref{eq:gaugebetaquad} for $v$ combined with \eqref{eq:superpotentialbeta} for its two link vertices imply that the sum of R-charges for all the matter fields in both $M$ and $N$ must vanish. A contradiction is therefore implied unless both $M$ and $N$ are constant, in which case $\tau_{\vec{G}}$ must be isomorphic to $\tau_+$.

Let $v$ be a chain in $\tau_{\vec{G}}$ with black/white link vertex such that edges $a$ and $d$ end on one white/black corner vertex while edges $b$ and $c$ end on the other white/black corner vertex of $v$ in $\tau_{\vec{G}}$. Up to an overall sign, the matter fields associated with these boundary edges must appear in the form $X_a X_d \, M +  X_b X_c \, N - X_a X_b X_c X_d \, P$ in $W_{\tau_{\vec{G}}}$, where $M$, $N$ and $P$ are monomials in the other matter fields (with $P$ being associated with the link of $v$). Condition \eqref{eq:gaugebetaquad} for $v$ combined with \eqref{eq:superpotentialbeta} for its link vertex imply that the sum of R-charges for all the matter fields in $P$ must vanish. Thus a contradiction must follow unless $P$ is constant, in which case the link vertex of $v$ must be exactly $4$-valent. Recall from section~\ref{sec:smoothquadtile} that this is the unique scenario in which a chain cannot be smoothly collapsed. The two white/black corner vertices of $v$ must therefore describe the links for two other chains $w$ and $x$ in $\tau_{\vec{G}}$, such that $w$ is adjacent to $v$ on edges $a$ and $d$ while $x$ is adjacent to $v$ on edges $b$ and $c$. Following the logic above for $w$ and $x$ implies an inconsistency must occur unless their links are also $4$-valent. Continuing in this manner, one finds that if $\tau_{\vec{G}}$ contains a chain and obeys \eqref{eq:gaugebetaquad} and \eqref{eq:superpotentialbeta} with all R-charges positive then it must be isomorphic to $\tau_{\vec{A}_{2p}}$ in figure~\ref{fig10}, for some $p>1$. 

%%%%%%%%%%%%%%    
\subsection{Duplication and lagging}
\label{sec:duplag}

To facilitate the subsequent consideration of generating moves for smooth quadrilateral tilings, let us first note the relatively trivial effect of the two simpler moves of duplicating and lagging an edge in a quadrilateral tiling.

As remarked in section~\ref{sec:bipartitetilingsT2}, the effect of duplicating an edge in tiling $\tau_{\vec{G}}$, by replacing an edge $a$ with two edges $b$ and $c$ connecting the same pair of vertices in $\tau_{\vec{G}}$, is well-known in the literature \cite{Hanany:2008fj,Davey:2009sr,Davey:2009bp} and encodes precisely the subdivision of arrow $a$ in $\vec{G}$ depicted in figure~\ref{fig1}. The only effect of this duplication on the superpotential $W_{\tau_{\vec{G}}}$ is to replace $X_a$ with $X_b X_c$ in both terms where it occurs. Moreover, by subdividing $a$ in $\vec{G}$, one obtains the same collection of circuits but with $a$ replaced by $bc$ in each of them. Thus $\cM_{\tau_{\vec{G}}}$ is unaffected by the duplication of an edge in $\tau_{\vec{G}}$. (However, it is worth mentioning that the effect of this duplication can be seen in the multiplicities of fields in the gauged linear sigma model description of $\cM_{\tau_{\vec{G}}}$ \cite{Feng:2000mi,Feng:2001xr,Hanany:2005ve,Franco:2005rj,Franco:2006gc}.) Despite seeming rather innocuous, notice that any such duplication must trigger an inconsistency in the associated superconformal field theory since \eqref{eq:gaugebeta} implies $R_b + R_c =0$ for the new face bounded by $b$ and $c$. Again, this is to be expected given that $b$ and $c$ cannot both be represented by straight lines in $\tau_{\vec{G}}$ without causing the new face to degenerate.      

A lagged edge in a quadrilateral tiling $\tau_{\vec{G}}$ (see figure~\ref{fig8}), containing a pair of bivalent vertices connected by an edge, corresponds to a special case of what is referred to as a \lq quadratic node tiling' in figure 5 of \cite{Davey:2009bp}. Of course, the existence of any bivalent vertex in the tiling implies an inconsistency in the associated superconformal field theory though it is worth describing the particularly trivial effect that de-lagging an edge in $\tau_{\vec{G}}$ has on $W_{\tau_{\vec{G}}}$ and $\cM_{\tau_{\vec{G}}}$. Consider lagging an edge $a$ in a quadrilateral tiling $\tau_{\vec{H}}$ as in figure~\ref{fig8} such that the new tiling $\tau_{\vec{G}}$ has face $v$ with boundary edge $b$ above the two bivalent vertices introduced by the lagging and face $w$ with boundary edge $c$ below. The superpotentials before and after lagging are of the form 
\begin{align}\label{eq:suplag}
W_{\tau_{\vec{H}}}&= X_a  (M(X) - N(X)) + W(X) \nonumber \\
W_{\tau_{\vec{G}}} &= X_{wv} ( Y_{vw} - X_{vw} ) + X_b \, X_c \, X_{vw} \, M( X^\prime ) - X_b \, X_c \, Y_{vw} \, N( X^\prime ) + W( X^\prime )~,
\end{align}
where $X_{vw}$, $X_{wv}$ and $Y_{vw}$ denote the matter fields for the three new edges connected to the two bivalent vertices in $\tau_{\vec{G}}$ while $M$ and $N$ are monomials in the remaining matter fields associated with the edges which $\tau_{\vec{H}}$ and $\tau_{\vec{G}}$ have in common. These remaining matter fields are written $\{ X \}$ in $\tau_{\vec{H}}$ and $\{ X^\prime \}$ in $\tau_{\vec{G}}$ while $W$ denotes the remaining superpotential terms encoded by the common edges and vertices in $\tau_{\vec{H}}$ and $\tau_{\vec{G}}$. By identifying trivially each $X^\prime = X$ and $X_b X_c X_{vw}$ in $W_{\tau_{\vec{G}}}$ with $X_a$ in $W_{\tau_{\vec{H}}}$, one sees that both superpotentials give the same F-term equations. Indeed $W_{\tau_{\vec{H}}}$ can obtained from $W_{\tau_{\vec{G}}}$ by simply integrating out the massive field $X_{wv}$. Moreover, since the lagging in figure~\ref{fig8} encodes the composite move in figure~\ref{fig6}, the set of gauge-invariant monomials corresponding to circuits in both $\vec{H}$ and $\vec{G}$ must agree at the respective critical points of $W_{\tau_{\vec{H}}}$ and $W_{\tau_{\vec{G}}}$, whence $\cM_{\tau_{\vec{H}}} \cong \cM_{\tau_{\vec{G}}}$. 

%%%%%%%%%%%%%%    
\subsection{Composite move I}
\label{sec:compmoveI}

Consider now the move shown in figure~\ref{fig12}, mapping $\tau_{\vec{H}} \in \eQ^{[t]}$ to $\tau_{\vec{G}} \in \eQ^{[t+2]}$. The superpotentials before and after the move are of the form
\begin{align}\label{eq:supcompI}
W_{\tau_{\vec{H}}} &= \phi_{vw} ( M(X) - N(X) ) + W(X) \nonumber \\
W_{\tau_{\vec{G}}} &= \phi_{xy} ( \phi_{yw} - \phi_{vx} ) + \phi_{vx} \, M( X^\prime ) - \phi_{yw} \, N( X^\prime ) + W( X^\prime )~,
\end{align}
where $\phi_{vw} := X_{vw} X_{wv}$ is gauge-invariant in $\tau_{\vec{H}}$, $\phi_{vx} := X_{vx} X_{xv}$, $\phi_{xy} := X_{xy} X_{yx}$ and $\phi_{yw} := X_{yw} X_{wy}$ are gauge-invariant in $\tau_{\vec{G}}$ while $M$ and $N$ are gauge-invariant monomials in the remaining matter fields they have in common. The F-term equation for $\phi_{xy}$ in $W_{\tau_{\vec{G}}}$ is $\phi_{vx} = \phi_{yw}$. After imposing this in $W_{\tau_{\vec{G}}}$ and identifying trivially each $X^\prime = X$ and $\phi_{vx} = \phi_{yw}$ in $W_{\tau_{\vec{G}}}$ with $\phi_{vw}$ in $W_{\tau_{\vec{H}}}$, one sees that both superpotentials in \eqref{eq:supcompI} give the same F-term equations. However, this is not the same as integrating out a massive field since the corresponding term in the superpotential is a quartic function of the matter fields for $\tau_{\vec{G}}$. Although $W_{\tau_{\vec{H}}}$ and $W_{\tau_{\vec{G}}}$ give the same F-term equations, $\cM_{\tau_{\vec{H}}}$ and $\cM_{\tau_{\vec{G}}}$ are not isomorphic since the set of gauge-invariant monomials corresponding to circuits in $\vec{H}$ and $\vec{G}$ obey different relations at the respective critical points. Each monomial for a circuit in $\vec{H}$ containing $X_{vw}$ but not $X_{wv}$ corresponds to the same monomial for a circuit in $\vec{G}$ but with $X_{vw}$ replaced by $X_{vx} X_{xy} X_{yw}$. Likewise, each monomial for a circuit in $\vec{H}$ containing $X_{wv}$ but not $X_{vw}$ corresponds to the same monomial for a circuit in $\vec{G}$ but with $X_{wv}$ replaced by $X_{wy} X_{yx} X_{xv}$.  The monomial $\phi_{vw}$ for $\tau_{\vec{H}}$ is identified with $\phi_{vx} = \phi_{yw}$ for $\tau_{\vec{G}}$ at the critical points and $\phi_{xy}$ is the only extra monomial for $\tau_{\vec{G}}$. These replacements give rise to new relations amongst the resulting monomials at the critical points, and this procedure defines the map from $\cM_{\tau_{\vec{H}}}$ to $\cM_{\tau_{\vec{G}}}$. We will see how this works in a simple example in a moment.

By taking the R-charges of all the matter fields corresponding to common edges in $\tau_{\vec{H}}$ and $\tau_{\vec{G}}$ to be the same in both, condition \eqref{eq:superpotentialbeta} stipulates that $R_{\phi_{vx}} = R_{\phi_{yw}} = 2- R_{\phi_{xy}}$ in $\tau_{\vec{G}}$ must equal $R_{\phi_{vw}}$ in $\tau_{\vec{H}}$. The remaining condition \eqref{eq:gaugebetaquad} for faces $x$ and $y$ in $\tau_{\vec{G}}$ is implied by those above. Thus, given an admissible solution corresponding to a strictly convex isoradial embedding for either $\tau_{\vec{H}}$ or $\tau_{\vec{G}}$, the R-charge assignments above define a strictly convex isoradial embedding for both. 

%%%%%%%%%%%%%%    
\subsubsection{Brane tiling for the necklace quiver}
\label{sec:necklace}

Recall from section~\ref{sec:smoothquadtile} that, for given $p>1,$ the unique smooth tiling $\tau_{\vec{A}_{2p}}$ in figure~\ref{fig10} encoding the necklace digraph $\vec{A}_{2p}$ in figure~\ref{fig11} constitutes an obstruction to the procedure of smoothly collapsing faces within the class $\eQ$ and can be generated by simply iterating composite move I on $\tau_+ \in \eQ^{[2]}$. Let us now compute $W_{\tau_{\vec{A}_{2p}}}$ and $\cM_{\tau_{\vec{A}_{2p}}}$ as a simple example of the construction described above. This superconformal quiver gauge theory was studied in \cite{Uranga:1998vf} as a generalisation of the construction in \cite{KlebanovWitten} (which corresponds to the $p=1$ case). It can be engineered straightforwardly in the T-dual IIA string theory setup via D4-branes stretched between $2p$ mutually orthogonal NS5-branes arranged around a circle.  

In terms of the labelling in figure~\ref{fig10}, the superpotential is given by
\begin{equation}\label{eq:supnecklace}
W_{\tau_{\vec{A}_{2p}}} = \sum_{j=0}^{p-1} \phi_{2j+1} \left( \phi_{2j} - \phi_{2j+2} \right)~,
\end{equation}
where $\phi_i := X_{i\, i+1} X_{i+1 \, i}$ with $i=1,...,2p$ modulo $2p$. The critical points of $W_{\tau_{\vec{A}_{2p}}}$ are at $\phi_i = \phi_{i+2}$ and let us denote $\gamma := \phi_1 = \phi_3 =...= \phi_{2p-1}$ and $\delta := \phi_2 = \phi_4 =...= \phi_{2p}$ at a critical point. In addition to $\phi_i$, the only other gauge-invariant monomials for circuits in $\vec{A}_{2p}$ are $\alpha := \prod_{i=1}^{2p} X_{i\, i+1}$ and $\beta := \prod_{i=1}^{2p} X_{i+1\, i}$. Collectively they are subject to the single relation $\alpha\beta = \prod_{i=1}^{2p} \phi_i$, whence
\begin{equation}\label{eq:modnecklace}
\cM_{\tau_{\vec{A}_{2p}}} = \{ (\alpha ,\beta ,\gamma ,\delta ) \in \CC^4 \, |\, \alpha\beta = (\gamma\delta )^p \}  ~.
\end{equation}
For $p=1$, \eqref{eq:modnecklace} corresponds to $\cM_{\tau_+}$ which describes the conifold. For $p>1$, $\cM_{\tau_{\vec{A}_{2p}}}$ corresponds to a quotient defined by the action of $\ZZ_p : (\alpha ,\beta ,\gamma ,\delta ) \mapsto ( e^{2\pi i/p} \, \alpha , e^{-2\pi i/p} \, \beta ,\gamma ,\delta )$ on $\cM_{\tau_+}$. The link of this toric Calabi-Yau cone corresponds to a Sasaki-Einstein manifold that often denoted $L^{ppp}$ in the physics literature. Henceforth we will denote by ${\sf{C}}(L)$ the toric Calabi-Yau cone whose Sasaki-Einstein link is $L$ (e.g. $\cM_{\tau_{\vec{A}_{2p}}} = {\sf{C}}( L^{ppp} )$).

Applying composite move I to $\tau_{\vec{A}_{2p}}$, as described above, with $v=2p-1$ and $w=2p$ gives $\tau_{\vec{A}_{2p+2}}$ with $v=2p-1$, $x=2p$, $y=2p+1$ and $w=2p+2$. At the respective critical points of $W_{\tau_{\vec{A}_{2p}}}$ and $W_{\tau_{\vec{A}_{2p+2}}}$, monomials $\phi_i$ are identified in terms of $\gamma$ and $\delta$ as above. Replacing $X_{2p-1\, 2p}$ with $X_{2p-1\, 2p} X_{2p\, 2p+1} X_{2p+1\, 2p+2}$ and $X_{2p\, 2p-1}$ with $X_{2p+2\, 2p+1} X_{2p+1\, 2p} X_{2p\, 2p-1}$ in the single monomial relation $\alpha\beta = \prod_{i=1}^{2p} \phi_i$ for $\tau_{\vec{A}_{2p}}$ gives $\alpha\beta = \prod_{i=1}^{2p+2} \phi_i$ for $\tau_{\vec{A}_{2p+2}}$. It is this replacement which increases by one the order of the cyclic group in the quotient of $\cM_{\tau_+}$. 

Conditions \eqref{eq:superpotentialbeta} and \eqref{eq:gaugebetaquad} for $\tau_{\vec{A}_{2p}}$ imply $R_{\phi_i} + R_{\phi_{i+1}} =2$. Thus $R_{\phi_1} = R_{\phi_3} =...= R_{\phi_{2p+1}} =: r$, $R_{\phi_2} = R_{\phi_4} =...= R_{\phi_{2p}} = 2-r$ and a-maximisation fixes $r=1$ with all the matter fields having the same R-charge equal to $\tfrac{1}{2}$ so that each face is a square in the isoradial embedding for $\tau_{\vec{A}_{2p}}$.

For an appropriate choice of basis in $H_1 ( T^2 , \ZZ )$, the zigzag cycles in $\tau_{\vec{A}_{2p}}$ can be taken to represent homology classes $\pm (1,0)$ and $\pm (0,p)$. As discussed in section~\ref{sec:vacuummodulispace}, since $\tau_{\vec{A}_{2p}}$ is consistent, these classes prescribe the toric diagram $\Delta_{\tau_{\vec{A}_{2p}}} = {\mbox{Conv}} ( (0,0) , (p,0) , (0,1) , (p,1) )$, corresponding to a lattice rectangle with area $p$ containing no interior points. Composite move I mapping $\tau_{\vec{A}_{2p}}$ to $\tau_{\vec{A}_{2p+2}}$ therefore just corresponds to adding two new boundary points at $(p+1,0)$ and $(p+1,1)$, increasing by one unit the area of the toric diagram. 

%%%%%%%%%%%%%%    
\subsection{Composite moves II and III}
\label{sec:compmoveIIandIII}

Consider now the move shown in figure~\ref{fig14}, mapping $\tau_{\vec{H}} \in \eQ^{[t]}$ to $\tau_{\vec{G}} \in \eQ^{[t+4]}$. Let us label the four boundary edges of the initial face in $\tau_{\vec{H}}$ in figure~\ref{fig14} $a$ (left), $b$ (right), $c$ (top) and $d$ (bottom), and preserve this labelling in $\tau_{\vec{G}}$. Relative to the central $4$-face $v$ in $\tau_{\vec{G}}$ in figure~\ref{fig14}, let us also label the four adjacent faces $i$ (left), $j$ (right), $k$ (top), $l$ (bottom). The superpotentials before and after the move are of the form
\begin{align}\label{eq:supcompII}
W_{\tau_{\vec{H}}} &= M(X) + N(X) - P(X) - Q(X) + W(X) \nonumber \\
W_{\tau_{\vec{G}}} &= X_{il} \, M( X^\prime )  + X_{jk} \, N( X^\prime ) - X_{jl} \, P( X^\prime ) - X_{ik} \, Q( X^\prime ) + W( X^\prime ) \nonumber \\
&\quad\; - X_{il} X_i X_l - X_{jk} X_j X_k + X_{jl} X_j X_l + X_{ik} X_i X_k~,
\end{align}
where $X_i := X_{vi}$, $X_j := X_{vj}$, $X_k := X_{kv}$ and $X_l := X_{lv}$ while $M$, $N$, $P$ and $Q$ are monomials in the matter fields corresponding to edges which $\tau_{\vec{H}}$ and $\tau_{\vec{G}}$ have in common. These monomials take the form $M(Y) = Y_a Y_d ...$, $N(Y) = Y_b Y_c ...$, $P(Y) = Y_b Y_d ...$ and $Q(Y) = Y_a Y_c ...$ with $Y=X$ in $\tau_{\vec{H}}$ and $Y= X^\prime$ in $\tau_{\vec{G}}$. The remaining superpotential terms in $W$ do not involve the matter fields associated with common edges $a$, $b$, $c$ and $d$. 

The F-term equations for $X_i$, $X_j$, $X_k$ and $X_l$ from $W_{\tau_{\vec{G}}}$ are
\begin{equation}\label{eq:FcompII}
X_{ik} X_k = X_{il} X_l \; , \quad X_{jk} X_k = X_{jl} X_l \; , \quad X_{ik} X_i = X_{jk} X_j \; , \quad X_{il} X_i = X_{jl} X_j~,
\end{equation}
which imply $X_{ik} X_{jl} = X_{jk} X_{il}$, or else $X_i = X_j = X_k = X_l =0$. At a generic point, the solutions of \eqref{eq:FcompII} are given by 
\begin{equation}\label{eq:FcompII2}
X_{ik} = \alpha X_j X_l \; , \quad X_{jl} = \alpha X_i X_k \; , \quad X_{jk} = \alpha X_i X_l \; , \quad X_{il} =  \alpha X_j X_k~,
\end{equation}
for any $\alpha \in \CC^*$. Substituting \eqref{eq:FcompII2} into \eqref{eq:supcompII} therefore shows that $W_{\tau_{\vec{H}}}$ can be recovered from $W_{\tau_{\vec{G}}}$ after integrating out $X_i$, $X_j$, $X_k$ and $X_l$ by identifying $\alpha X_j X^\prime_a = X_a$, $\alpha X_i X^\prime_b = X_b$, $X_l X^\prime_c = X_c$ and $X_k X^\prime_d = X_d$ (with a trivial identification of all the remaining matter fields). 

Each circuit in $\vec{H}$ can be mapped to a circuit in $\vec{G}$. For circuits which do not involve arrows $a$, $b$, $c$ or $d$, this proceeds trivially. Any other circuit containing $ac$, $ad$, $bc$ or $bd$ in $\vec{H}$ is mapped to a circuit in $\vec{G}$ by replacing these two arrows in the circuit in $\vec{H}$ with three arrows $aikc$, $aild$, $bjkc$ or $bjld$ respectively in $\vec{G}$. At the respective critical points of $W_{\tau_{\vec{H}}}$ and $W_{\tau_{\vec{G}}}$, this map is realised in the corresponding gauge-invariant monomials precisely as a consequence of the identification described in the paragraph above. Furthermore, the monomial for any other circuit in $\vec{G}$ can be expressed as a product of those just described at the critical points of $W_{\tau_{\vec{G}}}$. Thus $\cM_{\tau_{\vec{H}}} \cong \cM_{\tau_{\vec{G}}}$, as expected given the identification of composite move II with Seiberg duality after duplicating each boundary edge of the face it is applied to. 

Let us now determine if there can exist a solution of \eqref{eq:superpotentialbeta} and \eqref{eq:gaugebetaquad} for $\tau_{\vec{G}}$ with all R-charges positive (irrespective of whether $\tau_{\vec{H}}$ is admissible). Condition \eqref{eq:superpotentialbeta} for the four terms in $W_{\tau_{\vec{G}}}$ in the third line of \eqref{eq:supcompII} gives $R_{il} = 2- R_i - R_l$, $R_{jk} = 2- R_j - R_k$, $R_{jl} = 2- R_j - R_l$ and $R_{ik} = 2- R_i - R_k$ while condition \eqref{eq:gaugebetaquad} for face $v$ gives $R_i + R_j + R_k + R_l =2$. Substituting these expressions into condition \eqref{eq:gaugebetaquad} for faces $i$, $j$, $k$ and $l$ implies the respective contradictions $R_a + R_j =0$, $R_b + R_i =0$, $R_c + R_l =0$ and $R_d + R_k =0$. This inconsistency is therefore inevitable after applying composite move II to any smooth quadrilateral tiling. 

We will not need to repeat the analysis for composite move III in figure~\ref{fig15}, mapping $\tau_{\vec{H}} \in \eQ^{[t]}$ to $\tau_{\vec{G}} \in \eQ^{[t+6]}$ since it can be thought of as the composition of two moves we have already described, albeit not within the class $\eQ$. This follows by first lagging the edge in $\tau_{\vec{H}}$ in figure~\ref{fig15} according to figure~\ref{fig8} to produce an intermediate tiling which is not smooth. Applying composite move II to either of the new faces in this intermediate tiling gives $\tau_{\vec{G}}$. The superpotentials $W_{\tau_{\vec{H}}}$ and $W_{\tau_{\vec{G}}}$ are related by combining \eqref{eq:suplag} and \eqref{eq:supcompII}. They give the same F-term equations following the aforementioned identification of matter fields at each step and again one finds $\cM_{\tau_{\vec{H}}} \cong \cM_{\tau_{\vec{G}}}$. Since $\tau_{\vec{G}}$ here follows from composite move II in the final step, the conclusions in the paragraph above are still applicable, implying that it cannot admit a consistent assignment of R-charges. 

%%%%%%%%%%%%%%    
\subsubsection{Forbidden isolated faces}
\label{sec:forbiddenfaces}

The remarks above establish that any smooth quadrilateral brane tiling which contains a $4$-face cannot be consistent. More generally, consider a smooth quadrilateral tiling $\tau_{\vec{G}}$ which contains an isolated face $v$ with at least two trivalent corner vertices with opposite colours. Any two trivalent corner vertices with opposite colours must be connected by an edge on which $v$ is adjacent to another face $w$ in $\tau_{\vec{G}}$. Let us label the three edges connected to the white trivalent vertex as $b$, $d$, $j$ and the three edges connected to black trivalent vertex as $b$, $c$, $i$, such that $b$ is taken to be the edge on which $v$ and $w$ are adjacent. Relative to this labelling, $v$ is taken to be bounded by $b$, $c$, $d$ and another edge $a$ while $w$ is taken to be bounded by $b$, $i$, $j$ and another edge $k$. Condition \eqref{eq:superpotentialbeta} for the two terms $X_b X_d X_j - X_b X_c X_i$ in $W_{\tau_{\vec{G}}}$ defined by the white and black trivalent vertices in $\tau_{\vec{G}}$ implies $R_d + R_j = R_c + R_i = 2 - R_b$. Condition \eqref{eq:gaugebetaquad} for faces $v$ and $w$ in $\tau_{\vec{G}}$ implies $R_a + R_c + R_d = R_i + R_j + R_k = 2 - R_b$. Combining these two conditions therefore implies the contradiction $R_a + R_k =0$. 

This inconsistency is shown in figure 15 of \cite{Hanany:2005ss} and is also implied by the results of \cite{Balasubramanian:2008qf}. Any pair of trivalent vertices connected by an edge in a smooth quadrilateral tiling $\tau_{\vec{G}}$ defines in $W_{\tau_{\vec{G}}}$ precisely the \lq forbidden set' discussed in section 3.2 of \cite{Balasubramanian:2008qf}. The two trivalent vertices in $\tau_{\vec{G}}$ here encode two adjacent circuits in $\vec{G}$ corresponding to figure 4 in \cite{Balasubramanian:2008qf}.

%%%%%%%%%%%%%%    
\subsection{Smooth reconstruction}
\label{sec:smoothreconstruction}

Let us now consider the smooth reconstruction of a face, mapping $\tau_{\vec{H}} \in \eQ^{[t]}$ to $\tau_{\vec{G}} \in \eQ^{[t+1]}$, as defined in section \ref{sec:smoothquadtile}. Since there are many distinct ways in which this move can proceed, we shall only remark on a few generic features of interest. Let us focus on the special case shown in figure~\ref{fig9} for the reconstruction of an isolated face $v$ in $\tau_{\vec{G}}$ that we will take to be smooth. The superpotentials before and after the move are of the form
\begin{align}\label{eq:supsmoothrec}
W_{\tau_{\vec{H}}} &= X_\alpha X_\beta \, M(X) N(X)  - X_\alpha \, P(X) - X_\beta \, Q(X) + W(X) \nonumber \\
W_{\tau_{\vec{G}}} &= X_a X_d \, M( X^\prime )  + X_b X_c \, N( X^\prime ) - X_a X_c \, P( X^\prime ) - X_b X_d \, Q( X^\prime ) + W( X^\prime )~,
\end{align}
in terms of the labelling in figure~\ref{fig9} and where $M$, $N$, $P$ and $Q$ are monomials in the matter fields corresponding to edges which $\tau_{\vec{H}}$ and $\tau_{\vec{G}}$ have in common. The sum is replaced with a product of the two terms involving $P$ and $Q$ in both $W_{\tau_{\vec{H}}}$ and $W_{\tau_{\vec{G}}}$ when $v$ is a chain. 

The superpotentials $W_{\tau_{\vec{H}}}$ and $W_{\tau_{\vec{G}}}$ can be related via the auxiliary superpotential $W_{\tau_{\vec{F}}}$ for a quadrilateral tiling $\tau_{\vec{F}}$ that is defined as follows. Select one of the two black corner vertices of face $v$ in $\tau_{\vec{G}}$ in figure~\ref{fig9} together with the pair of boundary edges of $v$ which end on it. This defines the reconstruction of a new face in tiling $\tau_{\vec{F}}$. However, the reconstruction is not smooth and can be thought of as adding a single black bivalent vertex inside face $v$ in $\tau_{\vec{G}}$ such that the two new edges $e$ and $f$ connected to it are attached to the two white corner vertices of $v$ in $\tau_{\vec{G}}$. The auxiliary superpotential is therefore of the form
\begin{equation}\label{eq:supsmoothrec2}
W_{\tau_{\vec{F}}} = - X_e X_f  + X_e X^\prime_a X^\prime_d \, M( X^{\prime\prime} )  + X_f X^\prime_b X^\prime_c \, N( X^{\prime\prime} ) - X^\prime_a X^\prime_c \, P( X^{\prime\prime}  ) - X^\prime_b X^\prime_d \, Q( X^{\prime\prime}  ) + W( X^{\prime\prime}  )~.
\end{equation} 
By integrating out the massive fields for $\tau_{\vec{F}}$, one recovers precisely $W_{\tau_{\vec{H}}}$ after identifying $X^\prime_a X^\prime_c = X_\alpha$, $X^\prime_b X^\prime_d = X_\beta$ (with a trivial identification $X^{\prime\prime} =X$ of all the remaining matter fields). This relation between $W_{\tau_{\vec{H}}}$ and $W_{\tau_{\vec{G}}}$ can be understood in terms of the two different ways of removing a bivalent vertex from a quadrilateral tiling that were noted in section~\ref{sec:bivalentandpinched}. By construction, $\tau_{\vec{G}}$ is obtained from $\tau_{\vec{F}}$ by collapsing a face containing edges $e$ and $f$ on its boundary. On the other hand, integrating out the massive fields for $\tau_{\vec{F}}$ corresponds to pinching together the two white corner vertices of face $v$ in $\tau_{\vec{G}}$. Doing this produces a tiling which contains one face bounded by edges $a$ and $c$ and another face bounded by edges $b$ and $d$. This tiling is therefore not quadrilateral but is obtained from $\tau_{\vec{H}}$ by simply duplicating edges $\alpha$  and $\beta$. This is precisely the identification in the F-term equations for $W_{\tau_{\vec{H}}}$ and $W_{\tau_{\vec{F}}}$ described above.

Despite $W_{\tau_{\vec{H}}}$ and $W_{\tau_{\vec{F}}}$ having the same F-term equations,  $\cM_{\tau_{\vec{H}}}$ and $\cM_{\tau_{\vec{F}}}$ are not isomorphic since there is generically no bijection between the gauge-invariant monomials from circuits in $\vec{H} \in {\vec{\fF}}_2^{[t]}$ and $\vec{F} \in {\vec{\fF}}_2^{[t+2]}$ at the respective critical points. The superpotentials $W_{\tau_{\vec{G}}}$ and $W_{\tau_{\vec{F}}}$ do not have the same F-term equations but each circuit in $\vec{G}$ can be paired with a circuit in $\vec{F}$ by identifying $ac$, $ad$, $bc$ or $bd$ in circuits in $\vec{G}$ with $ac$, $aed$, $bfc$ or $bd$ in circuits in $\vec{F}$. After this identification, the only circuits in $\vec{F}$ that are not in $\vec{G}$ are those which contain the circuit $(ef)$.

There can never exist a solution of \eqref{eq:superpotentialbeta} and \eqref{eq:gaugebetaquad} with all R-charges positive for both $\tau_{\vec{H}}$ and $\tau_{\vec{G}}$. This is most easily proven using the characterisation of isoradial embeddings in terms of zigzag walks that was described in section~\ref{sec:isoradial}. If $\tau_{\vec{H}}$ is admissible then every maximally-extended zigzag walk in it is a cycle (i.e. a closed path with no edges or vertices repeated). In terms of figure~\ref{fig9}, the zigzag cycles in $\tau_{\vec{H}}$ which contain either $\alpha$ or $\beta$ define a zigzag walk in $\tau_{\vec{G}}$ of the form $({\circ} a {\bullet} c {\circ} ... {\bullet} a {\circ} d {\bullet} ... {\circ} b {\bullet} d {\circ} ... {\bullet} b {\circ} c {\bullet} ... )$ with each boundary edge of the reconstructed face $v$ repeated. Since the reconstruction of face $v$ in $\tau_{\vec{G}}$ is smooth, $\alpha$ and $\beta$ cannot be adjacent any zigzag cycle in $\tau_{\vec{H}}$. Therefore no zigzag cycle in $\tau_{\vec{H}}$ can contain both $\alpha$ and $\beta$ since this would require a repetition of the white vertex they both end on. Whence, if $\tau_{\vec{H}}$ is admissible then $\tau_{\vec{G}}$ is not. Conversely, if $\tau_{\vec{G}}$ is admissible, the zigzag cycles in it which contain an edge bounding $v$ are of the form $({\circ} a {\bullet} c {\circ} ...)$, $({\bullet} a {\circ} d {\bullet} ...)$, $({\bullet} b {\circ} c {\bullet} ...)$ and $({\circ} b {\bullet} d {\circ} ...)$. They define a zigzag walk in $\tau_{\vec{H}}$ of the form $({\circ} \alpha {\bullet} ... {\bullet} \beta {\circ} ... {\circ} \beta {\bullet} ... {\bullet} \alpha {\circ} ...)$ in which both $\alpha$ and $\beta$ are repeated. Whence, if $\tau_{\vec{G}}$ is admissible then $\tau_{\vec{H}}$ is not.

The remarks above lead to the following useful corollary. Given an admissible tiling $\tau_{\vec{H}} \in \eQ^{[t]}$, any $\tau_{\vec{G}} \in \eQ^{[t+s]}$ obtained from $\tau_{\vec{H}}$ via a sequence of smooth reconstructions of $s$ faces can be admissible only when $s$ is even and $\tau_{\vec{G}}$ contains precisely $\tfrac{s}{2}$ more white vertices and $\tfrac{s}{2}$ more black vertices than $\tau_{\vec{H}}$. In particular, after smoothly reconstructing a face such that an additional white/black corner vertex is introduced in the tiling, unless another smooth reconstruction is performed which involves a black/white corner vertex of this new face then the tiling must contain a zigzag walk that is not a cycle (wherein one of the boundary edges of the new face is repeated). Of course, performing any such balanced sequence of smooth reconstructions will not ensure $\tau_{\vec{G}}$ is admissible. The conclusion is that it is always inadmissible if the sequence is not balanced. Some instances of balanced sequences of smooth reconstructions within the class of admissible quadrilateral brane tilings will be discussed in section~\ref{sec:admissiblequadbranetilings}. The converse result also follows for any $\tau_{\vec{H}} \in \eQ^{[t]}$ obtained by smoothly collapsing $s$ faces in an admissible $\tau_{\vec{G}} \in \eQ^{[t+s]}$ which can be admissible only if $s$ is even with $\tau_{\vec{H}}$ containing precisely $\tfrac{s}{2}$ less white vertices and $\tfrac{s}{2}$ less black vertices than $\tau_{\vec{G}}$. Reconciling the results above with those obtained in sections~\ref{sec:smoothquadtile}, \ref{sec:bivalentandpinched}, \ref{sec:gridsandchains} and \ref{sec:compmoveIIandIII} implies that any admissible quadrilateral brane tiling must be smooth and contain an equal number of white and black vertices. 

%%%%%%%%%%%%%%    
\subsection{Seiberg duality}
\label{sec:seiberg}

\begin{figure}[h!]
\includegraphics[scale=1.2]{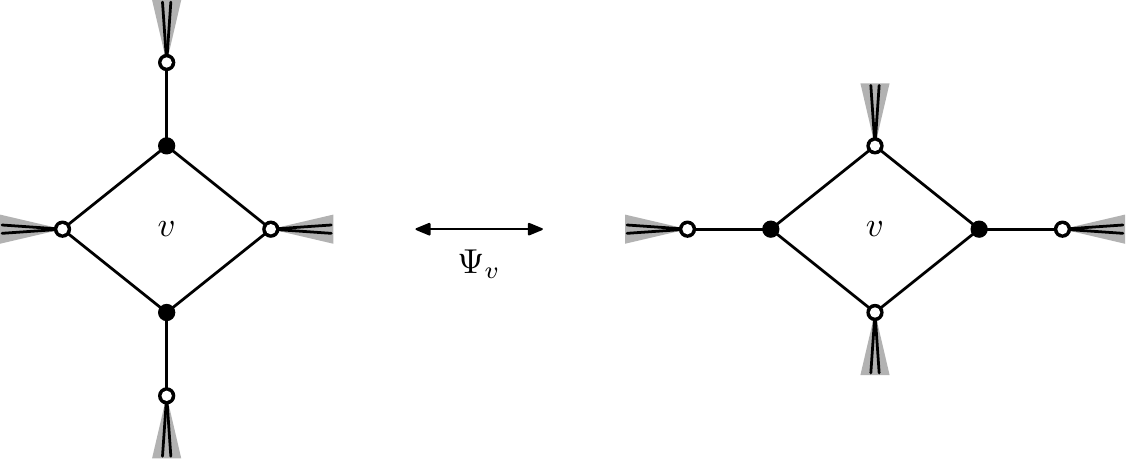}
\caption{Seiberg duality $\Psi_v$ mapping between $\tau_{\vec{G}}$ and $\tau_{\vec{G}^\prime}$ in $\eQ^{[t]}$.}
\label{fig16}
\end{figure}

Consider a tiling $\tau_{\vec{G}} \in \eQ^{[t]}$ encoding $\vec{G} \in {\vec{\fF}}_2^{[t]}$ which contains a face $v$ that encodes a vertex $v$ in $\vec{G}$ on which the mutation $\mu_v$ shown in figure~\ref{fig6a} can be applied to give $\vec{G}^\prime \in {\vec{\fF}}_2^{[t]}$. Since each arrow connected to $v$ ends on a different vertex in $\vec{G}$, the face $v$ must be adjacent to a different face in $\tau_{\vec{G}}$ on each of its four boundary edges. Furthermore, incoming and outgoing arrows at $v$ pair up in figure~\ref{fig6a} such that each pair forms part of a different circuit containing three arrows in $\vec{G}$. This means that adjacent boundary edges of face $v$ in $\tau_{\vec{G}}$ must also pair up such that each pair is adjacent on a trivalent corner vertex of $v$. The third edge attached to each trivalent vertex encodes the third arrow in each of the aforementioned circuits. These two trivalent corner vertices must therefore have the same colour. Thus face $v$ must be either isolated or a chain (with link vertex having the opposite colour to the two trivalent ones). In either case, no corner vertex with the opposite colour to the aforementioned trivalent ones can be trivalent or else it would necessarily encode a circuit in $\vec{G}$ that is forbidden in figure~\ref{fig6a}. An analogous treatment for face $v$ in $\tau_{\vec{G}^\prime} \in \eQ^{[t]}$ encoding $\vec{G}^\prime = \mu_v ( \vec{G} ) \in {\vec{\fF}}_2^{[t]}$ defines how the Seiberg duality map $\Psi_v$ applied to face $v$ relates $\tau_{\vec{G}}$ and $\tau_{\vec{G}^\prime}$. This is shown in figure~\ref{fig16} for the case where $v$ is isolated and its two trivalent corner vertices are coloured black. Each shaded wedge in figure~\ref{fig16} contains at least two edges. 
 
\begin{figure}[h!]
\includegraphics[scale=1.2]{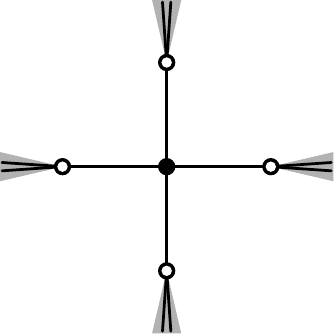}
\caption{Tiling $\tau_{\vec{H}} \in \eQ^{[t-1]}$ from which $v$ can be smoothly reconstructed in both $\tau_{\vec{G}}$ and $\tau_{\vec{G}^\prime}$.}
\label{fig17}
\end{figure}

If $v$ is isolated in $\tau_{\vec{G}}$, it can be collapsed in both ways to define a new quadrilateral tiling but only the collapse of $v$ which identifies its two trivalent corner vertices is smooth. If $v$ is a chain in $\tau_{\vec{G}}$, the unique collapse of $v$ must also identify its two trivalent corner vertices and this collapse is necessarily smooth since $v$ must be adjacent to different faces in $\tau_{\vec{G}}$ on each of its four boundary edges. In either case, the tiling $\tau_{\vec{H}} \in \eQ^{[t-1]}$ obtained from collapsing face $v$ in $\tau_{\vec{G}}$ encodes precisely the element $\vec{H} \in {\vec{\fF}}_2^{[t-1]}$ obtained from splitting vertex $v$ in $\vec{G}$ in the manner depicted in figure~\ref{fig6b}. The identified vertex is always $4$-valent in $\tau_{\vec{H}}$. An analogous treatment for face $v$ in $\tau_{\vec{G}^\prime} \in \eQ^{[t]}$ reveals that $v$ can only be smoothly collapsed in exactly one way to again give $\tau_{\vec{H}} \in \eQ^{[t-1]}$. The smooth collapse of $v$ in $\tau_{\vec{G}^\prime}$ must also identify a pair of trivalent vertices with the same colour as the pair which are identified in the smooth collapse of $v$ in $\tau_{\vec{G}}$. The result of this smooth collapse of $v$ for the example in figure~\ref{fig16} is shown in figure~\ref{fig17}.  
  
The map $\Psi_v$ between between $\tau_{\vec{G}}$ and $\tau_{\vec{G}^\prime}$ is therefore equivalent to the composite operation within $\eQ$ formed by smoothly collapsing face $v$ in $\tau_{\vec{G}}$ (or $\tau_{\vec{G}^\prime}$) to give $\tau_{\vec{H}}$ then smoothly reconstructing face $v$ in $\tau_{\vec{G}^\prime}$ (or $\tau_{\vec{G}}$). Rather than \lq urban renewal', this composite realisation of $\Psi_v$ is reminiscent of an \lq origami uranai' children's toy! Of course, as prescribed in \cite{Franco:2005rj}, $\Psi_v$ also follows by applying the \lq urban renewal' map to $v$ and then integrating out the massive fields associated with the two black bivalent vertices it produces.  

Let us label the four different faces in $\tau_{\vec{H}}$ around the central black vertex in figure~\ref{fig17} $i$ (south-west), $j$ (north-east), $k$ (north-west) and $l$ (south-east), which become the faces adjacent to $v$ in both $\tau_{\vec{G}}$ and $\tau_{\vec{G}^\prime}$ in figure~\ref{fig16}. The two superpotentials for $\tau_{\vec{G}}$ and $\tau_{\vec{G}^\prime}$ are of the form
\begin{align}\label{eq:supseibergdual}
W_{\tau_{\vec{G}}} &= - X_{jk} X_j X_k - X_{il} X_i X_l + X_k X_i \, M(X) + X_l X_j  \, N(X) + X_{jk} \, P(X) + X_{il} \, Q(X) + W(X) \nonumber \\
W_{\tau_{\vec{G}^\prime}} &=  - X^\prime_{ki} X^\prime_k X^\prime_i - X^\prime_{lj} X^\prime_l X^\prime_j + X^\prime_{ki} \, M( X^\prime ) + X^\prime_{lj} \, N( X^\prime ) + X^\prime_j X^\prime_k  \, P( X^\prime ) + X^\prime_i X^\prime_l \, Q( X^\prime ) + W( X^\prime )~,
\end{align}
where $X_i := X_{vi}$, $X_j := X_{vj}$, $X_k := X_{kv}$, $X_l := X_{lv}$ for $\tau_{\vec{G}}$ and $X^\prime_i := X^\prime_{iv}$, $X^\prime_j := X^\prime_{jv}$, $X^\prime_k := X^\prime_{vk}$, $X^\prime_l := X^\prime_{vl}$ for $\tau_{\vec{G}^\prime}$. The matter fields $X_{jk}$ and $X_{il}$ are associated with the third edge connected to each of the two black trivalent vertices in $\tau_{\vec{G}}$ while $X^\prime_{ki}$ and $X^\prime_{lj}$ are associated with the third edge connected to each of the two black trivalent vertices in $\tau_{\vec{G}^\prime}$. The monomials $M$, $N$, $P$ and $Q$ contain only the matter fields corresponding to edges which $\tau_{\vec{G}}$ and $\tau_{\vec{G}^\prime}$ have in common.

Consider now the consistency conditions \eqref{eq:superpotentialbeta} and \eqref{eq:gaugebetaquad} for both $\tau_{\vec{G}}$ and $\tau_{\vec{G}^\prime}$ and assume that the R-charges for all the matter fields corresponding to edges which $\tau_{\vec{G}}$ and $\tau_{\vec{G}^\prime}$ have in common are identical. Condition \eqref{eq:gaugebetaquad} for face $v$ in $\tau_{\vec{G}}$ is $R_i + R_j + R_k + R_l =2$ and substituting this into condition \eqref{eq:superpotentialbeta} for the first two terms in $W_{\tau_{\vec{G}}}$ in \eqref{eq:supseibergdual} implies $R_{jk} = R_i + R_l$ and $R_{il} = R_j + R_k$. Substituting these expressions into condition \eqref{eq:superpotentialbeta} for the next four terms in $W_{\tau_{\vec{G}}}$ in \eqref{eq:supseibergdual} implies $R_M = R_j + R_l$, $R_N = R_i + R_k$, $R_P = R_j + R_k$ and $R_Q = R_i + R_l$. On the other hand, condition \eqref{eq:gaugebetaquad} for face $v$ in $\tau_{\vec{G}^\prime}$ is $R^\prime_i + R^\prime_j + R^\prime_k + R^\prime_l =2$ and substituting this into condition \eqref{eq:superpotentialbeta} for the first two terms in $W_{\tau_{\vec{G}^\prime}}$ in \eqref{eq:supseibergdual} implies $R^\prime_{ki} = R^\prime_j + R^\prime_l$ and $R^\prime_{lj} = R^\prime_i + R^\prime_k$. Substituting these expressions into condition \eqref{eq:superpotentialbeta} for the next four terms in $W_{\tau_{\vec{G}^\prime}}$ in \eqref{eq:supseibergdual} implies $R_M = R^\prime_k + R^\prime_i$, $R_N = R^\prime_l + R^\prime_j$, $R_P = R^\prime_i + R^\prime_l$ and $R_Q = R^\prime_j + R^\prime_k$. The identification $R_i = R^\prime_j$, $R_j = R^\prime_i$, $R_k = R^\prime_l$ and $R_l = R^\prime_k$ establishes a bijection between the conditions \eqref{eq:superpotentialbeta} and \eqref{eq:gaugebetaquad} for $\tau_{\vec{G}}$ and $\tau_{\vec{G}^\prime}$ (with the remaining unchecked condition \eqref{eq:gaugebetaquad} for faces $i$, $j$, $k$ and $l$ automatically identified in $\tau_{\vec{G}}$ and $\tau_{\vec{G}^\prime}$ by this assignment). Given an admissible solution for $\tau_{\vec{G}}$, making this identification defines an admissible solution for $\tau_{\vec{G}^\prime}$ with $R^\prime_{ki} = R_k + R_i$ and $R^\prime_{lj} = R_l + R_j$. Conversely, given an admissible solution for $\tau_{\vec{G}^\prime}$, the identification defines an admissible solution for $\tau_{\vec{G}}$ with $R_{jk} = R^\prime_j + R^\prime_k$ and $R_{il} = R^\prime_i + R^\prime_l$. This proves that $\Psi_v$ can never map between admissible and inadmissible smooth quadrilateral tilings. Consequently, any two smooth quadrilateral tilings which are related via this elementary Seiberg duality must both be either admissible or inadmissible. This should be contrasted with the situation described in section~\ref{sec:compmoveIIandIII} where the manifestation of Seiberg duality involving composite move II never maps to a consistent tiling, no matter whether or not the initial tiling was itself consistent.  

As a final remark, consider $\Psi_v$ as shown in figure~\ref{fig16} such that $v$ is an isolated face in $\tau_{\vec{G}}$ with one of its black trivalent corner vertices connected by an edge to a white trivalent vertex. Face $v$ in $\tau_{\vec{G}^\prime}$ then has two black trivalent corner vertices and one white corner vertex that is exactly $4$-valent. From section~\ref{sec:forbiddenfaces}, any such $\tau_{\vec{G}}$ must be inconsistent and so the remarks above imply that $\tau_{\vec{G}^\prime}$ is also inconsistent. We therefore conclude that an isolated face containing two trivalent corner vertices with the same colour and one other $4$-valent corner vertex is always inconsistent.

%%%%%%%%%%%%%%    
\subsection{Examples}
\label{sec:examples}

The classification of brane tilings was initiated in \cite{Davey:2009bp} and contains a useful catalogue listing all the non-isomorphic tilings with $t \leq 6$ having an equal number of white and black vertices and no bivalent vertices. Amongst the smooth quadrilateral tilings in $\eQ$, the catalogue in \cite{Davey:2009bp} lists one tiling with $t=2$ (catalogue \#$(1.2)$), three tilings with $t=4$ (catalogue \#$(2.4){-}(2.6)$) and twelve tilings with $t=6$ (catalogue \#$(3.26){-}(3.37)$). 

Tiling \#$(1.2)$ corresponds to the unique tiling $\tau_+$ which encodes $\vec{G}_+$, depicted in the first row of figure~\ref{fig5}. Tiling \#$(2.4)$ corresponds to the unique tiling $\tau_{\vec{A}_4}$ (see figure~\ref{fig10} with $p=2$) which encodes the necklace digraph on four vertices, depicted in the first entry in the third row of figure~\ref{fig5}. Tilings \#$(2.5)$ and \#$(2.6)$ encode the digraphs shown in the second and fourth entries in the third row of figure~\ref{fig5}. The remaining two loopless $2$-regular eulerian digraphs on four vertices in the third and fifth entries in the third row of figure~\ref{fig5} can be encoded by quadrilateral tilings but they cannot be smooth. The fifth entry in the third row of figure~\ref{fig5} has all four vertices of the form shown in the third scenario in the second row of figure~\ref{fig3}, whence the existence of a smooth quadrilateral tiling for it is excluded by the discussion in section~\ref{sec:surjectivity}. The three consistent tilings \#$(1.2)$, \#$(2.4)$ and \#$(2.5)$ all have square faces in the isoradial embedding. The tiling \#$(2.6)$ is inconsistent, as expected since it contains a chain whose link is not $4$-valent (see section~\ref{sec:gridsandchains}) and also two trivalent vertices connected by an edge (see section~\ref{sec:forbiddenfaces}).

Tiling \#$(3.26)$ corresponds to the unique tiling $\tau_{\vec{A}_6}$ (see figure~\ref{fig10} with $p=3$)  which encodes the necklace digraph on six vertices. There exist precisely four distinct eulerian orientations on the complete tripartite graph on six vertices and each one is encoded by a different tiling \#$(3.28)$, \#$(3.34)$, \#$(3.36)$ and \#$(3.37)$. Tiling \#$(3.37)$ can be obtained by applying composite move II to a face in tiling \#$(1.2)$. As expected from section~\ref{sec:compmoveIIandIII}, the Higgs branches for tilings \#$(3.37)$ and \#$(1.2)$ are isomorphic to the conifold $\cM_+$. The Seiberg duality $\Psi_v$ described in section~\ref{sec:seiberg} can only be applied to vertices $v=3,4$ in tiling \#$(3.32)$ or vertex $v=2$ in tiling \#$(3.34)$. Identifying $\tau_{\vec{G}}$ with tiling \#$(3.32)$ and applying $\Psi_v$ to vertex $v=4$ then $\tau_{\vec{G}^\prime} = \Psi_v ( \tau_{\vec{G}} )$ is isomorphic to tiling \#$(3.34)$. As expected, these Seiberg-dual tilings have $\cM_{\vec{G}} \cong \cM_{\vec{G}^\prime} ( \cong  {\sf{C}}( L^{222} ) )$. The two tilings \#$(3.33)$ and \#$(3.35)$ are not isomorphic but encode the same loopless $2$-regular eulerian digraph on six vertices and are related via the construction described in section~\ref{sec:injectivity}. As expected from sections~\ref{sec:gridsandchains} and \ref{sec:forbiddenfaces}, the tilings \#$(3.29){-}(3.37)$ are all inconsistent since each one contains either a chain whose link is not $4$-valent or two trivalent vertices connected by an edge. In the isoradial embedding, all faces in the consistent tilings \#$(3.26)$ and \#$(3.27)$ are squares while in \#$(3.28)$ they are cyclic kites with an apex angle $\tfrac{\pi}{3}$.

%%%%%%%%%%%%%%    
\subsection{Maximal quadrilateral brane tilings}
\label{sec:maxquad}

In an isoradial embedding, each face in $\tau_{\vec{G}}$ is represented by a cyclic quadrilateral and so angles at opposite corners of each face are always supplementary (this is proposition 22 in book III of Euclid's {\emph{Elements}}). If $\tau_{\vec{G}}$ is maximal, in the sense defined in section~\ref{sec:amax}, each face must be bounded by two edges with R-charge $\tfrac{2}{p}$ that are connected to one corner vertex with degree $p$ and two edges with R-charge $\tfrac{2}{q}$ that are connected to another corner vertex with degree $q$. Both vertices must have the same colour, corresponding to opposite corners of the face, describing interior angles $\tfrac{2\pi}{p}$ and $\tfrac{2\pi}{q}$. The two other vertices on the remaining opposite corners of the face must both describe the same interior angle $\tfrac{\pi}{p} + \tfrac{\pi}{q}$. These two pairs of angles are only supplementary if $\tfrac{1}{p} + \tfrac{1}{q} = \tfrac{1}{2}$, whence $(p,q)$ must be either $(3,6)$, $(4,4)$ or $(6,3)$. Each face in a maximal quadrilateral brane tiling must therefore be either a square or a cyclic kite with apex angle $\tfrac{\pi}{3}$. Furthermore, it is clearly impossible to have a mixture of these two shapes in a maximal tiling. We therefore conclude that any maximal quadrilateral brane tiling must be based on a colouring of faces and a bipartite colouring of vertices in either a regular square tiling [4.4.4.4] or a deltoidal trihexagonal tiling [3.4.6.4] obtained by tessellating the respective shapes in the obvious way. The exact R-charge for each edge in any square tiling is $\tfrac{1}{2}$ and a generating function for the enumeration of square brane tilings was obtained in \cite{Hanany:2010cx}. In any deltoidal trihexagonal tiling, the exact R-charge is $\tfrac{2}{3}$ for edges ending on a trivalent vertex and $\tfrac{1}{3}$ for edges ending on a $6$-valent vertex. All the consistent quadrilateral brane tilings with less than eight vertices are also maximal \cite{Davey:2009bp} but there are consistent tilings with eight or more vertices that are not maximal.

%%%%%%%%%%%%%%    
\subsection{Admissible quadrilateral brane tilings}
\label{sec:admissiblequadbranetilings}

Let us now draw together the results from this section and those from section~\ref{sec:smoothquadtile} to describe some generic features of the class of admissible quadrilateral brane tilings. Any admissible quadrilateral brane tiling must be smooth and contain an equal number of white and black vertices. Any admissible $\tau_{\vec{G}} \in \eQ^{[t+2]}$ that is not isomorphic to either $\tau_+$ when $t=0$ or $\tau_{\vec{A}_{2p}}$ when $t=2p >0$ contains only isolated faces. Furthermore, neither $\tau_{\vec{G}}$ nor any Seiberg-dual tiling $\Psi_v ( \tau_{\vec{G}} ) \in \eQ^{[t+2]}$ can contain a pair of trivalent vertices connected by an edge. 

An isolated face $v$ in $\tau_{\vec{G}}$ can always be smoothly collapsed but the resulting tiling in $\eQ^{[t+1]}$ is not admissible. If it is possible to smoothly collapse a face $w$ in this inadmissible tiling to produce an admissible tiling $\tau_{\vec{H}} \in \eQ^{[t]}$, there must exist zigzag cycles in $\tau_{\vec{H}}$ and $\tau_{\vec{G}}$ which define the same zigzag walk with a repeated edge in the intermediate inadmissible tiling. Let $a$, $b$, $f$ and $g$ denote the boundary edges of $v$ in $\tau_{\vec{G}}$ appearing in anticlockwise order $(a {\circ} f {\bullet} b {\circ} g {\bullet} )$ around $v$ and assume its black corner vertices are to be identified in the smooth collapse (wherein $a$ is identified with $f$ and $b$ is identified with $g$). The identified edges are denoted by $a^\prime = a = f$ and $b^\prime = b = g$ in the intermediate tiling. Since the collapse of $v$ in $\tau_{\vec{G}}$ is smooth, neither of its white corner vertices can be trivalent. The zigzag cycles in $\tau_{\vec{G}}$ which contain an edge bounding $v$ are of the form $({\bullet} a {\circ} f {\bullet} ...)$, $({\circ} a {\bullet} g {\circ} ...)$, $({\bullet} b {\circ} g {\bullet} ...)$ and $({\circ} b {\bullet} f {\circ} ...)$ and they define a zigzag walk of the form $({\bullet} a^\prime {\circ} ... {\circ} b^\prime {\bullet} ... {\bullet} b^\prime {\circ} ...{\circ} a^\prime {\bullet} ...)$ in the intermediate tiling. Now let $\alpha$ and $\beta$ denote the pair of edges in $\tau_{\vec{H}}$ that are involved in the smooth reconstruction of face $w$ in the intermediate tiling and let $p$, $q$, $r$ and $s$ denote its boundary edges which appear in anticlockwise order $(p {\circ} s {\bullet} q {\circ} r {\bullet} )$ around $w$. If the identified corner vertices in the collapse of $w$ are white (with $\alpha = p = r$ and $\beta = q = s$), the zigzag cycles in $\tau_{\vec{H}}$ containing either $\alpha$ or $\beta$ define a zigzag walk of the form $({\bullet} p {\circ} s {\bullet} ... {\circ} q {\bullet} s {\circ} ... {\bullet} q {\circ} r {\bullet} ... {\circ} p {\bullet} r {\circ} ...)$ in the intermediate tiling. If the identified corner vertices in the collapse of $w$ are black (with $\alpha = q = r$ and $\beta = p = s$), the zigzag cycles in $\tau_{\vec{H}}$ containing either $\alpha$ or $\beta$ define a zigzag walk of the form $({\circ} q {\bullet} s {\circ} ... {\bullet} p {\circ} s {\bullet} ... {\circ} p {\bullet} r {\circ} ... {\bullet} q {\circ} r {\bullet} ...)$ in the intermediate tiling. In either case, it is impossible to have $\alpha$ and $\beta$ adjacent in a zigzag cycle in $\tau_{\vec{H}}$ since this would imply one of the corner vertices of $w$ in the intermediate tiling is bivalent which is impossible because it is smooth. For the zigzag walks in the intermediate tiling obtained from $\tau_{\vec{G}}$ and $\tau_{\vec{H}}$ above to agree, there must be an identification of their repeated edges. In particular $a^\prime$ and $b^\prime$ from $\tau_{\vec{G}}$ must be identified with opposite sides of face $w$ in the intermediate tiling. Consequently, the isolated faces $v$ and $w$ in $\tau_{\vec{G}}$ must be adjacent on their opposite sides. Therefore face $w$ in the intermediate tiling must be a chain whose link corresponds to the black corner vertex identified in the smooth collapse of $v$ in $\tau_{\vec{G}}$. The smooth collapse of $w$ in the intermediate tiling is therefore unique and must involve the identification of its white corner vertices to give $\tau_{\vec{H}}$. Whence $\alpha$ and $\beta$ must describe a non-contractible cycle on the torus. Since $\alpha$ and $\beta$ are not adjacent in any zigzag cycle in $\tau_{\vec{H}}$, the isolated face $v$ in $\tau_{\vec{G}}$ contains no trivalent corner vertices. 

The reverse operation can be applied to a pair of edges $\alpha$ and $\beta$ describing a non-contractible cycle on the torus in any admissible $\tau_{\vec{H}} \in \eQ^{[t]}$ which contains only isolated faces. The edges $\alpha$ and $\beta$ cannot be adjacent in any zigzag cycle in $\tau_{\vec{H}}$ or else it must contain a chain bounded by $\alpha$ and $\beta$ whose link is not $4$-valent, contradicting the assumption that $\tau_{\vec{H}}$ is admissible. The operation proceeds by first selecting a vertex of a given colour on which both $\alpha$ and $\beta$ end in $\tau_{\vec{H}}$ to define the smooth reconstruction of chain $w$ in the intermediate tiling. One then selects the link of $w$ and a pair of edges $f$ and $g$ on its opposite sides to define the smooth reconstruction of face $v$ in the admissible tiling $\tau_{\vec{G}} \in \eQ^{[t+2]}$. The two new vertices in $\tau_{\vec{G}}$ have opposite colours and the two new faces $v$ and $w$ are both isolated in $\tau_{\vec{G}}$ but adjacent to each other on edges $f$ and $g$ on their opposite sides. A cycle drawn on the torus which intersects only the new edges $f$ and $g$ in $\tau_{\vec{G}}$ is therefore homologous to the one formed by $\alpha$ and $\beta$ in $\tau_{\vec{H}}$. Let us label $a$ and $b$ the remaining opposite sides of $v$ and label $c$ and $d$ the remaining opposite sides of $w$ in $\tau_{\vec{G}}$ such that they appear in anticlockwise order $(a {\circ} f {\bullet} b {\circ} g {\bullet} )$ around $v$ and $(c {\bullet} f {\circ} d {\bullet} g {\circ} )$ around $w$.  Zigzag cycles not involving $\alpha$ or $\beta$ in $\tau_{\vec{H}}$ are the same in $\tau_{\vec{G}}$ while those of the form $({\bullet} \alpha {\circ} ...)$, $({\circ} \alpha {\bullet} ...)$, $({\bullet} \beta {\circ} ...)$ or $({\circ} \beta {\bullet} ...)$ in $\tau_{\vec{H}}$ become $({\bullet} a {\circ} f {\bullet} c {\circ} ...)$, $({\circ} a {\bullet} g {\circ} c {\bullet} ...)$, $({\bullet} b {\circ} g {\bullet} d {\circ} ...)$ or $({\circ} b {\bullet} f {\circ} d {\bullet} ...)$ in $\tau_{\vec{G}}$. Let us refer to the operation described above as {\emph{admissible move I}}. Admissible move I is the only move which maps $\eQ^{[t]} \rightarrow \eQ^{[t+2]}$ within the class of admissible quadrilateral brane tilings containing only isolated faces. If $\tau_{\vec{H}}$ is isomorphic to $\tau_{\vec{A}_{2p}}$, composite move I maps $\eQ^{[t]} \rightarrow \eQ^{[t+2]}$ to define the consistent tiling $\tau_{\vec{A}_{2p+2}}$. Every other move within the class of admissible quadrilateral brane tilings must add more than two new vertices to the tiling. The only tiling to which both the aforementioned moves can be applied is $\tau_+$. Applying composite move I to $\tau_+$ gives tiling \#$(2.4)$ while applying admissible move I to $\tau_+$ gives tiling \#$(2.5)$ in \cite{Davey:2009bp}. Applying composite move I to tiling \#$(2.4)$ gives tiling \#$(3.26)$ while applying admissible move I to tiling \#$(2.5)$ gives tiling \#$(3.27)$ in \cite{Davey:2009bp}. One can iterate admissible move I on $\tau_+$ to obtain any square tiling in $\eQ^{[2s]}$ that follows by tessellating a $2{\times} s$ rectangle. At each iteration, the cycle formed by $\alpha$ and $\beta$ is identified with a pair of adjacent edges traversing the side of the rectangle with length $2$. One can also apply admissible move I to the deltoidal trihexagonal tiling \#$(3.28)$ in \cite{Davey:2009bp} to obtain an admissible quadrilateral brane tiling with eight vertices that is not maximal. 

As discussed in section~\ref{sec:vacuummodulispace}, if $\tau_{\vec{H}} \in \eQ^{[t]}$ is consistent, the area $A ( \Delta_{\tau_{\vec{H}}} )$ of its toric diagram $\Delta_{\tau_{\vec{H}}}$ must equal $\tfrac{t}{2}$. The effect of admissible move I on the toric diagram $\Delta_{\tau_{\vec{H}}}$ can be derived from its effect on zigzag cycles in $\tau_{\vec{H}}$. Relative to a basis for $H_1 ( T^2 , \ZZ )$ in which the non-contractible cycle $(\alpha {\circ} \beta {\bullet} )$ represents homology class $(1,0)$, the zigzag cycles $({\bullet} \alpha {\circ} ...)$, $({\circ} \alpha {\bullet} ...)$, $({\bullet} \beta {\circ} ...)$ and $({\circ} \beta {\bullet} ...)$ in $\tau_{\vec{H}}$ can be taken to represent classes $(p^\prime ,1)$, $(p,1)$, $(- p^\prime ,-1)$ and $(-p,-1)$ respectively, for some $p, p^\prime \in \ZZ$. Any other zigzag cycle in $\tau_{\vec{H}}$ which does not contain $\alpha$ or $\beta$ must represent a class $( p^{\prime\prime},0)$, for some $p^{\prime\prime} \in \ZZ$ (with $p^{\prime\prime} =0$ only if there are no other zigzag cycles of this form). Identifying these classes with outward-pointing normal vectors for the boundary edges in $\Delta_{\tau_{\vec{H}}}$ implies the toric diagram can be represented as the convex hull $\Delta_{\tau_{\vec{H}}}  = {\mbox{Conv}} ( (0,0) , (1, - p^\prime ) ,  (2, - p^\prime -p ) , (0, p^{\prime\prime} ) , (1, p^{\prime\prime} -p ) ,  (2, p^{\prime\prime} - p^\prime -p )  )$, where $p \leq p^\prime$ and $p^{\prime\prime} \geq 0$ can be assumed without loss of generality. Therefore $\Delta_{\tau_{\vec{H}}}$ must contain precisely $p^{\prime\prime} + p^\prime -p-1$ interior points and $2( p^{\prime\prime} +2)$ lattice points on its boundary. Recall from Pick's theorem that any simple lattice polygon with $I$ interior points and $B$ boundary points has area $A = I + \tfrac{B}{2} -1$.  Whence, $A ( \Delta_{\tau_{\vec{H}}} ) = 2 p^{\prime\prime} + p^\prime -p = \tfrac{t}{2}$. Relative to the basis in homology we have chosen, the zigzag cycles $({\bullet} a {\circ} f {\bullet} c {\circ} ...)$, $({\circ} a {\bullet} g {\circ} c {\bullet} ...)$, $({\bullet} b {\circ} g {\bullet} d {\circ} ...)$ and $({\circ} b {\bullet} f {\circ} d {\bullet} ...)$ in $\tau_{\vec{G}}$ following admissible move I must represent classes $(p^\prime ,1)$, $(p-1,1)$, $(- p^\prime ,-1)$ and $(-p+1,-1)$ respectively. Any other zigzag cycle in $\tau_{\vec{G}}$ which does not contain $a$, $b$, $c$, $d$, $f$ or $g$ corresponds to a zigzag cycle in $\tau_{\vec{H}}$ which does not contain $\alpha$ or $\beta$, and so must represent the same homology class. The net effect of admissible move I on the toric diagram is therefore to decrease $p$ by one while leaving $p^\prime$ and $p^{\prime\prime}$ unchanged. Thus $A ( \Delta_{\tau_{\vec{G}}} ) = A ( \Delta_{\tau_{\vec{H}}} ) +1 = \tfrac{t+2}{2}$, as expected if $\tau_{\vec{G}} \in \eQ^{[t+2]}$ is also consistent. Furthermore, this increment is accounted for by $\Delta_{\tau_{\vec{G}}}$ containing precisely one more interior point than $\Delta_{\tau_{\vec{H}}}$ (whilst having the same number of lattice points on their boundaries). As was observed in section~\ref{sec:necklace}, the other possibility for increasing the area of the toric diagram by one unit via the addition of two boundary points is realised by composite move I acting on a consistent tiling. Any square tiling in $\eQ^{[2s]}$ that follows by tessellating a $2{\times} s$ rectangle has $p = 1-s$, $p^\prime = 1$ and $p^{\prime\prime} =0$ in the construction above and describes the toric diagram for $ {\sf{C}}( Y^{s0} )$ \cite{Benvenuti:2004dy,Franco:2005sm}. 

\begin{figure}[h!]
\includegraphics[scale=1.2]{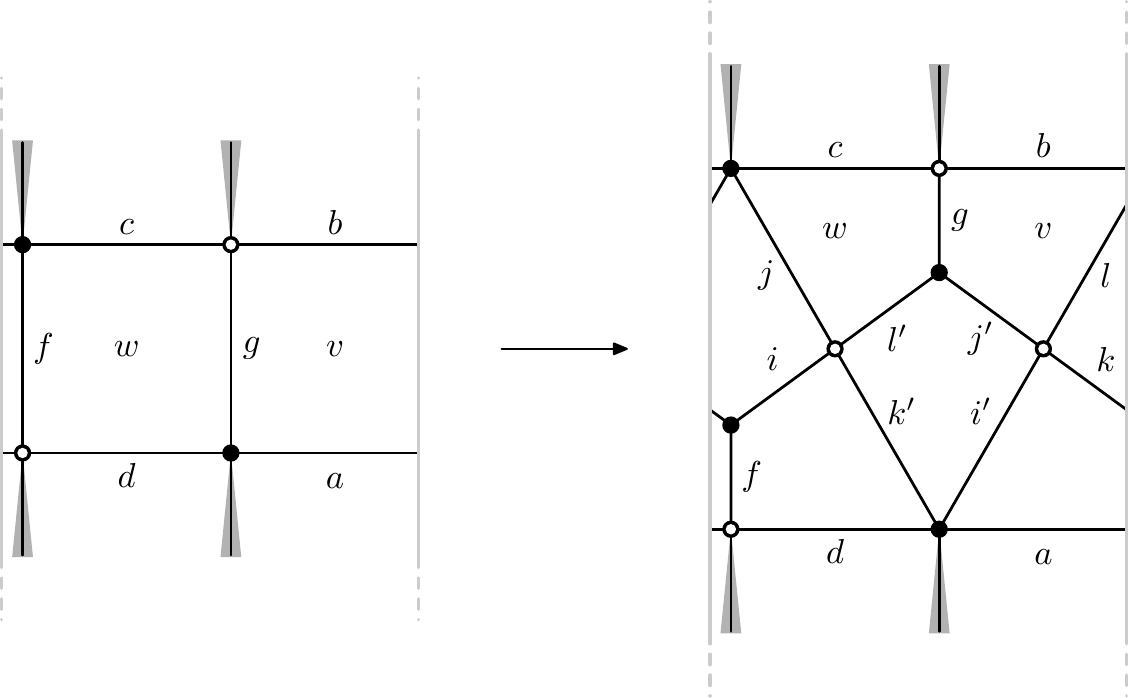}
\caption{Admissible move II mapping $\tau_{\vec{G}} \in \eQ^{[t+2]}$ to $\tau_{\vec{F}} \in \eQ^{[t+6]}$.}
\label{fig18}
\end{figure}

Let us take the construction above one step further. Consider an admissible tiling $\tau_{\vec{G}} \in \eQ^{[t+2]}$ which contains a pair of faces $v$ and $w$ that are adjacent on their opposite sides, with boundary edges labelled in anticlockwise order $(a {\circ} f {\bullet} b {\circ} g {\bullet} )$ around $v$ and $(c {\bullet} f {\circ} d {\bullet} g {\circ} )$ around $w$. By representing $v$ and $w$ locally as squares, a new tiling $\tau_{\vec{F}} \in \eQ^{[t+6]}$ can be obtained by replacing these two squares in $\tau_{\vec{G}}$ with six cyclic kites in $\tau_{\vec{F}}$ as shown in figure~\ref{fig18}. The vertical grey lines in figure~\ref{fig18} represent a periodic identification in the horizontal direction while $i$, $j$, $k$, $l$, $i^\prime$, $j^\prime$, $k^\prime$ and $l^\prime$ label the eight new edges in $\tau_{\vec{F}}$. This replacement will be referred to as {\emph{admissible move II}}.  
Admissible move II also follows from a particular sequence of smooth reconstructions to define the four new faces in $\tau_{\vec{F}}$. The first two smooth reconstructions correspond to the composite move described in section~\ref{sec:injectivity}, applied to edges $a$ and $d$ in $\tau_{\vec{G}}$. This defines an intermediate tiling in $\eQ^{[t+4]}$ which is not admissible since it contains two new chains whose links are not $4$-valent. By selecting pairs of edges which end on each of these two links and which do not bound a face in the intermediate tiling, one defines two more smooth reconstructions from which $\tau_{\vec{F}} \in \eQ^{[t+6]}$ follows with all four new faces isolated as in figure~\ref{fig18}. Zigzag cycles not involving $a$, $b$, $c$, $d$, $f$ or $g$ in $\tau_{\vec{G}}$ are the same in $\tau_{\vec{F}}$ while those of the form $({\bullet} a {\circ} f {\bullet} c {\circ} ...)$, $({\circ} a {\bullet} g {\circ} c {\bullet} ...)$, $({\bullet} b {\circ} g {\bullet} d {\circ} ...)$ or $({\circ} b {\bullet} f {\circ} d {\bullet} ...)$ in $\tau_{\vec{G}}$ become $({\bullet} a {\circ} f {\bullet} i {\circ} j {\bullet} c {\circ} ...)$, $({\circ} a {\bullet} i^\prime {\circ} j^\prime {\bullet} g {\circ} c {\bullet} ...)$, $({\bullet} b {\circ} g {\bullet} l^\prime {\circ} k^\prime {\bullet} d {\circ} ...)$ or $({\circ} b {\bullet} l {\circ} k {\bullet} f {\circ} d {\bullet} ...)$ in $\tau_{\vec{F}}$. These zigzag cycles represent the same homology classes in both $\tau_{\vec{G}}$ and $\tau_{\vec{F}}$. The only remaining zigzag cycles in $\tau_{\vec{F}}$ are $( {\circ} i {\bullet} k {\circ} i^\prime {\bullet} k^\prime )$ and $( {\circ} l {\bullet} j {\circ} l^\prime {\bullet} j^\prime )$ which represent homology classes $(-1,0)$ and $(1,0)$ relative to the basis for $H_1 ( T^2 , \ZZ )$ that was defined in the paragraph above. Whence, $\tau_{\vec{F}} \in \eQ^{[t+6]}$ is also admissible. If $\tau_{\vec{G}}$ is consistent, the effect of admissible move II on the toric diagram $\Delta_{\tau_{\vec{G}}}$ is to increase $p^{\prime\prime}$ by one while leaving $p$ and $p^\prime$ unchanged. Thus $A ( \Delta_{\tau_{\vec{F}}} ) = A ( \Delta_{\tau_{\vec{G}}} ) +2 = \tfrac{t+6}{2}$, as expected if $\tau_{\vec{F}} \in \eQ^{[t+6]}$ is also consistent. Furthermore, $\Delta_{\tau_{\vec{F}}}$ must contain precisely one more interior point and two more lattice points on its boundary than $\Delta_{\tau_{\vec{G}}}$. Applying admissible move II to $\tau_+$ gives the deltoidal trihexagonal tiling \#$(3.28)$ in \cite{Davey:2009bp}. Relative to the basis we have chosen for $H_1 ( T^2 , \ZZ )$, $\Delta_{\tau_+} = {\mbox{Conv}} ( (0,0) , (1,0) , (1, -1) , (2,-1) )$ with $p=0=p^{\prime\prime}$ and $p^\prime =1$ and so $\Delta_{\tau_{\vec{F}}} = {\mbox{Conv}} ( (0,0) , (1,-1) , (2,-1) , (0,1) , (1,1) , (2,0) )$ with $p=0$ and $p^\prime =1= p^{\prime\prime}$, which is the toric diagram for the complex cone over the del Pezzo surface obtained by blowing up three generic points in $\CC P^2$. 

%%%%%%%%%%%%%%%%%
 
\section*{Acknowledgments}

I would like to thank Amihay Hanany, Noppadol Mekareeya, Rak-Kyeong Seong and Giuseppe Torri for helpful discussions. This research was supported in part by grant ST/G000514/1 ``String Theory Scotland'' from the UK Science and Technology Facilities Council. 

\bibliographystyle{utphys}
\bibliography{Qtile}

\end{document}